\documentclass[
reprint, 
prd, 
aps, 
nofootinbib,
showpacs, 
superscriptaddress,
longbibliography]{revtex4-2}

\usepackage{graphicx}
\usepackage[usenames,dvipsnames]{xcolor}
\usepackage[utf8]{inputenc}
\usepackage[T1]{fontenc}
\usepackage{anyfontsize}
\usepackage{adjustbox}
\usepackage{placeins}
\usepackage{lipsum}
\usepackage{latexsym}
\usepackage{rotating}
\usepackage{mathtools}
\usepackage{amsmath, amssymb, amsthm, amsfonts}
\usepackage{mathrsfs}
\usepackage{bm}
\usepackage{enumerate}
\usepackage{aas_macros}
\usepackage{soul}
\usepackage{natbib}
\usepackage{hyperref}
\usepackage{orcidlink}
\usepackage{verbatim}
\usepackage[normalem]{ulem}
\hypersetup{colorlinks=true,citecolor=NavyBlue,linkcolor=NavyBlue,urlcolor=NavyBlue}

\newcommand{\lambdatilde}{\ensuremath{\bar{\lambda}}}

\newcommand{\lambdazero}{\ensuremath{\bar{\lambda}^{(0)}_0}}
\newcommand{\lambdak}{\ensuremath{\bar{\lambda}^{(k)}_0}}

\newcommand{\mchirp}{\ensuremath{\mathcal{M}_c}}
\newcommand{\hubble}{\ensuremath{H_0}}
\newcommand{\dl}{\ensuremath{D_L}}

\newcommand{\msun}{\ensuremath{M_{\odot}}}

\newcommand{\mdet}{\ensuremath{m_{\text{det}}}}

\newcommand{\msource}{\ensuremath{m_{\text{source}}}}
\newcommand{\bomega}{\ensuremath{\mathbf{\Omega}}}
\newcommand{\lcdm}{\ensuremath{\Lambda\text{CDM}}}

\newcommand{\hubbleunit}{\ensuremath{\text{km s}^{-1}\;\text{Mpc}^{-1}}}
\newcommand{\megaparsec}{\ensuremath{\mathrm{Mpc}}}
\newcommand{\gigaparsec}{\ensuremath{\mathrm{Gpc}}}

\newcommand{\pbilby}{PARALLEL\_BILBY}
\newcommand{\dynesty}{DYNESTY}

\newcommand{\ICASU}{\affiliation{Illinois Center for Advanced Studies of the Universe, Department of Physics, University of Illinois at Urbana-Champaign, Urbana, Illinois 61801, USA}}
\newcommand{\CAPS}{\affiliation{Center for AstroPhysical Surveys, National Center for Supercomputing Applications, Urbana, Illinois 61801, USA}}
\newcommand{\LIGOLAB}{\affiliation{LIGO Laboratory and Kavli Institute for Astrophysics and Space Research,
Massachusetts Institute of Technology, 185 Albany Street, Cambridge, Massachussets 02139, USA}}
\newcommand{\PHYSUVA}{\affiliation{Department of Physics, University of Virginia, Charlottesville, Virginia 22904, USA}}

\newcommand{\UCHICAGO}{\affiliation{Department of Physics, Department of Astronomy and Astrophysics, Enrico Fermi Institute, and Kavli Institute for Cosmological Physics, University of Chicago, Chicago, Illinois 60637, USA}}

\begin{document}

\title{Breaking bad degeneracies with Love relations: \\ Improving gravitational-wave measurements 
through universal relations}

\author{Yiqi~Xie~\orcidlink{0000-0002-8172-577X}}
\ICASU

\author{Deep~Chatterjee~\orcidlink{0000-0003-0038-5468}}
\ICASU
\CAPS
\LIGOLAB

\author{Gilbert~Holder}
\ICASU

\author{Daniel~E.~Holz~\orcidlink{0000-0002-0175-5064}}
\UCHICAGO

\author{Scott~Perkins~\orcidlink{0000-0002-5910-3114}}
\ICASU
\CAPS

\author{Kent~Yagi~\orcidlink{0000-0002-0642-5363}}
\PHYSUVA

\author{Nicol\'as~Yunes~\orcidlink{0000-0001-6147-1736}}
\ICASU

\date{\today}

\begin{abstract}
The distance-inclination degeneracy limits gravitational-wave parameter estimation of compact binary mergers.
Although the degeneracy can be partially broken by including higher-order modes or precession, these effects are suppressed in binary neutron stars.
In this work, we implement a new parametrization of the tidal effects in the binary neutron-star waveform, exploiting the binary Love relations, that breaks the distance-inclination degeneracy.
The binary Love relations prescribe the tidal deformability of a neutron star as a function of its source-frame mass in an equation-of-state insensitive way and, thus, allows direct measurement of the redshift of the source. 
If the cosmological parameters are assumed to be known, the redshift can be converted to a luminosity distance, and the distance-inclination degeneracy can thus be broken.
We implement this new approach, studying a range of binary neutron-star observing scenarios using Bayesian parameter estimation on synthetic data.
In the era of the third-generation detectors, for observations with signal-to-noise ratios ranging from 6 to 167, we forecast up to an $\sim70\%$ decrease in the $90\%$ credible interval of the distance and inclination and up to an $\sim50\%$ decrease in that of the source-frame component masses. 
For edge-on systems, our approach can result in moderate ($\sim50\%$) improvement in the measurements of distance and inclination for binaries with
a signal-to-noise ratio as low as 10. 
This prescription can be used to better infer the source-frame masses and, hence, refine population properties of neutron stars, such as their maximum mass, impacting nuclear astrophysics. When combined with the search for electromagnetic counterpart observations, the work presented here can be used to put improved bounds on the opening angle of jets from binary neutron-star mergers.
\end{abstract}

\maketitle

\section{Introduction}\label{sec:introduction}
The field of gravitational-wave (GW) astronomy has seen great advances in the past decade. The 2015 discovery
of GWs from a binary black hole (BBH) merger, GW150914, marked a spectacular confirmation of general relativity~\cite{LIGOScientific:2016aoc}.
Since then, the number of detected compact binary coalescences (CBCs) has seen an exponential increase
with each new observing run of the advanced Laser Interferometer Gravitational-wave Observatory (LIGO)~\cite{advanced_ligo} and
the advanced Virgo observatory~\cite{advanced_virgo}. The latest catalog from the LIGO-Virgo Collaboration, GWTC-3,
reports 90 confirmed CBC events~\cite{LIGOScientific:2021djp}. In addition, independent analyses of the data have also been carried out
and reported by other groups~\cite{Nitz:2018imz,Magee:2019vmb,Zackay:2019tzo,Nitz:2020oeq,Venumadhav:2019lyq,Zackay:2019btq,Nitz:2021uxj,Nitz:2021zwj}. 
The trend will likely continue with future observing runs, as the existing GW
detectors are upgraded to design sensitivity, and as additional detectors, such as KAGRA~\cite{KAGRA:2018plz} and LIGO-India~\cite{Unnikrishnan:2013qwa},
are added to the global network.

The prospect of doing multimessenger astrophysics is one of the most exciting areas in the GW field. The detection of the first binary neutron-star (BNS) merger, 
GW170817~\cite{LIGOScientific:2017vwq}, along with the simultaneous observation of the short gamma-ray burst (GRB), GRB170817A~\cite{Goldstein:2017mmi,LIGOScientific:2017zic}, and the
kilonova, AT2017~gfo~\cite{Smartt:2017fuw}, had a rich science impact across several areas of physics. However, detecting such 
electromagnetic (EM) counterparts of GW sources is extremely challenging, with no success since GW170817. 
Among other science goals, the prospect of measuring 
cosmological parameters independent of the established probes, such as type Ia supernovae (SNe~Ia)~\cite{Riess:2021jrx} and
the cosmic microwave background (CMB)~\cite{Planck:2018vyg}, is one of the promises of GW multimessenger astronomy. 
Since GWs from CBCs are \emph{standard sirens}~\citep{schutz_1986,holz_hughes_2005}, they allow direct measurement of the luminosity distance.
When this is combined with an independent measurement of the cosmological redshift, either through bright sirens directly
from a counterpart~\cite{2006PhRvD..74f3006D,2017Natur.551...85A}, or statistically from dark sirens coupled with a galaxy catalog~\cite{schutz_1986, 2012PhRvD..85b3535T,Fishbach_2019,Soares_Santos_2019,Mukherjee:2020hyn,2020PhRvD.101l2001G}, or through spectral sirens exploiting properties of the GW population~\cite{1993ApJ...411L...5C,2012PhRvD..85b3535T,2022PhRvL.129f1102E}, it allows determination of the Hubble
constant \hubble. There is currently a $\geq 5\sigma$ tension in $\hubble$ between the
local-Universe SNe and the early-Universe CMB values~\cite{Riess:2021jrx,Freedman:2021ahq,Anand:2021sum}.

Finding a counterpart to a BNS merger leads to more constrained measurements of {\hubble} from GW data as compared to a statistical measurement. However, there are several detection uncertainties that impact the measurement. For example, the \emph{distance-inclination} degeneracy impacts the measurement of the distance to the source, because a face-on source at a farther distance produces a similar signal amplitude as an inclined source at closer distances. This directly affects the measured value of \hubble~\cite{Markovic:1993cr,Cutler_1994,Usman:2018imj}.
Although it is possible to break the distance-inclination degeneracy through extraction of the inclination angle from higher-order modes~\cite{London:2017bcn} or precession~\cite{Vitale:2018wlg}, these techniques have limited application to BNS, for which the higher-order modes are suppressed because the component masses are nearly equal~\cite{Varma:2014jxa} and the precession is suppressed because the spins are small compared with the orbital angular momentum~\cite{Kidder:1995zr}.

Other prescriptions to measure \hubble, not involving any EM information, have also been proposed in the literature~\cite{Fishbach_2019,Soares_Santos_2019}. In particular, \citet{Chatterjee:2021xrm} (hereafter C21) showed how to apply the \emph{binary Love} relations in merging neutron stars (NSs) to measure \hubble.
They use the technique proposed by ~\citet{messenger_read_2012} to extract source-frame masses from the tidal deformability of NSs, in combination with the binary Love relations discovered by 
~\citet{yagi_yunes_2016,yagi_yunes_2017}~(hereafter YY17), to construct a NS equation of state (EOS) insensitive
parametrization (see Ref.~\cite{Carson:2019rjx} for updated binary Love relations after GW170817). 
This parametrization can then be used to directly measure the redshift of the source from GW data. C21 also forecasted that combining the \hubble~measurements from BNS systems without electromagnetic counterparts could lead to $\sim 2\%$ measurement uncertainty in \hubble~in the era of the third-generation (3G) GW detectors.

Here, we report another application of the binary Love relations---to constrain the above-mentioned
distance-inclination measurement. In brief, this can be thought of as a corollary to the prescription mentioned
in C21. Instead of using the binary Love relations to measure \hubble, here we show a complementary
use case when {\hubble} is well constrained. In the traditional parametrization of a GW signal, the distance
is measured from the amplitude of the waveform~\cite{PhysRevD.49.1723,Cutler_1994,van_der_Sluys_2008,Veitch_2012,Veitch_2015}. In C21, it was shown that the redshift is measurable from the 
matter effect in the phase of the BNS inspiral. In the limiting case of fixing the value of \hubble, the phase
contribution of the matter terms also captures information about the distance. Hence, the distance enters
in both the amplitude and the phase of the GW signal, instead of only the amplitude.

We perform Bayesian parameter estimation on synthetic BNS signals and show that, in the 3G detector era, the use of the binary Love relations will significantly improve the GW parameter estimation by breaking the distance-inclination degeneracy. In particular, we forecast up to $\sim 70\%$ decrease in the 90\% credible interval (CI) of the estimated distance and inclination angle and up to $\sim 50\%$ decrease in that of the source-frame masses. Additionally, for edge-on systems, our approach will make it possible to put reasonable constraints on the distance and the inclination angle with signal-to-noise ratios (SNRs) as low as 10. 

In the remainder of this paper, we present the detailed calculations that lead to the conclusions discussed above. In Sec.~\ref{sec:blove}, we provide a brief review of the binary Love relations. 
In Sec.~\ref{sec:measure_with_blove}, we describe the parametrization and show how the distance appears in both the amplitude and phase of the GW signal. 
In Sec.~\ref{sec:compute_setup}, we describe our computational setup and show that the distance-inclination estimation will be improved using our approach in the era of 3G detectors. 
In Sec.~\ref{sec:improve_dliota}, we do a parameter sweep across systems and report the most promising systems for which the distance-inclination estimation will be improved. 
In Sec.~\ref{sec:improve_masses}, we report improvement in source-frame mass estimation using our approach. 
In Sec.~\ref{sec:robstness}, we show that the improvements are robust to relaxing the assumptions made by previous sections, such as the accuracy of the binary Love relations and the cosmology. We also show that the Fisher analysis is not applicable to our study. 
Finally, we conclude in Sec.~\ref{sec:conclusion}. Henceforth, we use geometric units in which $G=1=c$.

\section{Binary Love Relations}
\label{sec:blove}
The GWs emitted by the quasicircular inspiral of a compact binary can be described under the post-Newtonian (PN) formalism~\cite{Blanchet_1995}. In this scheme, the waveform is solved for in powers of the velocity, which can be related to the GW frequency through the PN version of Kepler's third law. At each PN order, the coefficients of the expansion are functions of the binary parameters, such as the component masses and spins of the compact objects.
For a BNS system, the tidal interaction between the component stars leaves distinctive imprints in the GW emission during the late inspiral phase. 
This effect enters the GW phase first at the 5PN order, leading to an earlier merger~\cite{2008PhRvD..77b1502F}.
The BNS parameters responsible for the tidal emission are the electric-type, quadrupolar tidal deformability of each NS, $\lambdatilde_A=(2k_{2,A}/3)C_A^{-5}$, where $C_A=M_A/R_A$ is the compactness of NS $A\;(A=1,2)$ in the binary, with mass $M_A$ and radius $R_A$, while $k_{2,A}$ is its relativistic Love number~\cite{Hinderer_2008}. 

If the NS EOS is known, the radius and the Love number (and the tidal deformability) of the NS can be solved for as functions of its mass. While calculating the correct EOS of NSs from first principles is difficult, there are certain EOS-insensitive relations have been derived among some NS observables, such as the moment of inertia, the quadruple moment, and the tidal deformability~\cite{Yagi_2013,Yagi_2013_PRD} (see also~\cite{Yagi:2016bkt,Doneva:2017jop,Yunes:2022ldq} for reviews). 
In the context of GW astrophysics, these imply EOS-insensitive binary Love relations, presented in YY17:
\begin{enumerate}
\item a relation between the symmetric and antisymmetric combination of the individual tidal deformabilities, $\lambdatilde_s=(\lambdatilde_1+\lambdatilde_2)/2$ and $\lambdatilde_a=(\lambdatilde_1-\lambdatilde_2)/2$;
\item a relation between the waveform tidal parameters $\bar{\Lambda}$ and $\delta\bar{\Lambda}$ appearing at 5PN and 6PN order, respectively; and
\item a relation between the coefficients of the Taylor expansion of the tidal deformability $\lambdatilde(M)$ about some mass $m_0$.
\end{enumerate}
Here, we are concerned with the third item in the list, which we will refer to as the $\lambdazero$--$\lambdak$ relation. 

The $\lambdazero$--$\lambdak$ relation is embedded in the following Taylor expansion of $\lambdatilde(M)$:
\begin{align}
    \lambdatilde(M)=\sum_{k=0}^\infty\frac{\lambdak}{k!}\left(1-\frac{M}{m_0}\right)^k,
    \label{eq:lambda_m}
\end{align}
where $\lambdak=(-1)^kM^k\,(d^k\lambdatilde/dM^k)$, evaluated at the reference mass $M=m_0$, are the coefficients of expansion.
The $\lambdazero$--$\lambdak$ relation states that each $\lambdak$ can be related to $\lambdazero$ in an EOS-insensitive way. As shown by YY17, the relation can be generally modeled as,
\begin{align}
    \lambdak
    =\frac{\Gamma\left(k+\frac{10}{3-\bar{n}}\right)}{\Gamma\left(\frac{10}{3-\bar{n}}\right)} \lambdazero \left[1+\sum_{i=1}^3 a_{i,k}(\lambdazero)^{-i/5}\right], \label{eq:lambda_k_relativistic}
\end{align}
where $\bar{n}$ is the mean effective polytropic index, and $a_{i,k}$ are numerical coefficients to be fitted given a set of possible NS EOSs. 
Here, we follow the C21 implementation, i.e., choosing $\bar{n}=0.8$ and fitting $a_{i,k}$ up to $k=3$ using 29 NS EOSs that are consistent with recent LIGO/Virgo and Neutron Star Interior Composition Explorer observations (see Table I in Ref.~\cite{Chatterjee:2021xrm} for the fitted values of $a_{i,k}$).
Including $k>3$ terms will enhance the accuracy of the Taylor expansion in Eq.~\eqref{eq:lambda_m}, but the universality of the $\lambdazero$--$\lambdak$ relation deteriorates for these terms. 
C21 noted that the expansion to $k=3$ is sufficient to accurately represent $\lambdatilde(M)$ with less than 10\% loss of universality in the range $M_A\in(1.2,\,1.5)\,\msun$ for $m_0=1.4\msun$, which we will also choose as the reference mass in this work. 

The $\lambdatilde(M)$ function in Eqs.~\eqref{eq:lambda_m} and \eqref{eq:lambda_k_relativistic} has only one parameter left free, namely $\lambdazero$, that models the individual differences among those possible NS EOSs. The value of $\lambdazero$ can, therefore, be constrained by observational data. 
For example, using GW170817 and its EM counterpart, C21 measured $\lambdazero$ at 90\% confidence to be $191^{+113}_{-134}$ by directly applying the $\lambdazero$--$\lambdak$ relation; similarly, Ref.~\cite{LIGOScientific:2018cki} found $\lambdazero$ at 90\% confidence to be $190^{+390}_{-120}$ by applying the $\lambdatilde_a$--$\lambdatilde_s$ relation and converting the result into a linear expansion of $\lambdatilde(M)M^5$. 
Future observing runs of LIGO/Virgo/KAGRA with coincident operation of next-generation telescope facilities, such as the Rubin Observatory~\cite{Ivezic_2019}, is expected to yield more multimessenger BNS events. These events allow for more accurate measurements of $\lambdazero$, and stacking data from multiple observations (even those without electromagnetic counterparts) further reduces the uncertainty. 
In the following sections, we will assume that $\lambdazero$ is a fixed constant when we discuss BNS parameter estimation with the $\lambdazero$--$\lambdak$ relation.

\section{GW Measurements with the Binary Love Relations}
\label{sec:measure_with_blove}
The parameters of a GW signal are measured using Bayesian inference. The result is represented by a posterior distribution:
\begin{align}
    p(\pmb{\Theta}|d^\mathrm{GW})\propto p(d^\mathrm{GW}|\pmb{\Theta})\,p(\pmb{\Theta}),
\end{align}
where $\pmb{\Theta}$ is the set of parameters, $d^\mathrm{GW}$ is the GW data, $p(d^\mathrm{GW}|\pmb{\Theta})$ is the likelihood of getting $d^\mathrm{GW}$ from the GW signal, parametrized by $\pmb{\Theta}$, in noisy data, and $p(\pmb{\Theta})$ is the prior distribution. 
For a review of GW parameter estimation, see Ref.~\cite{creighton2011gravitational.ch7}.

For BBH coalescences, the GW signal is described by 15 parameters (when one neglects eccentricity), which include intrinsic ones, such as the masses $m_A$ and the spin vectors $\pmb{a}_A$, and extrinsic ones, such as the luminosity distance $\dl$ and the inclination angle $\iota$ (see, for example, Refs.~\cite{PhysRevD.49.1723,Cutler_1994,van_der_Sluys_2008,Veitch_2012,Veitch_2015}).
BNS coalescences use the same set of parameters, plus two additional ones to account for matter effects, namely, the tidal deformability of each NS $\lambdatilde_A$. These tidal parameters enter the phase of the waveform first at 5PN and then 6PN order as
\begin{align}
    \Psi_{\text{tid}}
    =&-\frac{3}{128\eta x^{5/2}} \bigg[\frac{39}{2}\bar{\Lambda}x^5 + \bigg(\frac{3115}{64}\bar{\Lambda} \notag\\
    &- \frac{6595}{364}\sqrt{1-4\eta}\, \delta\bar{\Lambda}\bigg)x^6 + \mathcal{O}(x^7)\bigg], \label{eq:psi_tid}
\end{align}
where $x=\left[\pi(m_1+m_2)f\right]^{2/3}$ is the PN expansion parameter, $f$ is the GW frequency, and $\eta=m_1m_2/(m_1+m_2)^2$ is the symmetric mass ratio. 
The coefficients $\bar{\Lambda}$ and $\delta\bar{\Lambda}$ are related to the tidal deformability parameters $\lambdatilde_A$, via
\begin{align}
    \bar{\Lambda}
    =&f(\eta)\left(\frac{\lambdatilde_1+\lambdatilde_2}{2}\right) + g(\eta)\left(\frac{\lambdatilde_1-\lambdatilde_2}{2}\right), \notag \\
    \delta\bar{\Lambda}
    =&\delta f(\eta)\left(\frac{\lambdatilde_1+\lambdatilde_2}{2}\right) + \delta g(\eta)\left(\frac{\lambdatilde_1-\lambdatilde_2}{2}\right), \label{eq:tidal_coeffs}
\end{align}
where the exact expressions for $\{f, g, \delta f, \delta g\}$ are given in Sec.~2.2 of YY17. 

Because of cosmic expansion, the GW signal is redshifted in the observed frame of the detectors. Hence, the masses measured above differ from the true masses of the binary. By convention, the former is referred to as the detector-frame mass $\mdet{}_A$, while the latter is referred to as the source-frame mass $\msource{}_A$, and they are related by $\mdet{}_A=\msource{}_A(1+z)$, where $z$ is the redshift.
For NSs, the source-frame mass is the mass parameter that enters the $\lambdatilde(M)$ function. 
Therefore, given a universal $\lambdatilde(M)$, or, equivalently, an EOS, one can replace the tidal deformability parameters in the GW waveform by
\begin{align}
    \lambdatilde_A = \lambdatilde\left(\frac{\mdet{}_A}{1+z}\right),
\end{align}
which, in turn, changes the parametrization of $\Psi_\mathrm{tid}$ from $(\mdet{}_1,\mdet{}_2,\lambdatilde_1,\lambdatilde_2)$ to $(\mdet{}_1,\mdet{}_2,z)$. 
The rise of $z$ as an independent measurable parameter enables enhanced cosmological inferences using only GW observations, which has been explored with $\lambdatilde(M)$ derived from both specific EOSs~\cite{messenger_read_2012} and EOS-insensitive relations~\cite{Chatterjee:2021xrm}.

In this work, we derive the $\lambdatilde(M)$ function from the EOS-insensitive binary Love relations. Additionally, we assume that the distance-redshift relation, i.e., the cosmology, is well constrained and given to us by, e.g., Planck observations~\cite{Planck:2018vyg}. Combining these two, the tidal deformability parameters can be expressed as follows:
\begin{align}
    \lambdatilde_A
    =\lambdazero + \sum_{k=1}^3 \frac{\lambdak}{k!} \left[1 - \frac{\mdet{}_A/m_0}{1+z(D_L;\hubble,\bomega)}\right]^k\,,
    \label{eq:lambda_rep}
\end{align}
where $\lambdak = \lambdak(\lambdazero)$ are given by Eq.~\eqref{eq:lambda_k_relativistic}. 
Since {\lambdazero} is expected to be a universal constant, which was estimated with GW170817 (e.g.,~in C21) and will be further constrained by future measurements, we fix its value when reporting our main results in Secs.~\ref{sec:improve_dliota} and \ref{sec:improve_masses}. We will then show in Sec.~\ref{sec:robstness} that relaxing the constraint on {\lambdazero} does not impact our main results.
The distance-redshift relation $z(D_L)$ is given by a flat {\lcdm} model with Hubble constant $\hubble$ and other cosmological parameters $\bomega$, which we fix to the Planck values\footnote{We use the Astropy implementation~\cite{astropy:2013,astropy:2018}.} measured using CMB anisotropies~\cite{Planck:2018vyg}. 
The statistical uncertainties of these Planck values are percent level and, therefore, negligible for measuring BNS parameters in this work. 
However, we note that local-Universe measurements suggest other $\hubble$ values that are several $\sigma$ away from the Planck value of $\hubble$, which is known as the ``Hubble tension'' (see, for example, Refs.~\cite{Riess:2021jrx,Freedman:2021ahq,Anand:2021sum}). We will discuss the impact of this discrepancy in Sec.~\ref{sec:robstness}. 

\begin{figure}[t]
    \centering
    \includegraphics[width=0.48\textwidth]{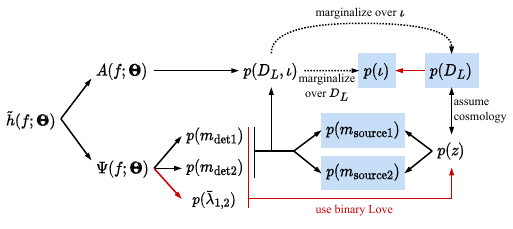}
    \caption{
    Flow chart of measuring the BNS parameters mentioned in Sec.~\ref{sec:measure_with_blove}. The waveform $\tilde{h}$ splits into the amplitude $A$ and phase $\Psi$ and leads to inferences denoted by the marginalized posteriors $p(\cdot)$. The blue shaded boxes highlight parameters that we expect to improve. The black arrows represent the major paths along which information is extracted and combined in the traditional approach to constrain these boxed parameters. The dotted line represents the traditional schema of obtaining marginalized posteriors on $\dl$ and $\iota$. The red arrows are paths added in our approach to aid the inference, enabled by the use of the binary Love relations. Note that the chart is simplified---elements that have low impact on constraining the boxed parameters are omitted. }
    \label{fig:blove_measure_flow}
\end{figure}

In Fig.~\ref{fig:blove_measure_flow}, we provide a visual representation of the flow of ideas underlying this work. Traditionally, in GW parameter estimation, one extracts the parameters of the binary following the black arrows in the figure, where the detector-frame masses are mostly determined using the GW phase. Combining the latter with the GW amplitude, one can extract the distance and inclination angle. The distance relates to the redshift of the source, assuming a cosmological model. Then, the source-frame masses are inferred using both the detector-frame masses and the redshift. 
In these steps, the extraction of the distance and inclination angle from the GW amplitude is limited because of a degeneracy in the way they affect the GW amplitude~\cite{Markovic:1993cr,Cutler_1994,Usman:2018imj}.
Instead of following this traditional approach, we will here use the fact that the GW phase also carries information about the distance through the tidal parameters, according to Eq.~\eqref{eq:lambda_rep}. This additional information may help break the distance-inclination degeneracy and tighten the constraints on both parameters, as well as lead to a more accurate determination of the source-frame masses, which is depicted with red arrows in the figure.
Hence, we expect that our use of the binary Love relations may improve the estimation of certain BNS parameters, such as the luminosity distance, inclination angle, and the source-frame masses.

\section{Computational Setup and Detector-Design Choice}
\label{sec:compute_setup}
We compare the GW parameter estimation on synthetic BNS signals with and without the binary Love relations. 
Without loss of generality, we fix the source to have component masses $\msource{}_{1\rm inj}=1.46\,\msun$ and $\msource{}_{2\rm inj}=1.27\,\msun$, which are similar to that of GW170817~\cite{LIGOScientific:2018hze}. 
We assume that the true NS EOS can be characterized by $\lambdazero{}_{\rm inj}=200$. As a consequence, the tidal deformability parameters of the BNS are $\lambdatilde_{1\rm inj}=183$ and $\lambdatilde_{2\rm inj}=322$, given the source-frame masses. 
We also neglect the spins of the binary, as they are expected to be small for NSs and have little impact in our analysis (see Appendix.~\ref{app:effect_of_spins}).
Such a BNS source is then used to simulate GW signals detected at different distances, inclination angles, etc.

We assume that the systems described in the previous paragraph are BNSs i.e., the analysis does not apply to compact objects with similar masses but not classified as neutron stars. We use the {\pbilby} inference library~\cite{Smith_2020} with the IMRPhenomPv2\_NRTidal waveform~\cite{PhysRevD.99.024029} and the {\dynesty} sampler~\cite{dynesty}.
For each injection, we use a 128\,s signal duration and model the noise through the spectral noise density of various detectors~\cite{observing_scenarios,Abbott_2017}.
In particular, we do not inject the signal in specific realizations of noise because we wish to study averaged \textit{statistical} errors that are independent of a given noise artifact. 
We sample the masses in terms of the detector-frame chirp mass $\mchirp^\mathrm{det}=(\mdet{}_1+\mdet{}_2)\eta^{3/5}$ and the mass ratio $q=\mdet{}_2/\mdet{}_1=\msource{}_2/\msource{}_1$, each with a uniform prior. We fix the spins to be zero and do not sample over them.
In Appendix.~\ref{app:effect_of_spins}, we show that presence of intrinsic spin and the use of precessing spin priors do not impact the main result of the paper.
We put a prior on the luminosity distance that is uniform in comoving volume, given by the same cosmology used in Eq.~\eqref{eq:lambda_rep} for the injections.
When the binary Love relations are used, the tidal deformability parameters are determined using the masses and the distance through Eq.~\eqref{eq:lambda_rep} and are, therefore, not sampled. 
In contrast, when the relation is not used, we use a uniform prior on $\lambdatilde_A$ in $[0,5000]$ to reflect our ignorance of the NS EOS.
For all other parameters sampled, we use the same priors as in Ref.~\cite{LIGOScientific:2016vlm}.

We consider observing the simulated signals using three detectors located and orientated in the same way as LIGO-Hanford, LIGO-Livingston, and Virgo, respectively. This three-detector configuration (HLV for later reference) is sufficient for distinguishing face-on and face-off inclinations. 
The sensitivity required for the network to demonstrate improvements in the parameter estimation using the binary Love relations is then to be determined in our study. 
As explained in Sec.~\ref{sec:measure_with_blove}, the expected improvements rely on resolving the tidal effects in the GW signal, which are weak until the late inspiral, at frequencies of $\gtrsim 400\,\mathrm{Hz}$. 
However, the HLV detectors are most sensitive to GWs inside their sensitivity buckets, at $\sim 100\,\mathrm{Hz}$. 
Detectors with advanced designs, therefore, have two benefits. One is that they are generally better at capturing weak effects in the signal. The other is that they allow detection of more distant sources, whose late-inspiral tidal imprints are more redshifted toward the bucket of the sensitivity band.

What detector network should we choose to carry out our analysis? To answer this question, let us consider the accuracy to which the inclination angle and the luminosity distance can be estimated with and without the binary Love relations using a second-generation (2G) and a hypothetical 3G network. More precisely, the 2G network will be composed of HLV detectors with A+ (O5) noise curves~\cite{observing_scenarios}\footnote{\url{https://dcc.ligo.org/LIGO-T2000012/public}}, while the 3G network will again be composed of HLV-like detectors but with the noise curve of Cosmic Explorer (CE)~\cite{Abbott_2017}\footnote{\url{https://dcc.ligo.org/LIGO-T1500293/public}}. 
In reality, the 3G network may be a stand-alone Einstein Telescope (ET)~\cite{Punturo:2010zza} or a combination of CE and ET. 
Our work demonstrates the general level of sensitivity of these approaches, and our conclusions are expected to be qualitatively robust to different network configurations.

\begin{figure*}[htbp]
    \centering
    \includegraphics[width=0.98\textwidth]{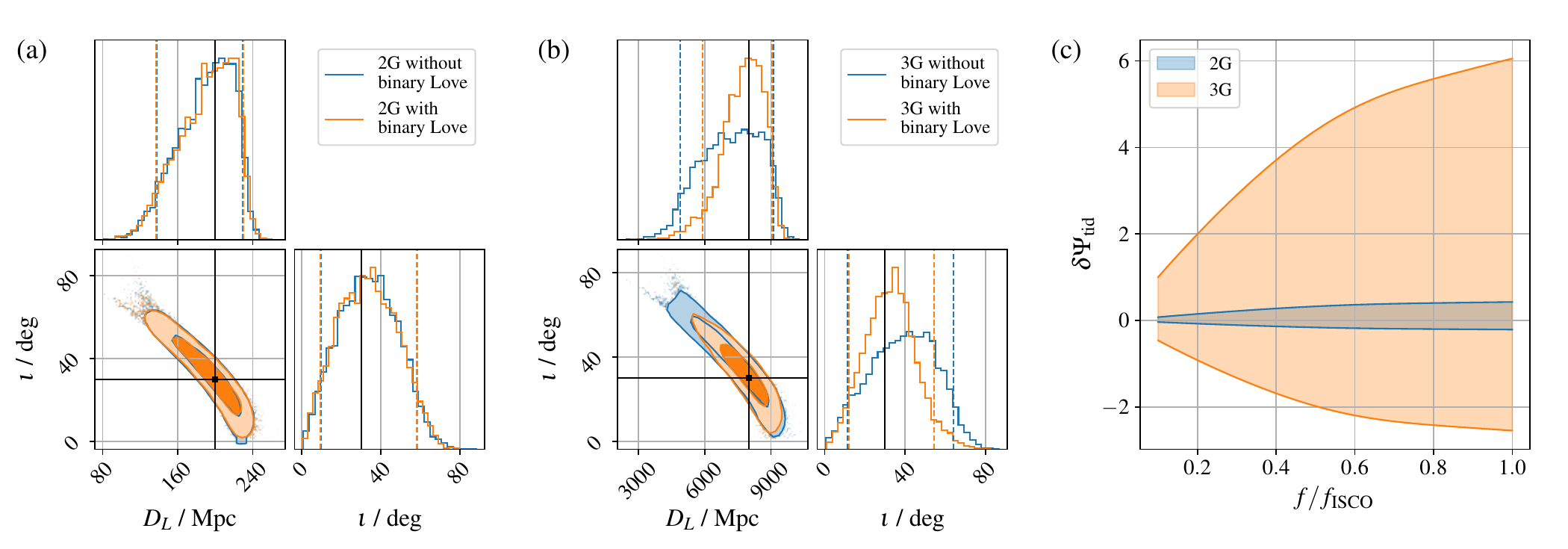}
    \caption{Comparison between 2G detectors and 3G detectors. The 2G network is composed of HLV detectors at A+ (O5) sensitivity, and the 3G network is composed of HLV-like detectors with CE sensitivity. The observed signal is synthesized with a BNS source similar to that of GW170817. The injected inclination angle is $30^\circ$ for both networks. However, the injected luminosity distances are $200\,\megaparsec$ for the 2G detectors and $8\,\gigaparsec$ for the 3G detectors, respectively, so that the SNRs are both about 30. 
    (a) [(b)] shows the corner plots of the $\dl$ and $\iota$ estimate from the 2G (3G) detection, with and without the use of the binary Love relations. The vertical dashed lines in the 1D histograms mark the 90\% credible intervals, the contours in the 2D histograms represent 50\% and 90\% of the posterior samples, and the black lines correspond to the injected values. 
    Note that for the 2G observation, the estimation is not affected by the use of the binary Love relations. However, for the 3G observation, improvement shows up as the posterior peaks get closer to the injected values and the 90\% CIs shrink. 
    (c) shows the $\dl$ variability of the tidal phase with respect to the injected tidal phase as a function of the GW frequency, i.e.~$\delta\Psi_\mathrm{tid}(\dl)=\Psi_\mathrm{tid}(\pmb{\Theta}=\pmb{\Theta}_{{\rm inj} \neq \dl},\dl)-\Psi_{\mathrm{tid}}(\pmb{\Theta}=\pmb{\Theta}_{\rm inj})$.
    The shaded regions show the variation of $\delta\Psi_\mathrm{tid}$ associated with the 90\% CIs of $\dl$ posteriors. In each observation scenario, we use the detector-frame innermost stable circular orbit (ISCO) frequency, $f_\mathrm{ISCO}=(1/6)^{3/2}/[\pi(\mdet{}_1+\mdet{}_2)]$, as a frequency cutoff. 
    Observe that for the 3G observation, the spread of $\delta\Psi_\mathrm{tid}$ is wider, which means more tidal information is used to extract $\dl$. 
    }
    \label{fig:compare_detectors}
\end{figure*}

Let the BNS (injected) source be at $\iota{}_\mathrm{inj}=30^\circ$ (the inclination angle at which detections are most likely to be made~\cite{schutz_2011}) and at $\dl{}_\mathrm{inj} = 200\,\megaparsec$ for the 2G network and $\dl{}_\mathrm{inj} = 8\,\gigaparsec$ for the 3G network (so that both SNRs are near $30$). All other extrinsic parameters are kept the same between the 2G and 3G study, although we have checked that this does not affect the conclusions. 
In Fig.~\ref{fig:compare_detectors}, we show corner plots for the inclination angle and the luminosity distance with the 2G (a) and 3G networks (b). Observe that, while the 3G network allows for an improvement in the $\dl$--$\iota$ measurement, this is not so for the 2G network.

The reason for this is that the impact of the tidal effects on the phase for a 3G network is much larger than for a 2G network, as shown in Fig.~\ref{fig:compare_detectors}(c). This panel shows the $\dl$ variability of the tidal phase with respect to the injected tidal phase as a function of the GW frequency, i.e.,~$\delta\Psi_\mathrm{tid}(\dl)=\Psi_\mathrm{tid}(\pmb{\Theta} = \pmb{\Theta}_{{\rm inj} \neq \dl},\dl)-\Psi_{\mathrm{tid}}(\pmb{\Theta} = \pmb{\Theta}_{\rm inj})$. To estimate the $\dl$ variability, we evaluate the tidal phase $\Psi_\mathrm{tid}$ with the posterior of $\dl$, setting all other parameters $\pmb{\Theta}_{{\rm inj} \neq \dl}$ to their injected values. 
Because the 3G detector can see systems that are much farther out than the 2G detector, the impact of the $\dl$ posterior on $\delta\Psi_\mathrm{tid}$ is much greater, having, therefore, a greater impact in parameter estimation and, in particular, allowing for an improvement in the extraction of both $\dl$ and $\iota$. 
Given that the binary Love relations do not improve the estimation of $D_L$ or $\iota$ with the 2G network, henceforth, we will carry out all future studies with the 3G configuration.

\section{Improvements in the Distance and Inclination with the Binary Love Relations}
\label{sec:improve_dliota}
We now study how the improvement in the estimation of the luminosity distance and the inclination angle (due to the use of the binary Love relations) varies with the value of the injected $\dl{}_{\rm inj}$ and $\iota_{\rm inj}$. In particular, we set up a $\dl{}_{\rm inj}$--$\iota_{\rm inj}$ grid, letting $\dl{}_{\rm inj}$ vary between $1$ and $32\,\gigaparsec$ and $\iota$ vary between $0^\circ$ and $90^\circ$ (leading to SNRs in $6$--$167$). 
We have checked that the other half of the inclination range, $90^\circ$--$180^\circ$, gives almost the mirrored pattern of $0^\circ$--$90^\circ$, which is not particularly interesting. 
Again, when analyzing this injection grid, we fix all other extrinsic parameters, such as the sky location and the arrival time, since they do not significantly impact our results.

\begin{figure*}[htbp]
    \centering
    \includegraphics[width=0.98\textwidth]{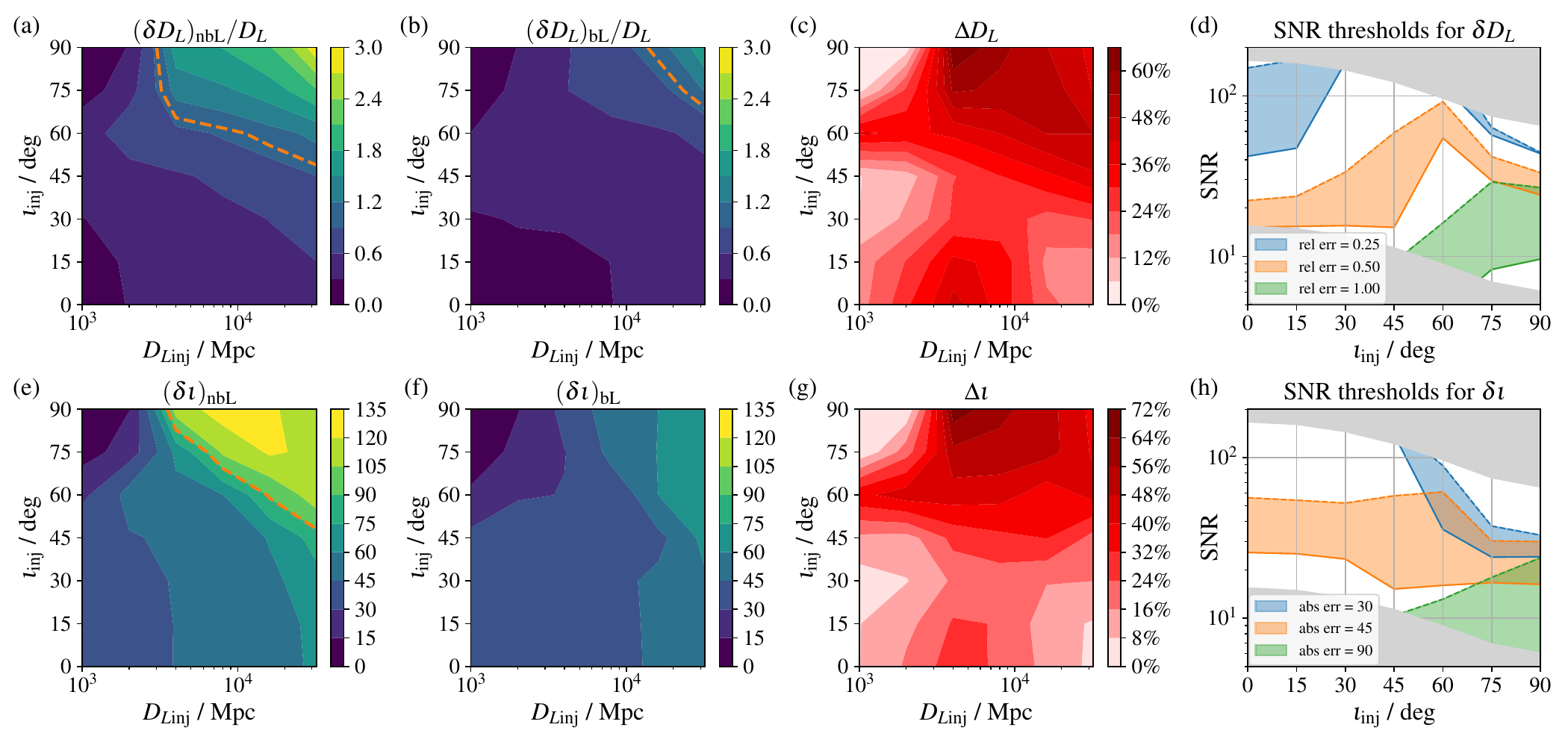}
    \caption{
    Impact of the binary Love relations in $\dl$--$\iota$ estimation on the $\dl{}_{\rm inj}$--$\iota_{\rm inj}$ grid. 
    The signals are synthesized using a GW170817-like source detected by a 3G HLV-like network. 
    (a) [(b)] shows the relative error of $\dl$ measurement with (without) the binary Love relations, $(\delta\dl)_{\rm bL}/\dl$ [$(\delta \dl)_{\rm nbL}/\dl$]. The error $\delta\dl$ is evaluated as the width of the 90\% CI. Note that (a) and (b) share the same set of contour levels under the same color bar. The dashed orange contour denotes $\delta\dl/\dl=100\%$, beyond which the error is considered unacceptable.
    (c) shows the relative improvement in $\dl$ from the binary Love relations, defined as $\Delta\dl=[1-(\delta\dl)_{\rm bL}/(\delta\dl)_{\rm nbL}]$. 
    (d) shows the minimal optimal SNRs for constraining $\delta\dl/\dl$ to the levels specified in the legend. The solid (dashed) lines correspond to estimation with (without) the binary Love relations, and the difference between each pair of them is marked by a shade of the same color. The regions not covered by our $\dl{}_{\rm inj}$--$\iota_{\rm inj}$ grid are left gray. 
    (e)--(h) are the same analysis repeated for the $\iota$ estimation and presented in absolute errors. The dashed contour in (e) corresponds to $\delta\iota=90^\circ$, whose counterpart in (f) is beyond the $\dl{}_{\rm inj}$--$\iota_{\rm inj}$ grid.
    We note that the estimation is always improved throughout the grid.}
    \label{fig:improve_dliota}
\end{figure*}

In Figs.~\ref{fig:improve_dliota}(a), \ref{fig:improve_dliota}(b) and \ref{fig:improve_dliota}(e), and \ref{fig:improve_dliota}(f), we show the measurement uncertainties of $\dl$ and $\iota$ in terms of their 90\% CIs, denoted by $\delta \dl$ and $\delta \iota$, respectively, as functions of $\dl{}_{\rm inj}$ and $\iota_{\rm inj}$ with [$(\delta D_L)_{\rm bL},(\delta \iota)_{\rm bL}$] and without [$(\delta D_L)_{\rm nbL},(\delta \iota)_{\rm nbL}$] using the binary Love relations.
Note that the accuracy to which these parameters can be measured deteriorates as $\dl{}_\mathrm{inj}$ increases and $\iota_\mathrm{inj}$ approaches $90^\circ$, because this corresponds to a decrease in SNR. 
However, the region in the injected $\dl{}_{\rm inj}$--$\iota_{\rm inj}$ plane inside which measurements with a reasonable uncertainty [$\delta \dl/\dl<100\%$ or $\delta\iota<90^\circ$, denoted with a dashed line in Figs.~\ref{fig:improve_dliota}(a), \ref{fig:improve_dliota}(b) and \ref{fig:improve_dliota}(e), and \ref{fig:improve_dliota}(f)] are possible is greatly increased when we use the binary Love relations.

The impact of the binary Love relations in parameter estimation can be more easily assessed by looking at the ``improvement'' or ``deterioration'' in the estimation of $\dl$ and $\iota$. We define the relative fractional improvement via
\begin{align}
    \Delta D_L &= 100\% \times \left[1-\frac{(\delta D_L)_{\rm bL}}{(\delta D_L)_{\rm nbL}}\right]\,,
    \\
    \Delta \iota &= 100\% \times \left[1-\frac{(\delta \iota)_{\rm bL}}{(\delta \iota)_{\rm nbL}}\right]\,.
\end{align}
Positive values of $\Delta D_L$ and $\Delta \iota$ correspond to an improvement in parameter estimation. 
As suggested by Figs.~\ref{fig:improve_dliota}(c) and \ref{fig:improve_dliota}(g), the binary Love relations always \textit{improve} the estimation of $\dl$ and $\iota$ throughout the $\dl{}_{\rm inj}$--$\iota_{\rm inj}$ grid chosen. 
The greatest improvement is found at about $(4\,\gigaparsec, 90^\circ)$, with $\Delta\dl\approx70\%\approx\Delta\iota$. 
Aside from that, a secondary improvement region is found at about $(4\,\gigaparsec, 0^\circ)$, with $\Delta\dl\approx40\%$ and $\Delta\iota\approx30\%$. 

Another way to understand and visualize the improvement in parameter estimation due to the use of the binary Love relations is to study the minimum SNR required to achieve a certain measurement uncertainty. In Figs.~\ref{fig:improve_dliota}(d) and \ref{fig:improve_dliota}(h), we show that the SNR threshold is cut in almost half due to the use of the binary Love relations, when $\delta\dl/\dl=50\%$ or $\delta\iota=45^\circ$ is targeted. 
Also note that when edge-on systems are measured, the SNR threshold for $\delta\dl/\dl=100\%$ or $\delta\iota=90^\circ$ drops from $~30$ to $\lesssim10$, which confirms the move of the dashed lines in Figs.~\ref{fig:improve_dliota}(a), \ref{fig:improve_dliota}(b), \ref{fig:improve_dliota}(e), and \ref{fig:improve_dliota}(f).
Since one expects to detect many more events at low SNR than at high SNR, the use of the binary Love relations therefore allows us to extract meaningful astrophysical information from a much larger set of events. 

The detailed pattern in Figs.~\ref{fig:improve_dliota}(c) and \ref{fig:improve_dliota}(g) is complicated and deserves more discussion. 
First, note that, as $\dl$ increases, the improvement first also increases, reaching a maximum around 3--10\,\gigaparsec, and then the improvement decreases. 
This pattern is related to the contrast between the uncertainty of $\dl$ constrained by the waveform amplitude, $(\delta\dl/\dl)_\mathrm{ampl}$, and the uncertainty of $z(\dl)$ constrained by $\Psi_\mathrm{tid}$, $(\delta z/z)_\mathrm{tid}$. 
The former is roughly proportional to the inverse of the SNR~\cite{Cutler_1994} and, hence, constantly increases as $\dl{}_{\rm inj}$ increases. 
The latter is affected by not only the SNR, but also the redshifting of the high-frequency tidal imprint and the detectors' sensitivity band~\cite{messenger_read_2012}. The two effects compete against each other, and the increase of $(\delta z/z)_\mathrm{tid}$ is suppressed before the distance becomes so large that the SNR effect starts to dominate. 
Therefore, we expect that, at small distances, because $(\delta\dl/\dl)_\mathrm{ampl}$ increases with distance while $(\delta z/z)_\mathrm{tid}$ does not, the use of $\Psi_\mathrm{tid}$ to tighten the constraint of $\dl$ should be more effective as $\dl{}_{\rm inj}$ increases. 
After some critical $\dl{}_{\rm inj}$, $(\delta\dl/\dl)_\mathrm{ampl}$ and $(\delta z/z)_\mathrm{tid}$ increase at similar rates, so $\Psi_\mathrm{tid}$ becomes less helpful. 
Messenger and Read~\cite{messenger_read_2012} showed that, for 3G detectors measuring BNSs using a certain EOS, the critical point for $(\delta z/z)_\mathrm{tid}$ to increase is around $z_{\rm inj}\sim 1$, or $\dl{}_{\rm inj}\sim 7\,\gigaparsec$ according to the cosmology they assumed. This explains our observation of the maximum improvement around 3--10\,\gigaparsec.

\begin{figure}[htbp]
    \centering
    \includegraphics[width=0.48\textwidth]{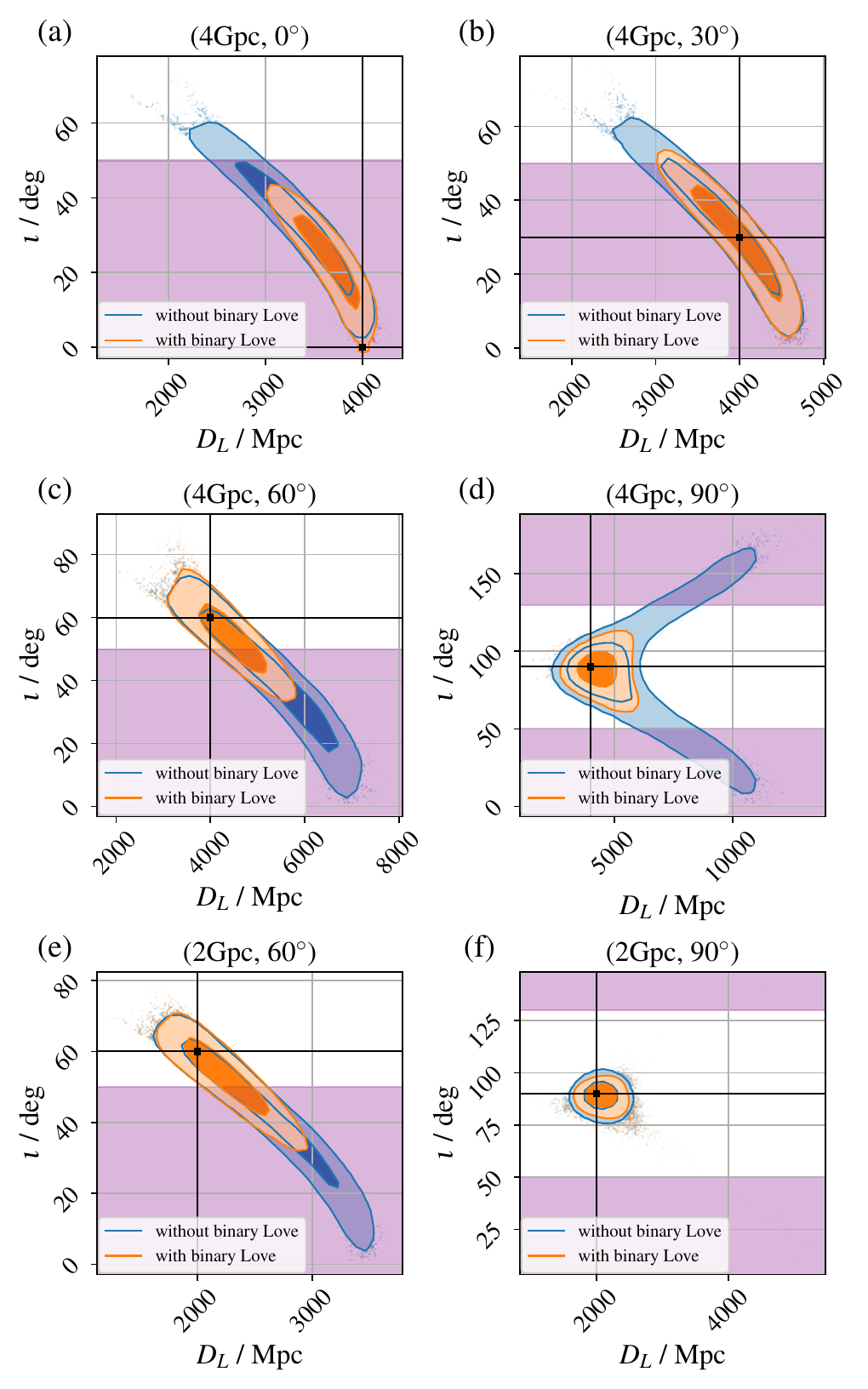}
    \caption{
    Typical outcomes of the distance-inclination degeneracy, in terms of 2D histograms of the $\dl$--$\iota$ joint posteriors. Cases are taken from the $\dl{}_{\rm inj}$--$\iota_{\rm inj}$ grid, including (a) $(4\,\gigaparsec, 0^\circ)$, (b) $(4\,\gigaparsec, 30^\circ)$, (c) $(4\,\gigaparsec, 60^\circ)$, (d) $(4\,\gigaparsec, 90^\circ)$, (e) $(2\,\gigaparsec, 60^\circ)$, and (f) $(2\,\gigaparsec, 90^\circ)$.
    The 2D histograms follow the same format as in Figs.~\ref{fig:compare_detectors}(a) and \ref{fig:compare_detectors}(b). We use purple to shade the regions where the distance-inclination degeneracy is strong ($\iota_{\rm inj}<50^\circ$ or $\iota_{\rm inj}>130^\circ$). 
    Observe that, when the binary Love relations are not used, the posterior distribution tends to skew towards the strong-degeneracy regions [except for the small-distance, edge-on case in (f)]. The binary Love relations improve the estimation by reducing that skewness. }
    \label{fig:degeneracy_cases}
\end{figure}

The pattern along the $\iota_{\rm inj}$ direction is more complicated. In general, the improvement is greater when $\iota_{\rm inj} \approx 0^\circ$ or $90^\circ$. The latter is more significant, except that when $\dl{}_{\rm inj}$ is small, the improvement near $\iota_{\rm inj}=90^\circ$ is suppressed.  
This can be explained by the specific effects of the distance-inclination degeneracy at different injected inclination angles. In Fig.~\ref{fig:degeneracy_cases}, we show a typical set of examples of this degeneracy, portrayed as an elongated shape in the marginalized 2D posterior.
The distance-inclination degeneracy is strongest when the inclination angle is small, and the two GW polarizations have almost the same amplitude. 
In particular, the difference in the two amplitudes does not exceed 10\% as long as $\iota_{\rm inj}\lesssim50^\circ$ and gradually vanishes as $\iota_{\rm inj} \to 0^\circ$~\cite{Usman:2018imj}.
In other words, for small injected inclination angles [$\iota_{\rm inj}\lesssim 50^\circ$, especially $\iota_{\rm inj}\rightarrow 0^\circ$; see Figs.~\ref{fig:degeneracy_cases}(a) and \ref{fig:degeneracy_cases}(b)], the degeneracy has a more negative impact on the $\dl$--$\iota$ measurement, and, thus, more of an improvement can be made there when additional information from $\Psi_\mathrm{tid}$ is provided. 

For large inclination angles [$50^\circ\lesssim\iota_{\rm inj}<90^\circ$; see Figs.~\ref{fig:degeneracy_cases}(c) and \ref{fig:degeneracy_cases}(d)], the degeneracy causes the likelihood function to form a tail that reaches out to small inclination angles. This tail has been known to be responsible for misclassifying some edge-on systems as face on in the worst cases (see, for example, Refs.~\cite{Usman:2018imj,Chen:2018omi}). 
Therefore, for large injected inclination angles, the additional information on $\dl$ from $\Psi_\mathrm{tid}$ can significantly improve the measurement by eliminating these tails in the likelihood. 
Because the tail can become longer when $\iota_{\rm inj}\rightarrow 90^\circ$, the potential improvement can be even greater there. 

The tail argument can also explain why the improvement near $\iota_{\rm inj}=90^\circ$ is suppressed when $\dl{}_{\rm inj}$ is small. 
When the injected distance is small, the SNR is high. Therefore, for nearly edge-on systems [$70^\circ\lesssim\iota_{\rm inj}<90^\circ$; see Fig.~\ref{fig:degeneracy_cases}(f)], the tails are suppressed by the high SNR and are not captured by the 90\% CI, leaving little space for $\Psi_\mathrm{tid}$ to improve the parameter estimation.
For medium to large injected inclination angles [$50^\circ\lesssim\iota_{\rm inj}\lesssim70^\circ$; see Fig.~\ref{fig:degeneracy_cases}(e)], however, the tails are less suppressed by the SNR. This is because these angles are rather close to the degenerate region and the tails are firmly attached to the likelihood peaks. Therefore, the improvement from $\Psi_\mathrm{tid}$ for these medium to large angles can still show up at small distances.

\section{Improvements in the Component Masses with the Binary Love Relations}
\label{sec:improve_masses}
\begin{figure*}[htbp]
    \centering
    \includegraphics[width=0.98\textwidth]{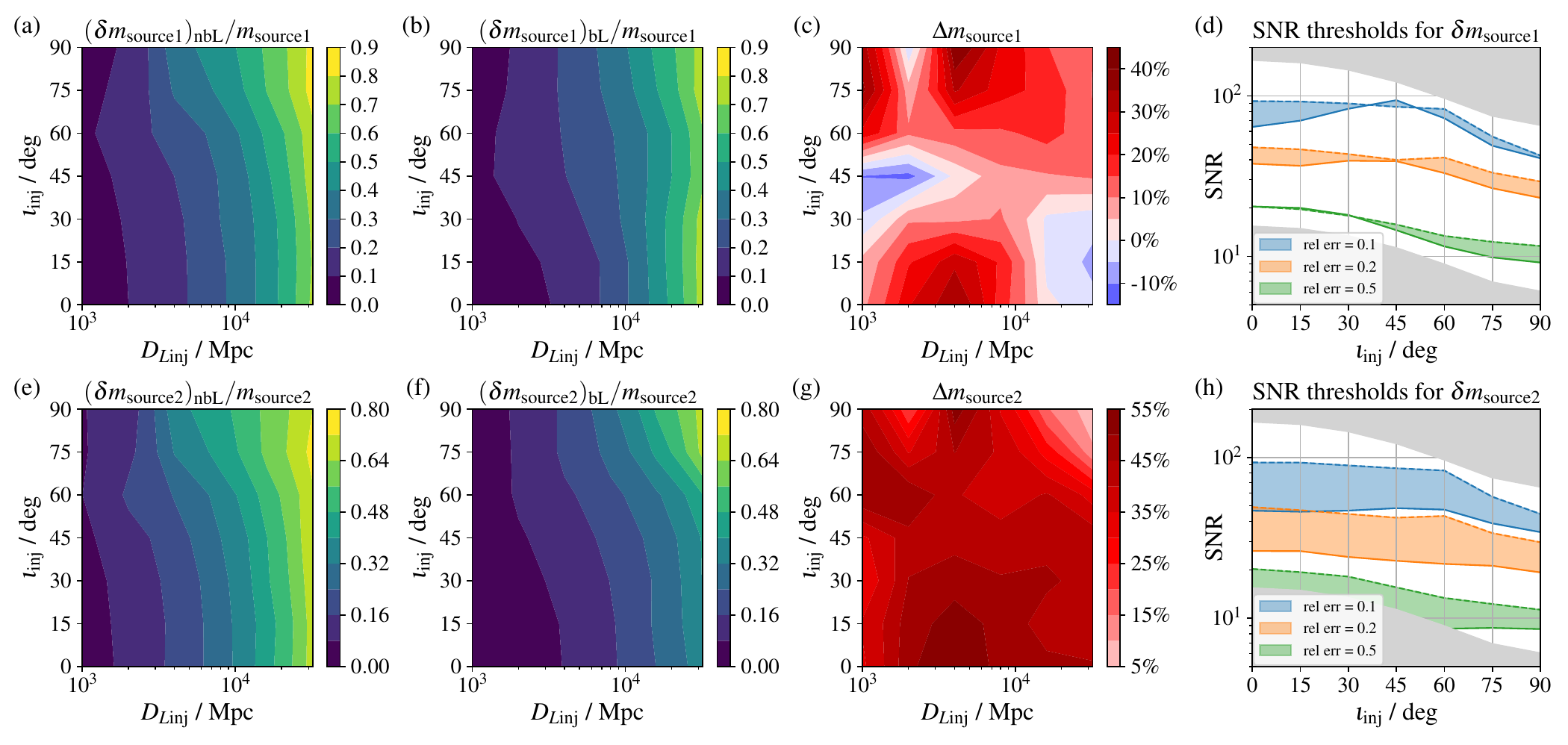}
    \caption{
    Impact of the binary Love relations in the source-frame mass estimation on the $\dl{}_{\rm inj}$--$\iota_{\rm inj}$ grid. 
    The signals are synthesized using a GW170817-like source detected by a 3G HLV-like network. 
    The format of each subplot follows the same as in Fig.~\ref{fig:improve_dliota}.
    We see that improvement happens in most cases, and the highest level of improvement is close to that for $\dl$. However, the improvement in the primary mass is relatively less significant, and deterioration [blue regions in (c) and solid lines above the dashed ones in (d)] can sometimes take place.
    }
    \label{fig:improve_masses}
\end{figure*}

In this section, we study the impact of binary Love relations to  measurements of NS masses.
In Figs.~\ref{fig:improve_masses}(a), \ref{fig:improve_masses}(b), \ref{fig:improve_masses}(e), and \ref{fig:improve_masses}(f), we show the measurement uncertainties of $\msource{}_1$ and $\msource{}_2$ in terms of their 90\% CIs, denoted by $\delta\msource{}_1$ and $\delta\msource{}_2$, respectively, as functions of $\dl{}_{\rm inj}$ and $\iota_{\rm inj}$ with and without using the binary Love relations.
Similar to the distance-inclination measurement, the accuracy of these measured masses also generally deteriorates as $\dl{}_\mathrm{inj}$ increases and $\iota_\mathrm{inj}$ approaches $90^\circ$, in correspondence to the decrease of SNR. 
Note that without the binary Love relations, the masses are already measured with relative errors lower than 100\% on this $\dl{}_\mathrm{inj}$--$\iota_\mathrm{inj}$ grid. However, we can still see improvement, as Figs.~\ref{fig:improve_masses}(b) and \ref{fig:improve_masses}(f) have more dark (low-uncertainty) area than Figs.~\ref{fig:improve_masses}(a) and \ref{fig:improve_masses}(e) do.

Again, to see the impact of the binary Love relations, we define the relative fractional improvement via
\begin{align}
    \Delta\msource{}_A &= 100\% \times \left[1-\frac{(\delta\msource{}_A)_{\rm bL}}{(\delta\msource{}_A)_{\rm nbL}}\right]\,.
\end{align}
As suggested by Fig.~\ref{fig:improve_masses}(g), the binary Love relations improve the estimation of the secondary mass $\msource{}_2$ throughout the $\dl{}_{\rm inj}$--$\iota_{\rm inj}$ grid chosen. The high improvement regions are around $(4\,\gigaparsec, 0^\circ)$, $(4\,\gigaparsec, 90^\circ)$, and $(1\,\gigaparsec, 75^\circ)$, where $\Delta\msource{}_2\approx55\%$ is reached.
For the primary mass $\msource{}_1$, as suggested by Fig.~\ref{fig:improve_masses}(c), there is a similar trend of improvement as for $\msource{}_2$, but the level of improvement is generally weaker, reaching $\Delta\msource{}_1\approx45\%$ in the high improvement regions and even going negative (down to $\sim-10\%$) at about $(2\,\gigaparsec, 45^\circ)$ and $(32\,\gigaparsec, 15^\circ)$. 

Correspondingly, in Fig.~\ref{fig:improve_masses}(h), we see that the SNR thresholds are almost cut in half due to the use of the binary Love relations, when $\delta\msource{}_2/\msource{}_2=20\%$ or even $\delta\msource{}_2/\msource{}_2=10\%$ is targeted. However, as suggested by Fig.~\ref{fig:improve_masses}(d), the decrease in SNR thresholds for the $\msource{}_1$ measurement is relatively less significant, and an increase is observed when $\delta\msource{}_1/\msource{}_1=10\%$ is targeted.

The generally weaker improvement in the primary mass is expected as the $\lambdatilde(M)$ function is less sensitive to larger NS masses. In our case, using the $\lambdatilde(M)$ function determined in Sec.~\ref{sec:blove}, we have
\begin{align}
    \frac{d\lambdatilde(M)}{dM}\bigg|_{\msource{}_1}&=-230\,\msun^{-1},\notag\\
    \frac{d\lambdatilde(M)}{dM}\bigg|_{\msource{}_2}&=-1771\,\msun^{-1},
\end{align}
the latter of which is larger in absolute value by one order of magnitude. Therefore, the binary Love relations tend to put a tighter constraint on the secondary mass, leaving the primary mass with less of an improvement. 

We also note that the high improvement regions for both mass parameters at about $(4\,\gigaparsec, 0^\circ)$ and $(4\,\gigaparsec, 90^\circ)$ overlap with the regions for which the estimation of $\dl$ also improves the most. This is because one major source of uncertainty when determining the source-frame masses at large distances is the redshift, which is a function of the distance assuming the cosmology. The better constraint on the distance means better constraint on the redshift and, thus, also means better constraint on the source-frame masses.

\begin{figure}[htbp]
    \centering
    \includegraphics[width=0.48\textwidth]{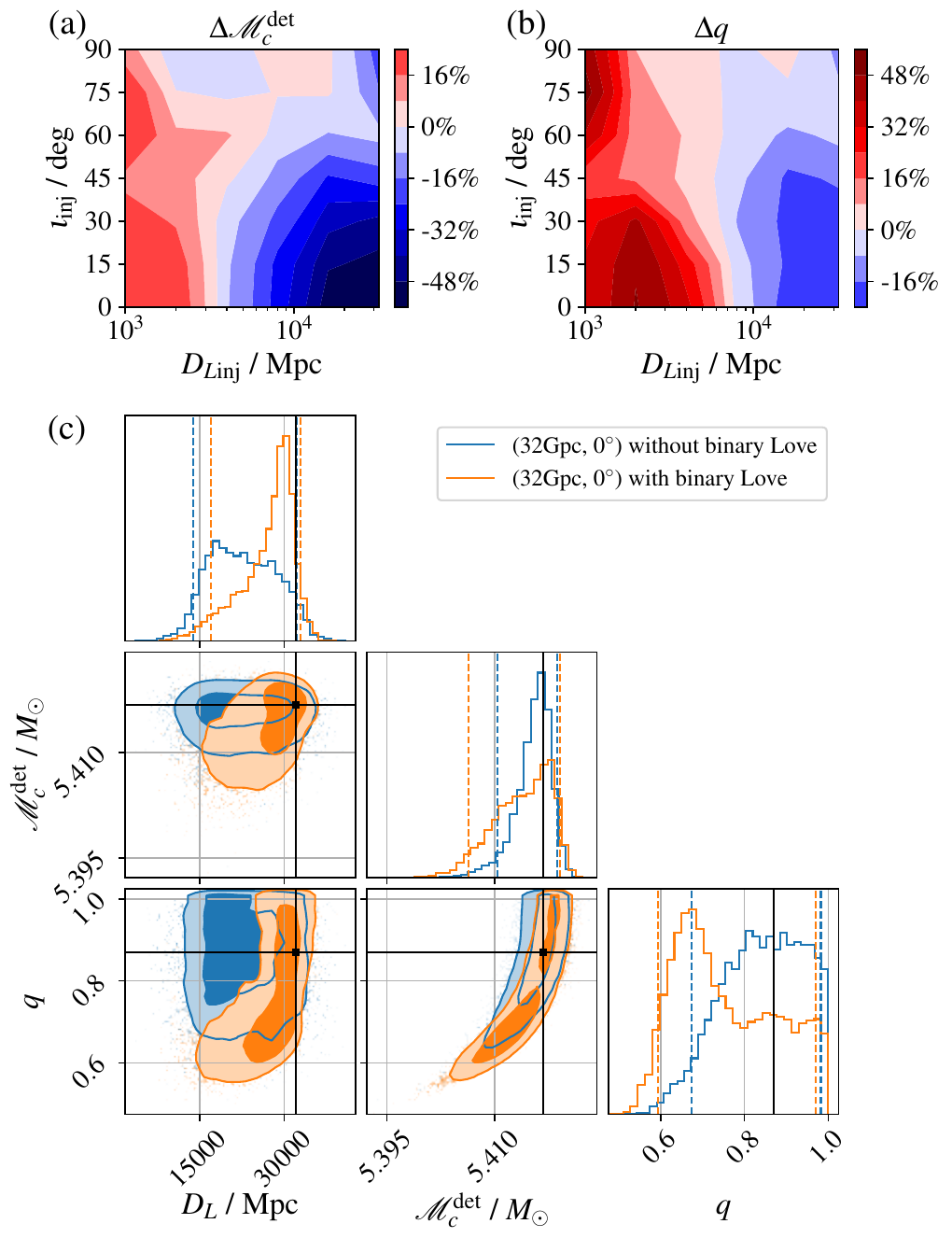}
    \caption{
    Impact of the binary Love relations in $\mchirp^\mathrm{det}$--$q$ estimation on the $\dl{}_{\rm inj}$--$\iota_{\rm inj}$ grid. The signals are synthesized using a GW170817-like source detected by a 3G HLV-like network.
    (a) and (b) show the relative decrease in the 90\% CIs, which follow the same format as in Figs.~\ref{fig:improve_dliota}(c) and \ref{fig:improve_dliota}(g). 
    We see deterioration in these parameters especially for large distances and small inclinations. 
    (c) shows the corner plot of $\mchirp^\mathrm{det}$, $q$, and $\dl$ for $(\dl{}_{\rm inj},\iota_{\rm inj})=(32\,\gigaparsec, 0^\circ)$. The plot follows the same format as in Figs.~\ref{fig:compare_detectors}(a) and \ref{fig:compare_detectors}(b).
    Note that the binary Love relations add bias to the estimate, presenting as a new peak in the $\mchirp^\mathrm{det}$--$q$ posterior, aside from the original one that covers the injected values.
    }
    \label{fig:improve_mcq}
\end{figure}

Some other features in Figs.~\ref{fig:improve_masses}(c) and \ref{fig:improve_masses}(g) can be attributed to the interplay between the distance $\dl$ and the detector-frame masses $\mdet{}_A$ as they jointly determine $\Psi_\mathrm{tid}$. 
In the actual parametrization of the waveform, the detector-frame masses $\mdet{}_A$ are reexpressed using the detector-frame chirp mass $\mchirp^\mathrm{det}$ and the mass ratio $q$. We present the improvement in these two parameters in Figs.~\ref{fig:improve_mcq}(a) and \ref{fig:improve_mcq}(b). 
We note that the high improvement in $\mchirp^\mathrm{det}$ and $q$ for small $\dl{}_{\rm inj}$ and large $\iota{}_{\rm inj}$ is responsible for the high improvement in $\msource{}_A$ at about $(1\,\gigaparsec, 75^\circ)$, which is not explained by the improvement in $\dl$ alone. 
Similarly, the deterioration in the estimation of $\mchirp^\mathrm{det}$ and $q$ for large $\dl{}_{\rm inj}$ and small $\iota{}_{\rm inj}$ is related to the deterioration in the estimation of $\msource{}_1$ near $(32\,\gigaparsec, 15^\circ)$.

To study the origin of this deterioration in the estimation of $\mchirp^\mathrm{det}$ and $q$, we have investigated the posterior of $\mchirp^\mathrm{det}$, $q$, and $\dl$. Taking $(\dl{}_{\rm inj},\iota_{\rm inj}) = (32\,\gigaparsec, 0^\circ)$ as an example; those posteriors are shown in Fig.~\ref{fig:improve_mcq}(c). 
Observe that, when the binary Love relations are used, a new peak arises in the 2D histogram of $\mchirp^\mathrm{det}$--$q$, in the lower left corner of the original one that covers the injected parameters. 
This means that the information from $\Psi_\mathrm{tid}$ with the aid of the binary Love relations favors smaller $\mchirp^\mathrm{det}$ and $q$ for large $\dl{}_{\rm inj}$ and small $\iota{}_{\rm inj}$, which deteriorates the measurement of the mass parameters. 

We note that the other region inside which the estimation of $\msource{}_1$ deteriorates, at about $(\dl{}_{\rm inj},\iota_{\rm inj}) = (2\,\gigaparsec, 45^\circ)$, does not have a counterpart in $\dl$, $\mchirp^\mathrm{det}$, or $q$ alone, although the improvements in these parameters are not high in that region. 
This is likely the result from competition between $\msource{}_1$ and $\msource{}_2$. As previously mentioned, the $\lambdatilde(M)$ function prefers improvements of the smaller $\msource{}_2$ mass. 
For the $(\dl{}_{\rm inj},\iota_{\rm inj}) = (2\,\gigaparsec, 45^\circ)$ injection, given that the total space for improvement from $\dl$, $\mchirp^\mathrm{det}$, and $q$ is small, 
the deterioration in the estimation of $\msource{}_1$ is likely responsible for the preferred improvement in the estimation of $\msource{}_2$. 

As a final remark, let us compare and contrast our results to those of \citet{Chatziioannou:2018vzf}. The latter also studied the impact of the binary Love relations in the estimation of the mass ratio, but concluded that the difference was negligible. 
In that work, the authors studied simulated signals detected by a network of 2G detectors. As we showed in Sec.~\ref{sec:compute_setup}, when 2G detections are made at moderate SNRs, the tidal effects in the signal are not strong enough to impact the estimation of the nontidal parameters; therefore, there is no conflict between our results and theirs.
Furthermore, in Fig.~8 in Ref.~\cite{Chatziioannou:2018vzf}, all the mass ratios estimated with the binary Love relations are smaller (although not significantly smaller) than those estimated without the relations. Our Fig.~\ref{fig:improve_mcq}(c) actually shows an enhanced version of this trend. Therefore, the deterioration reported here could be seen as an enhanced version of that observed in Ref.~\cite{Chatziioannou:2018vzf} as one may reasonably expect when going from 2G to 3G observations.

\section{Robustness of Forecasts}\label{sec:robstness}
In previous sections, we have made several assumptions to arrive at a forecast of how much improvement can be achieved in the measurement of various parameters. 
In this section, we investigate the robustness of these forecasts by relaxing some of our assumptions, which includes the accuracy of $\lambdazero$, the ignorance of the Hubble tension, and the universality of the binary Love relations. 
Because our main result is presented in terms of the improvements in 90\% CIs of BNS parameters, it is expected that a statistical uncertainty in $\lambdazero$ will widen the CIs in measurements that use binary Love relations and lead to weaker improvements than those presented in previous sections. 
However, as discussed in Sec.~\ref{sec:improve_dliota}, the improvements in $\dl$ and $\iota$ are primarily achieved by resolving the distance-inclination degeneracy, which appears as a large bias in many measurement cases. Therefore, we may expect that a systematic bias in $\lambdazero$ could affect our main conclusions, too. 
This is also the reason why we should consider the Hubble tension and the loss of universality with the binary Love relations---the former implies a bias in $\hubble$, and the latter implies a bias in the EOS.

Another interesting factor that can affect our main results is the timing accuracy. The timing at GW detection is usually accurate but not made use of in the parameter estimation of CBCs. We will show that, when the binary Love relations are used in parameter estimation, the timing information can impact the estimation of other parameters.

We end this section with a discussion of the usefulness of Fisher analysis in this work. We will show that our work is an example in which a Fisher analysis fails because of the nontrivial geometry of the likelihood, and, therefore, a full posterior analysis using sampling methods is necessary to produce accurate forecasts.

\subsection{Effect of uncertainty in $\lambdazero$}\label{subsec:robust_lam00}
Equation~\eqref{eq:lambda_rep} implies that uncertainty in $\lambdazero$ affects parameter estimation when using the binary Love relations. 
The current constraint obtained by C21 using GW170718 and its EM counterpart suggests that $\lambdazero=191^{+113}_{-134}$ to 90\% confidence. This error bar will shrink in the future by stacking observation of multimessenger BNS events. In particular, for $N$ similar observations the uncertainty in $\lambdazero$ should shrink by roughly $\sqrt{1/N}$.
Let us then imagine a future in which LIGO, Virgo, and KAGRA are operating jointly with the Rubin Observatory. According to Ref.~\cite{Andreoni:2021epw}, with 20 Rubin pointings one could expect $N=19$ EM and GW coincident events during the fifth GW observing run. If this were to occur, these coincident observations alone would reduce the 90\% CI of $\lambdazero$ to about $(113+134)/\sqrt{19} \approx 57$, before the 3G GW detectors start to operate and our proposed approach starts to help in parameter estimation.

We investigate these effects by taking the $(\dl{}_{\rm inj},\iota_{\rm inj}) = (4\,\gigaparsec, 30^\circ)$ injection as an example. We use the same computational setup as in Sec.~\ref{sec:compute_setup}, except that for the parameter estimation study, instead of assuming that $\lambdazero{}_{\rm inj}=200$, we impose two $\lambdazero$ priors to account for two types of uncertainty. 
One type is statistical in nature and we study it through a Gaussian prior on $\lambdazero$ with a mean of 200 and a standard deviation of 34 (in correspondence to a 90\% CI of 57). The other type is systematic and we study it through a delta function prior on $\lambdazero$ that is peaked at 234 instead of 200.

\begin{figure}[htbp]
    \centering
    \includegraphics[width=0.48\textwidth]{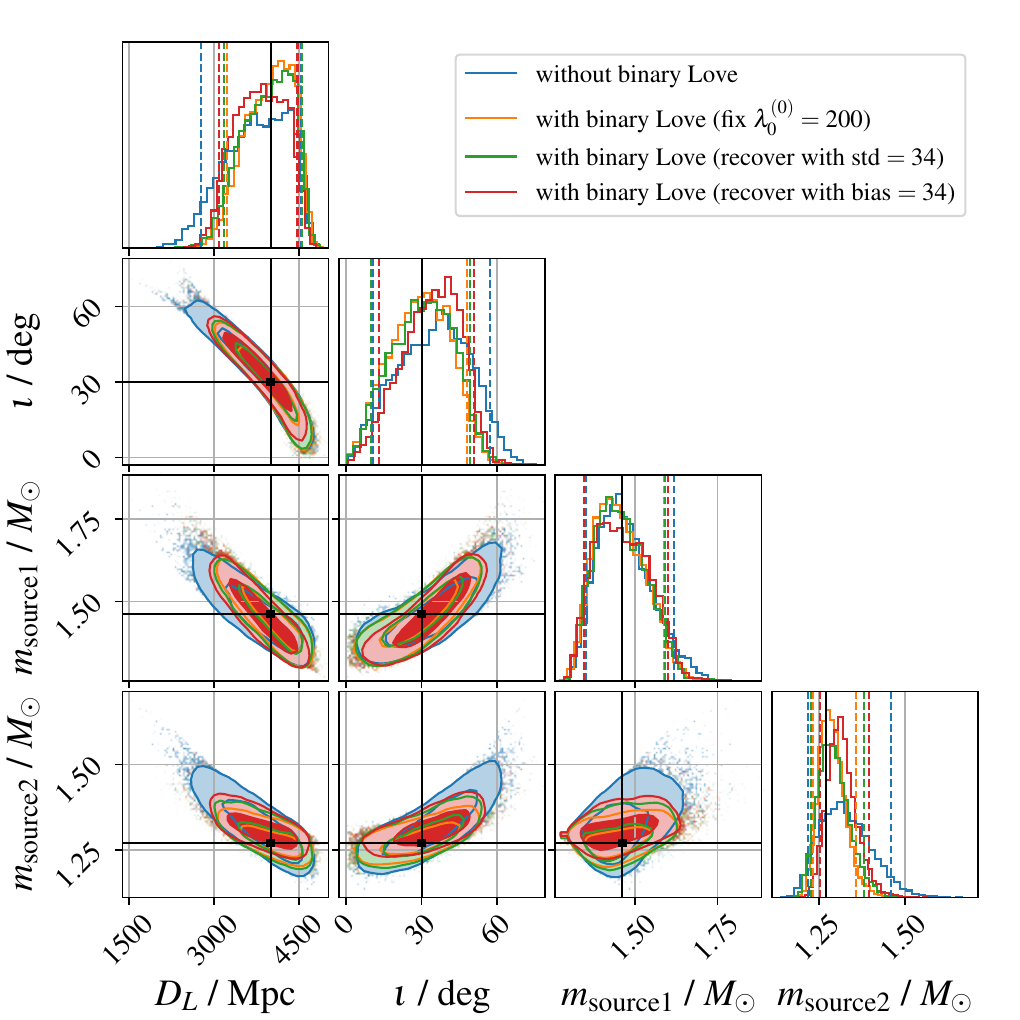}
    \caption{
    Effect of uncertainty in $\lambdazero$ on the estimation of distance, inclination, and source-frame masses. 
    The signal is synthesized using a GW170817-like source with $(\dl{}_{\rm inj},\iota_{\rm inj})=(4\,\gigaparsec,30^\circ)$ and $\lambdazero{}_{\rm inj}=200$ and detected by a 3G HLV-like network.
    The corner plots show posteriors recovered using a model that directly samples on $\lambdatilde_A$ (blue), a model that uses binary Love relations and correctly fixes $\lambdazero=200$ (orange), a model that uses binary Love relations but samples on $\lambdazero$ with a Gaussian prior whose mean is 200 and standard deviation is 34 (green), and a model that uses binary Love relations but fixes $\lambdazero$ at 234 instead of 200 (red). 
    Observe that the posteriors in green and red are close to the posterior in orange, compared with their differences from the posterior in blue. This means that neither the statistical error nor systematic bias in $\lambdazero$ significantly affects the level of improvement.}
    \label{fig:robust_lam00}
\end{figure}

The parameter estimation results, in terms of the posterior corner plots of $\dl,\,\iota,\,\msource{}_1$, and $\msource{}_2$, are shown in Fig.~\ref{fig:robust_lam00}.
Observe that, when the binary Love relations are used, the marginalized posteriors are insensitive to the statistical error or the systematic bias added to $\lambdazero$. 
Compared to the posteriors obtained without the binary Love relations, the posteriors obtained with the binary Love relations show the same level of improvement as before.

\subsection{Effect of uncertainty in the binary Love relations}\label{subsec:robust_lam0k}
In previous sections, we have assumed that the binary Love relations are perfectly EOS independent. However, a certain loss of universality exists as one varies the EOS, and this can in principle affect parameter estimation. 
To study this, we investigate a $(\dl{}_{\rm inj},\iota_{\rm inj}) = (4\,\gigaparsec, 30^\circ)$ injection with an assumed EOS and attempt to extract it with a model that uses the binary Love relations instead of assuming a particular EOS.
For the assumed EOS we choose MPA1 because it has the largest residual among all EOSs used to fit the $\lambdazero$--$\lambdak$ relation in C21 
(see Appendix~\ref{app:blove_residual} for more details about this residual.)
To avoid confusion, we fix $\lambdazero$ to be the exact tidal deformability of a $1.4\,\msun$ neutron star with a MPA1 EOS, since the effect of the uncertainty of $\lambdazero$ on parameter estimation was discussed in Sec.~\ref{subsec:robust_lam00}.

\begin{figure}[htbp]
    \centering
    \includegraphics[width=0.48\textwidth]{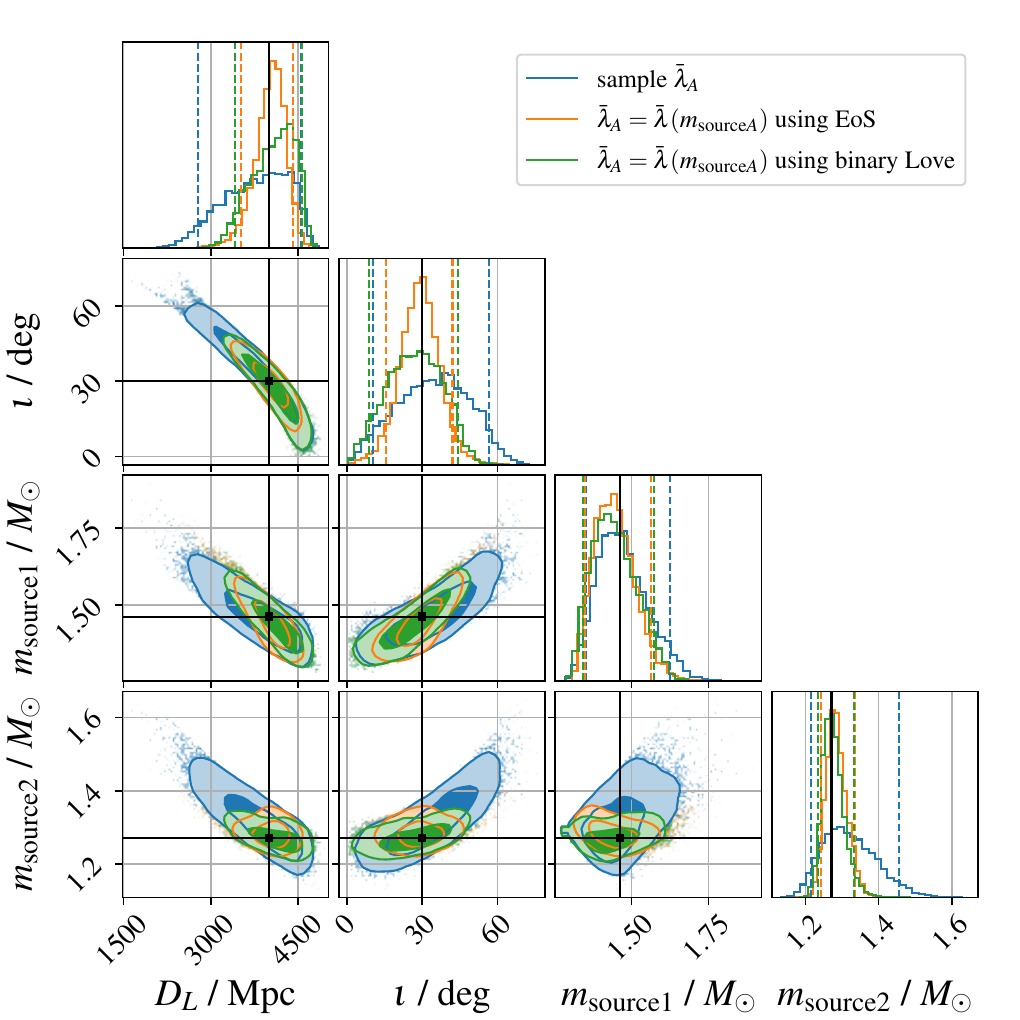}
    \caption{Effect of uncertainty in the binary Love relations on the estimation of distance, inclination and source-frame masses. 
    The signal is synthesized using a GW170817-like source with $(\dl{}_{\rm inj},\iota_{\rm inj})=(4\,\gigaparsec,30^\circ)$, and detected by a 3G HLV-like network. The injected tidal deformability parameters are computed using the MPA1 EOS from the injected source-frame masses. 
    The corner plots show posteriors recovered using a model that directly samples on $\lambdatilde_A$ (blue), a model that writes $\lambdatilde_A$ as a function of $\msource{}_A$ with the correct MPA1 EOS (orange), and a model that writes $\lambdatilde_A$ as a function of $\msource{}_A$ with the binary Love relations and $\lambdazero$ fixed to that of a $1.4\,\msun$ according to MPA1 (green). 
    Observe that the posterior in green is close to the posterior in orange, compared with their differences from the posterior in blue. This means that the EOS sensitivity of the binary Love relations does not significantly affect the level of improvement.}
    \label{fig:robust_lam}
\end{figure}

Corner plots for $\dl,\,\iota,\,\msource{}_1$, and $\msource{}_2$ are shown in Fig.~\ref{fig:robust_lam}. These plots show the accuracy of parameter estimation when (i) the model does not use the binary Love relations and samples on $\lambdatilde_A$ directly (blue), (ii) the model does not use the binary Love relations but the tidal deformabilities are computed using the (``correct'') MPA1 EOS from the sampled source-frame masses (orange), and (iii) the model does use the binary Love relations and we fix $\lambdazero$ to that of a $1.4\,\msun$ with a MPA1 EOS (green). Observe that the posteriors using the binary Love relations are very similar to those found when using the correct EOS (especially in terms of the 90\% CIs and their peak likelihoods). Therefore, the improvement in parameter estimation due to the binary Love relations is not affected by the EOS sensitivity of the relations themselves. 

\subsection{Effect of uncertainty in $\hubble$}
Equation~\eqref{eq:lambda_rep} implies that uncertainty in $\hubble$ also affects parameter estimation when using the binary Love relations. 
In previous sections, we used the Planck measurement of $\hubble=67.66\,\hubbleunit$~\cite{Planck:2018vyg} in our simulations. Late time cosmological observations make use of local-Universe supernovae, which generally gives an $\hubble$ value around $73\,\hubbleunit$ (see, for example, Refs~\cite{Riess:2021jrx,Freedman:2021ahq,Anand:2021sum}). 

\begin{figure}[htbp]
    \centering
    \includegraphics[width=0.48\textwidth]{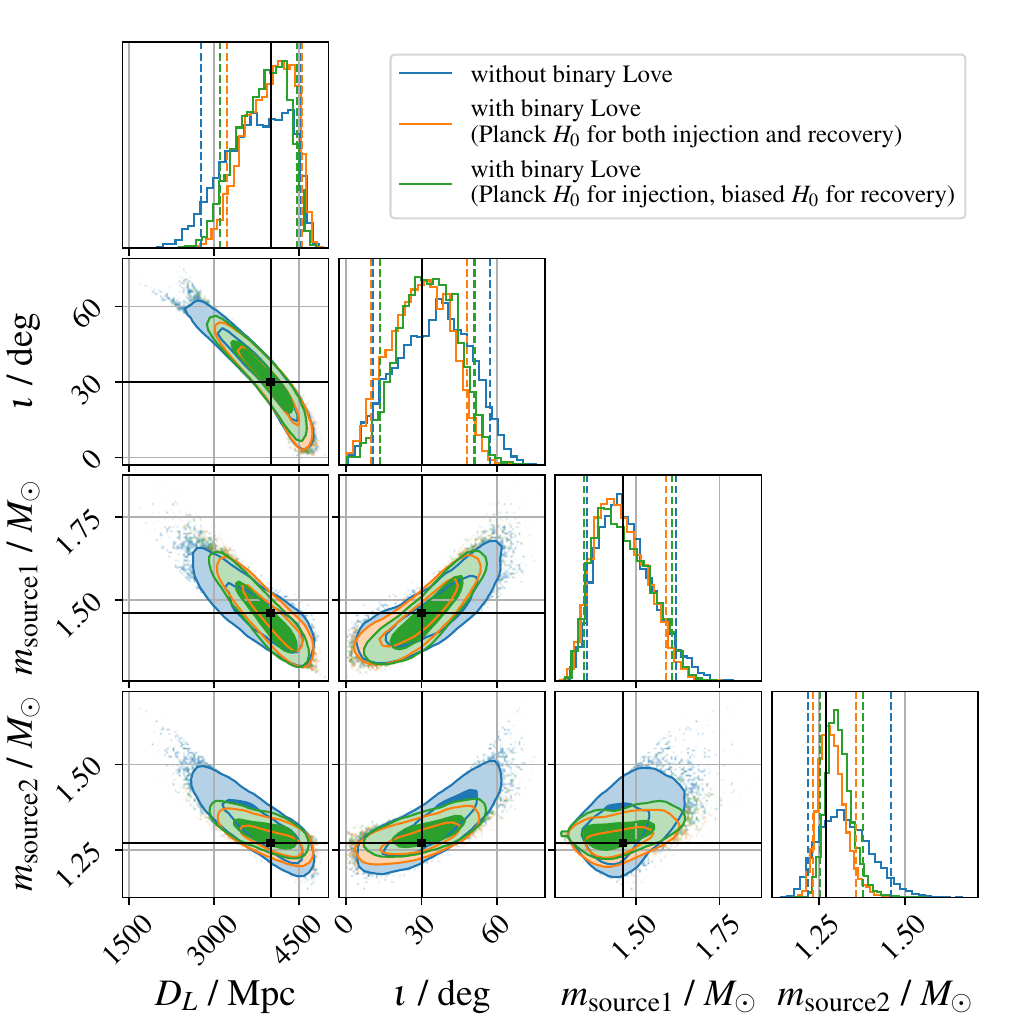}
    \caption{Effect of uncertainty in the Hubble constant on the estimation of distance, inclination, and source-frame masses. 
    The signal is synthesized using a GW170817-like source with $(\dl{}_{\rm inj},\iota_{\rm inj})=(4\,\gigaparsec,30^\circ)$, and detected by a 3G HLV-like network. Also, the Planck $\hubble=67.66\,\hubbleunit$ is assumed in the injection. 
    The corner plots show posteriors recovered using a model that directly samples on $\lambdatilde_A$ (blue), a model that uses the binary Love relations with the Planck $\hubble$ (orange), and a model that uses the binary Love relations with $\hubble=73\,\hubbleunit$ which is the typical result from local-Universe measurements (green). 
    Observe that the posterior in green is close to the posterior in orange, compared with their differences from the posterior in blue. This means that the Hubble tension does not significantly affect the level of improvement.}
    \label{fig:robust_h0}
\end{figure}

We now investigate whether our use of the binary Love relations to better estimate the parameters of the binary is affected by an error in our assumed value of the Hubble constant. We consider an injection at
$(\dl{}_{\rm inj},\iota_{\rm inj}) = (4\,\gigaparsec, 30^\circ)$ with the Planck value of $\hubble$, and  extract it with three models: one that does not use the binary Love relations, one that does use them and fixes $\hubble$ to the Planck value, and one that uses $\hubble=73\,\hubbleunit$ instead. The corner plots for $\dl,\,\iota,\,\msource{}_1$, and $\msource{}_2$ for these three models are shown in Fig.~\ref{fig:robust_h0}. 
Observe that the ``Hubble tension''  causes a tiny shift in the peak of the marginalized posteriors obtained using the binary Love relations; this shift, however, is small and fits completely within the 90\% CI. Clearly then, this effect does not affect the overall improvement as compared to parameter estimation without the binary Love relations. 

\subsection{Effect of better timing accuracy}
Another factor that can affect our result is the accuracy in the estimation of the arrival time of the signal, or ``timing accuracy'' for short. 
In the GW model, the arrival time of the signal affects the frequency-domain GW by adding $2\pi ft_c$ to the phase, where $t_c$ is the time of arrival at the geocenter. 
In terms of PN expansions, this $2\pi ft_c$ term is of 4PN relative order, which is close to the 5PN relative order term where the tidal effects first appear. Therefore, an error in the time of arrival could impact estimation of the tidal parameters and, hence, also affect the estimation of distance, inclination, and source-frame masses when the binary Love relations are used. 
In previous sections, we followed the standard LIGO prior setup and used a wide, flat prior for $t_c$ that covers $t_{c\rm inj}\pm0.1$\,s. With this prior in hand, we then carried out parameter estimation, including $t_c$ in our parameter array of the BNS GW model. 
This procedure assumes that $t_c$ is determined only by matching the signal to the BNS GW model. However, in reality, $t_c$ can also be estimated at detection by maximizing the SNR over $t_c$. If the SNR is high, then the timing accuracy at detection may also be high, and one can then use this to set a tighter $t_c$ prior when one later carries out BNS parameter estimation. 

\begin{figure}[htbp]
    \centering
    \includegraphics[width=0.48\textwidth]{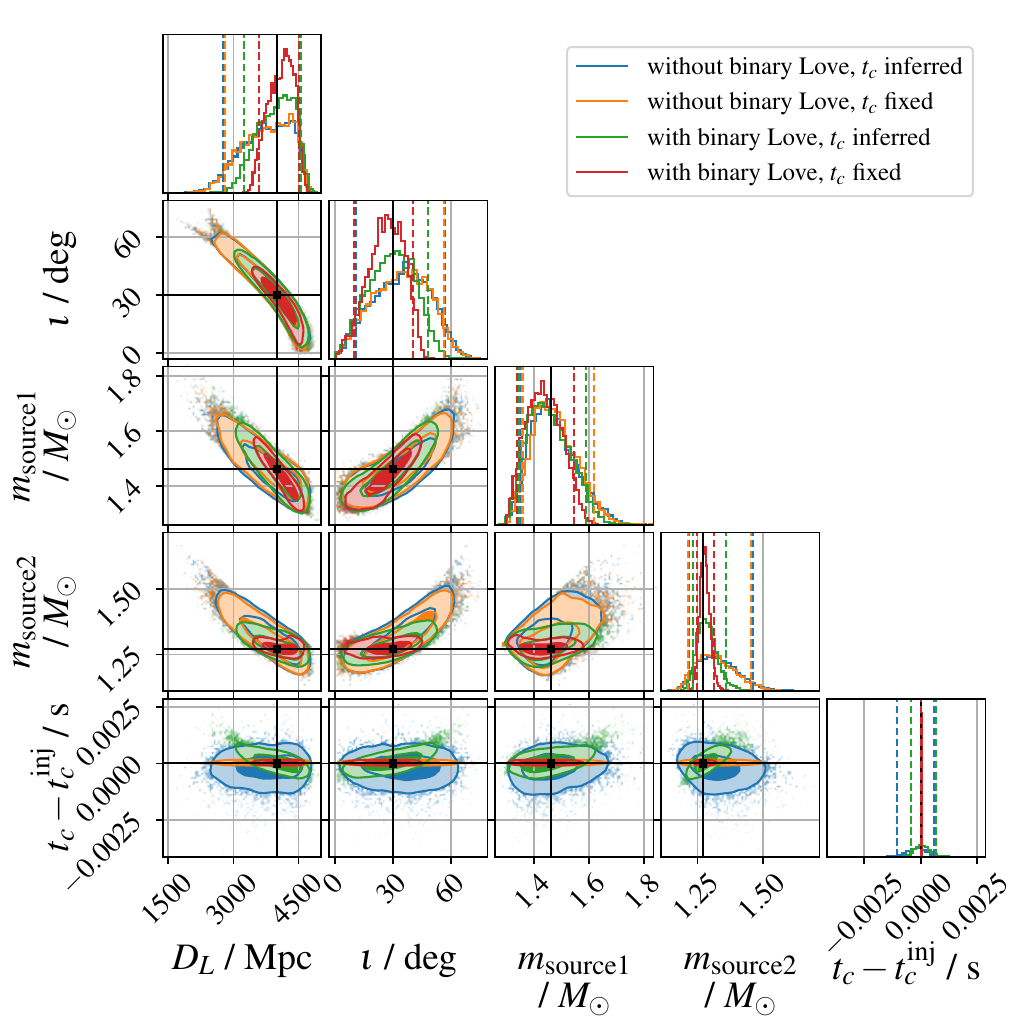}
    \caption{Effect of better timing accuracy on the estimation of distance, inclination, and source-frame masses. 
    The signal is synthesized using a GW170817-like source with $(\dl{}_{\rm inj},\iota_{\rm inj})=(4\,\gigaparsec,30^\circ)$, and detected by a 3G HLV-like network. 
    The corner plots show posteriors recovered using a model that directly samples on $\lambdatilde_A$ and uses a flat prior for $t_c$ covering $t_{c\rm inj}\pm0.1\,{\rm s}$ (blue), a model that directly samples on $\lambdatilde_A$ and fixes $t_c=t_{c\rm inj}$ (orange), a model that uses the binary Love relations and a flat prior for $t_c$ covering $t_{c\rm inj}\pm0.1\,{\rm s}$ (green), and a model that uses the binary Love relations and fixes $t_c=t_{c\rm inj}$ (red). 
    Observe that the improvement in distance, inclination, and source-frame masses from blue to green is smaller than that from orange to red. 
    This means that a better timing accuracy will allow the binary Love relations to lead to an even larger improvement on the estimation of these parameters.}
    \label{fig:robust_tc}
\end{figure}

Let us then investigate whether this tighter prior leads to a better estimate of the system parameters when using the binary Love relations. To study this, we focus on a $(\dl{}_{\rm inj},\iota_{\rm inj})=(4\,\gigaparsec, 30^\circ)$ injection using two models, each with two priors: one with the same wide and flat prior on $t_c$ as before, and one with a delta-function $t_c$ prior (centered at the injected value).  
The corner plots for $\dl,\,\iota,\,\msource{}_1,\,\msource{}_2$, and $t_c$ using these four cases is shown in Fig.~\ref{fig:robust_tc}. 
Observe that when the binary Love relations are used, the tighter $t_c$ prior leads to narrower posteriors on $\dl,\,\iota,\,\msource{}_1$, and $\msource{}_2$. Meanwhile, when the binary Love relations are not used, the tighter $t_c$ prior does not change the posteriors. 
Comparing the green with the blue in the last row in Fig.~\ref{fig:robust_tc}, we also confirm that the difference is made because the binary Love relations add to the correlation between $t_c$ and the other parameters.
Therefore, better timing accuracy will allow the binary Love relations to lead to an even larger improvement on parameter estimation. 

\subsection{Failure of the Fisher analysis}
Fisher analysis has been widely used in the GW community to generate fast estimates of measurement errors. This method approximates the posterior as a single-peaked Gaussian distribution, whose inverse covariance matrix is constructed from second derivatives of the log-likelihood. 
In this work, we find that the Fisher approximation is insufficient to predict the accuracy to which parameters can be estimated, and instead, we have to run a full posterior analysis using numerical sampling methods such as nested sampling. The reasons for this, as we show below, is that the likelihood surface is not single peaked (there are secondary peaks that are important), and the tallest peak of the likelihood is not a Gaussian (there are long tails in the distribution). 

\begin{figure}[htbp]
    \centering
    \includegraphics[width=0.48\textwidth]{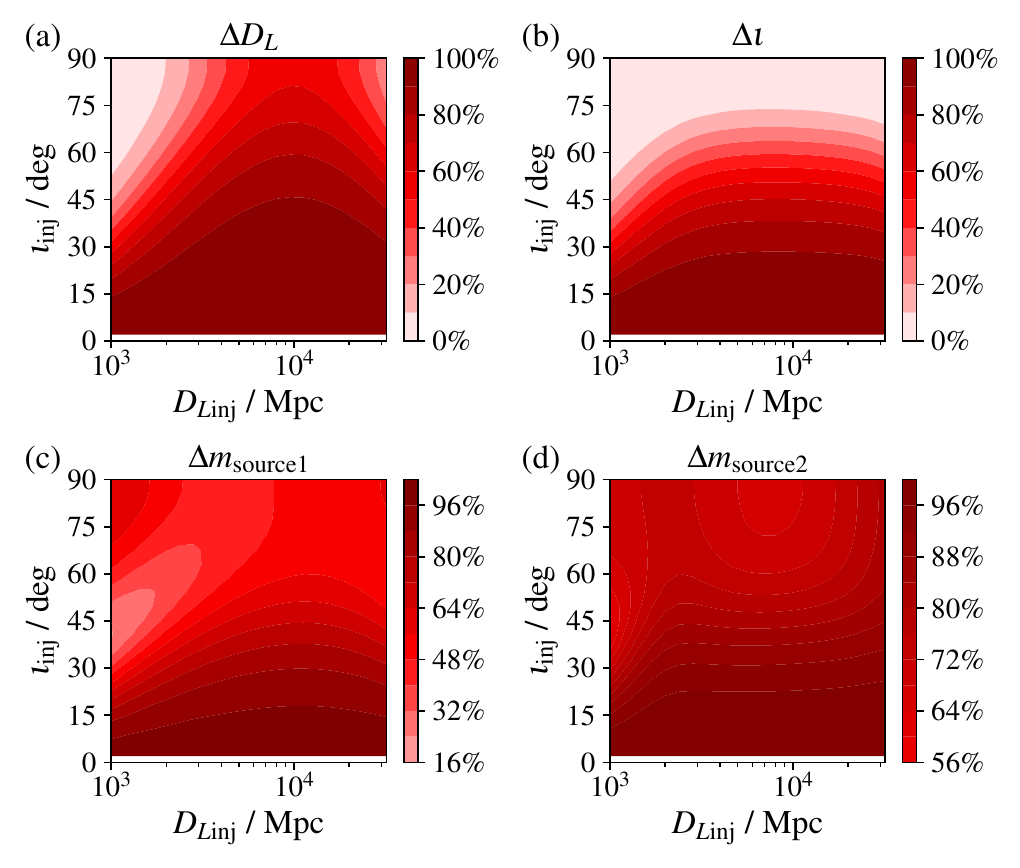}
    \caption{Relative improvement in the estimation of (a) the distance, (b) the inclination and (c),(d) the source-frame masses, suggested by Fisher analysis. 
    The signals are synthesized using a GW170817-like source detected by a 3G HLV-like network. 
    We let $\dl{}_{\rm inj}$ and $\iota_{\rm inj}$ vary within the same ranges as for the grid described in Sec.~\ref{sec:compute_setup}, except that we skipped $\iota<2^\circ$ to avoid numerical issue in calculations, leaving a white band at the bottom of each plot. 
    (a)--(d) are Fisher counterparts of the full posterior results in Figs.~\ref{fig:robust_fisher}(a)--(d) with Figs.~\ref{fig:improve_dliota}(c) and \ref{fig:improve_dliota}(g) and Figs.~\ref{fig:improve_masses}(c).
    Observe that the patterns given by the full posterior analysis are poorly reproduced here. 
    }
    \label{fig:robust_fisher}
\end{figure}

In Fig.~\ref{fig:robust_fisher} we show Fisher estimates of the relative fractional improvement in the accuracy to which parameters can be measured when using the binary Love relations. The analysis is performed using the GW Fisher analysis package GWBENCH~\cite{Borhanian:2020ypi} on the same injection grid as that used in Sec.~\ref{sec:compute_setup}.
Comparing Figs.~\ref{fig:robust_fisher}(a)--(d) with Figs.~\ref{fig:improve_dliota}(c) and \ref{fig:improve_dliota}(g) and Figs.~\ref{fig:improve_masses}(c) and \ref{fig:improve_masses}(g), we see that the patterns given by the full posterior analysis are poorly reproduced by the Fisher analysis. 
In particular, the Fisher results fail to show the significant improvement for edge-on systems and the negative improvement in the primary source-frame mass. 
As has been discussed in Secs.~\ref{sec:improve_dliota} and \ref{sec:improve_masses}, the improvement for edge-on systems is related to the tail in the likelihood, and the deterioration in the mass estimate is related to a secondary peak in the posterior. Neither feature can be captured by the single-peaked and Gaussian approximation inherent to Fisher theory. Thus, our study provides an example in which Fisher analysis fails~\cite{Vallisneri:2007ev}.

\section{Conclusions}\label{sec:conclusion}
In this study, we present an application of the binary Love relations to constrain the distance-inclination
degeneracy in GW parameter estimation, finding significant improvement in the estimation of parameters including distance, inclination, and source-frame mass. This work is closely related to the measurement of $\hubble$ using
binary Love relations reported in C21~\cite{Chatterjee:2021xrm}. The binary Love relations allow the NS tidal deformability $\lambdatilde$
to be written as a function of the source-frame mass $\msource$ in an EOS-insensitive way that is controlled by a
constant, $\lambdazero$. The value of $\lambdazero$ is universal, given $m_0$, and will be constrained by
stacking future multimessenger BNS observations. When this is combined together with precise measurement of cosmological
parameters, i.e., the distance-redshift relation, one can reparametrize the BNS waveform by replacing the tidal deformability
parameters $\lambdatilde_A$ with the detector-frame masses $\mdet{}_A$ and the luminosity distance $\dl$. 
This reparameterization allows $\dl$ information to enter not only in the amplitude, but also the phase of the waveform through
the tidal term, $\Psi_\mathrm{tid}$. Hence, it allows better distance-inclination measurements than the traditional approach of
inferring $\dl$ and $\iota$ solely from the amplitude.

We demonstrate this prescription by performing Bayesian parameter estimation on synthetic GW signals and showing relative improvement 
in $\dl$--$\iota$ and source-frame masses in the era of 3G detectors. 
In particular, we find that the improvement peaks for face-on and edge-on BNS systems at $\dl \sim 4\,\gigaparsec$, with up to $\sim 70\%$
decrease in the 90\% CI of $\dl$--$\iota$, and up to $\sim 50\%$ decrease in that of the source-frame masses. 
The use of the binary Love relations also makes it possible to put reasonable constraints on $\dl$ and $\iota$ for edge-on systems with SNR as low as 10. 
On the other hand, the SNR threshold for constraining the relative error of $\dl$ to below $50\%$ is halved. 
A similar decrease in the SNR threshold is observed for constraining the absolute error of $\iota$ below $45^\circ$ and the relative error of $\msource{}_2$ below $20\%$, respectively.
The improvement in $\msource{}_1$ is weaker, and a small deterioration is observed in certain situations. This is because
$\lambdatilde(M)$ is less sensitive to the larger, primary mass and the fact that the use of the binary Love relations can weaken the
estimation of $\mchirp{}^{\rm det}$ and $q$ in certain circumstances (see Fig.~\ref{fig:improve_mcq}). A detailed investigation into the
reason for this deterioration is left for future study.
We report that the uncertainty in $\lambdazero$, the uncertainty in the NS EOS, and uncertainty in the Hubble parameter do not significantly affect
our application of the binary Love relation to constrain $\dl$--$\iota$. In addition, if the time of arrival is well measured, the improvement
reported in the main results is further enhanced. We have also shown that our results cannot be accurately reproduced by Fisher analysis; a full
Bayesian analysis is necessary.

Our prescription has direct application to measuring the source-frame parameters of BNS systems from GW data alone.
This is relevant since not all future BNS mergers are expected to have observable electromagnetic counterparts.
An improved measurement of luminosity distance, along with known cosmological parameters, leads to a better inference of the redshift,
which, in turn, leads to improved estimates of the source-frame masses $\msource{}_A$ of future BNS systems. The increased number of detections expected in the near future, combined with these improved mass measurements, will, in turn, result in improved estimation of the population
properties of NSs and, in particular, the BNS mass distribution. Moreover, when combined with a search for GRB counterparts, our
procedure of improving inclination measurements will help in joint EM-GW observations~\cite{2020ApJ...895..108F}. For example, our approach will allow for improved constraints on the jet
opening angle for cases where a GRB is also detected, since the half opening angle of
the merger jet has to be comparable to the inclination angle of the BNS. On the other hand, in the absence of a GRB observation,
the improved inclination angle estimation can be used to assist subthreshold searches.

Other approaches to address the distance-inclination degeneracy have been reported that use higher-order modes and precession~\cite{London:2017bcn,Vitale:2018wlg}. In comparison, our approach of using the tidal effects is novel, and is specialized for BNSs since the
higher-order modes and precession are suppressed owing to the near-equal mass ratio of BNS systems as well as their low spin. While higher-order modes and precession
of BNSs may be an avenue for high SNR detections in the 3G era (see, for example, Ref.~\cite{CalderonBustillo:2020kcg}), for the majority of detections made at moderate-to-low SNRs, without a counterpart,
the improvement from the tidal effects in conjunction with the binary Love relations presented in this work will be crucial.

Our approach uses the binary Love relations as a substitute for the NS EOS to implement the $\lambdatilde(M)$ function. This strategy is advantageous, because there are currently many possible EOSs that are consistent with observations, and the binary Love relations offer a tractable representation for all of them. 
However, if and when the EOS is better determined, one may use the selected EOS to directly construct $\lambdatilde(M)$, and one should then expect similar improvements in parameter estimation as those presented in this work.
We also note that the binary Love relations will lose universality if strong phase transitions are considered~\cite{Tan:2021nat,Carson:2019rjx}. However, the existence of these phase transitions in NSs is still under investigation, and a discussion incorporatinng phase transitions is beyond the scope of this work. 

\begin{figure*}[htbp]
    \centering
    \includegraphics[width=0.98\textwidth]{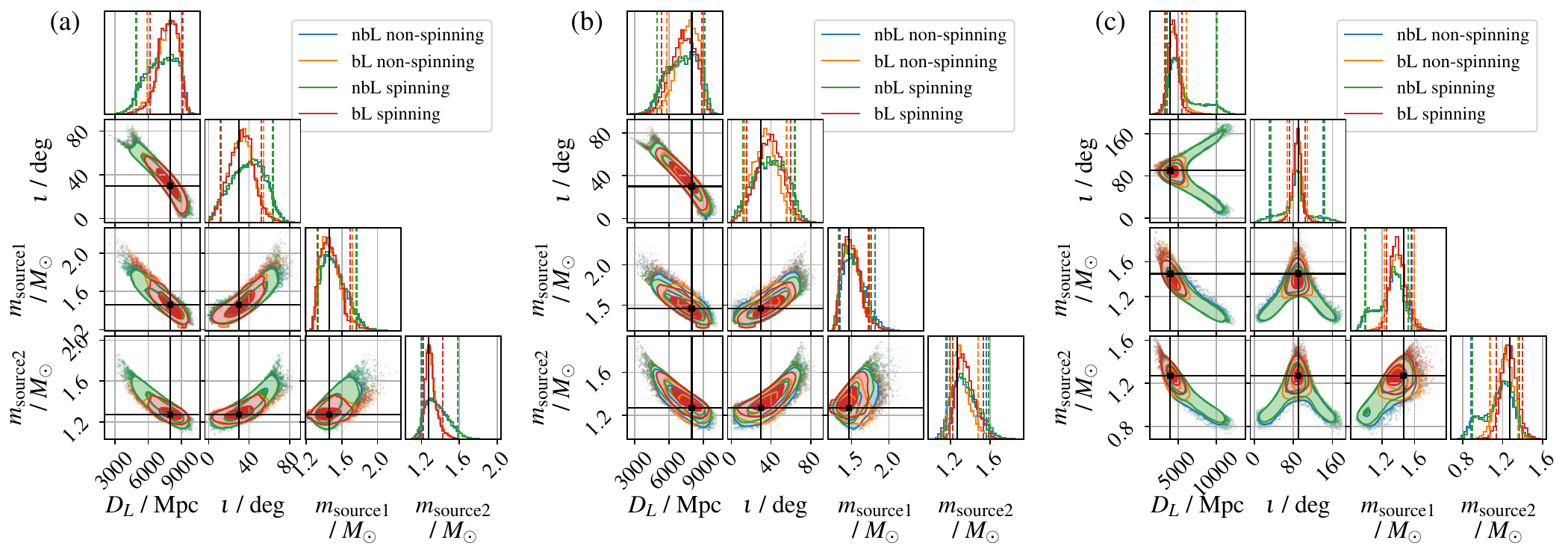}
    \caption{In this figure, we show the effect of spins on the estimation of distance, inclination, and source-frame mass measurements. The simulated signals are from a GW170817-like source at $(\dl{}_{\rm inj},\iota_{\rm inj})=(8\,\gigaparsec,30^\circ)$ for (a) and (b), and at $(\dl{}_{\rm inj},\iota_{\rm inj})=(4\,\gigaparsec,90^\circ)$ for (c). The source is detected by a 3G HLV-like network. In (a), we fix the spins to the injected values during sampling. In (b) and (c), we use fully precessing priors with spin magnitudes up to $\chi_a = 0.05$.  In each panel, we compare a nonspinning injection with one with both aligned spins components $\chi_{a,\rm inj} = 0.02$. For each case, a pair of posteriors are shown with the label ``bL'' indicating the use of binary Love relations and the label ``nbL'' indicating the absence. We find in (a), that the measurements are similar between the spinning versus nonspinning case. In (b), there are some differences coming from correlations between mass, spins, and distance during recovery. However, in (c), we see that this difference between use of fixed spin versus precessing spin priors is small compared to the overall improvement for edge-on inclined sources, which are most promising in terms of the use of binary Love relations.}
    \label{fig:robust_spins}
\end{figure*}

\acknowledgments
G. H., Y. X., and N.Y. acknowledge support from NSF Grant No.~2009268. 
K.Y. acknowledges support from NSF Grants No.~PHY-1806776 and No.~PHY-2207349, a Sloan Foundation Research Fellowship, and the Owens Family Foundation.
D.C. acknowledges support from the Illinois Survey Science Fellowship from the Center for AstroPhysical Surveys at NCSA,
University of Illinois Urbana-Champaign, and the support from subawards to MIT under NSF Grants No.~OAC-2117997 and No.~PHY-1764464. 
G.H. has support from the Canadian Institute for Advanced Research and is supported by Brand Fortner. 
D.E.H. acknolwedges support from NSF AST-2006645, as well as support from the Kavli Institute for Cosmological Physics through an endowment from the Kavli Foundation and its founder Fred Kavli. 
D.E.H. also gratefully acknowledges support from the Marion and Stuart Rice Award.
S.P. acknowledges partial support by the Center for AstroPhysical Surveys (CAPS) at the National Center for Supercomputing Applications (NCSA), University of Illinois Urbana-Champaign.
This work made use of the Illinois Campus Cluster, a computing resource that is operated by the Illinois Campus Cluster Program (ICCP) in conjunction
with NCSA, and is supported by funds from the University of Illinois at Urbana-Champaign.
The authors thank Philippe Landry for reviewing the document and providing helpful feedback. This document is given the LIGO DCC No.~P2200299.

\begin{figure*}[ht]
    \centering
    \includegraphics[width=0.85\textwidth]{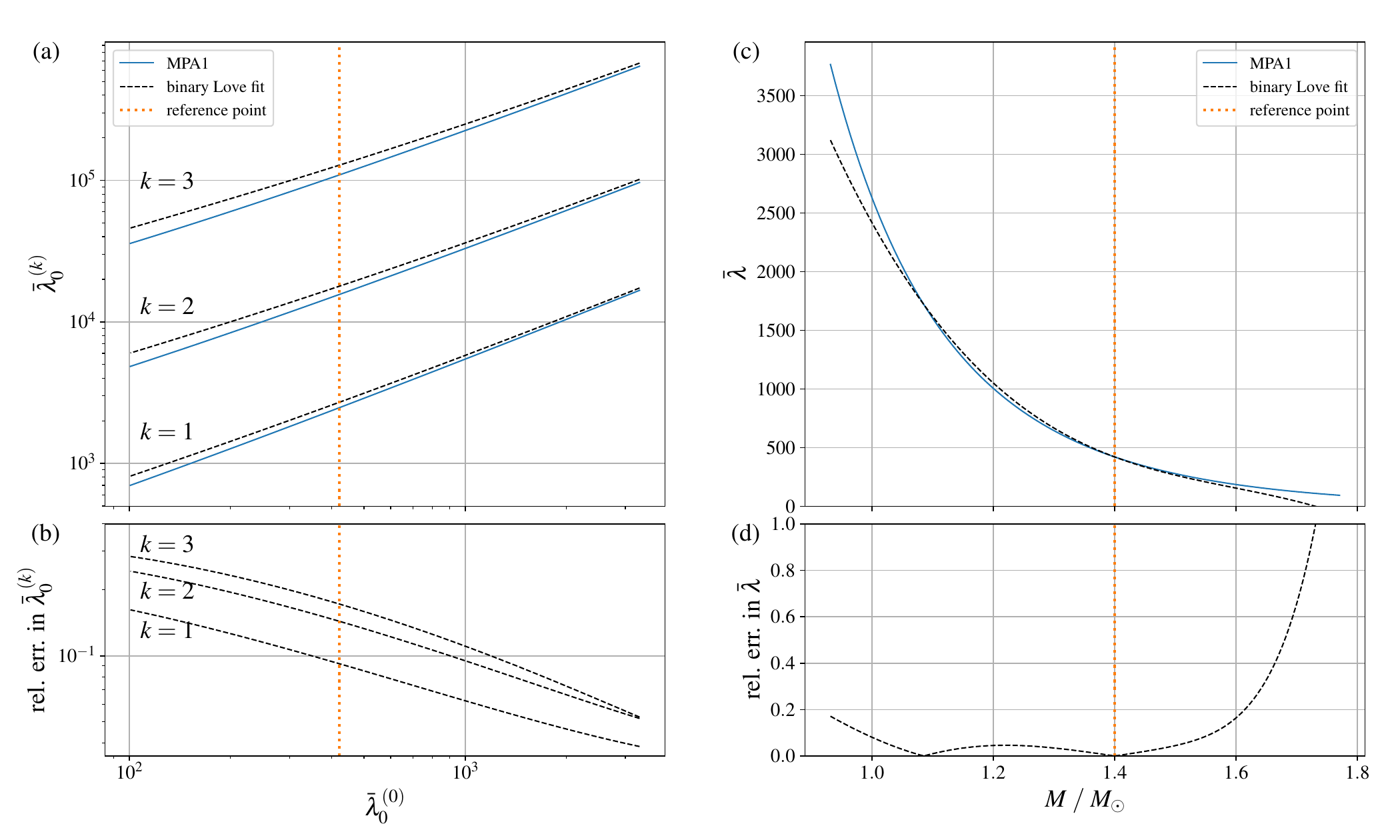}
    \caption{Comparison between the MPA1 EOS and its binary Love fit, in terms of (a),(b) the $\lambdazero$--$\lambdak$ relation and (c),(d) the $\lambdatilde(M)$ function. The orange dotted lines correspond to the reference point used to reconstruct the $\lambdatilde(M)$ function through Eq.~\eqref{eq:lambda_m}, which is at $m_0=1.4\,\msun$ and $\lambdazero=\lambdatilde_\mathrm{MPA1}(m_0)=422$. }
    \label{fig:blove_res_mpa1}
\end{figure*}

\appendix
\section{EFFECT OF SPINS}\label{app:effect_of_spins}
In this section, we compare the measurement of distance, inclination, and mass coming from a non-spinning binary versus one with nonzero spin. In Fig.~\ref{fig:robust_spins} we extend the result shown Fig.~\ref{fig:compare_detectors}(b) by doing a comparison between a spinning and a nonspinning BNS in the 3G era. In Fig.~\ref{fig:robust_spins}(a) we explore the spinning case, setting the true dimensionless aligned spin $\chi_a = 0.02$, and fixing the spins to their injected values during parameter estimation. We find that there is little impact between the spinning and the nonspinning cases. However, in Fig.~\ref{fig:robust_spins}(b) we recover the same injection with a fully precessing prior with spin magnitudes up to $\chi_a = 0.05$. We note that the results deteriorate slightly from the non-spinning case, but overall still lead to a more constrained measurement by the use of the binary Love relation. We expect some deterioration due to the correlation of luminosity distance, which now appears in the phase of the waveform [see Eq.~(\ref{eq:psi_tid}, \ref{eq:lambda_rep})] with the other intrinsic parameters, like the masses and spins which are recovered at lower PN order. However, we find that this effect is small compared to the improvement obtained using the binary Love relation for close to edge-on sources, which is part of the parameter space where the technique is most promising. In Fig.~\ref{fig:robust_spins}(c), we redo the injection for an inclined edge-on system at 4 Gpc [the same injection as Fig.~\ref{fig:degeneracy_cases}(d)] with a fully precessing spin prior. We find a~$\sim 70\%$ ($\sim 50\%$) improvement in the $90\%$ confidence interval for the distance and inclination (source-frame masses), for both the fixed versus precessing spin prior cases.

\section{BINARY LOVE FITTING RESIDUAL FROM THE MPA1 EOS}\label{app:blove_residual}
In Sec.~\ref{subsec:robust_lam0k}, we compared parameter estimation results using an EOS and its binary Love fit. In particular, we chose MPA1 for the EOS because this EOS leads to the largest residual between any EOSs we studied and fit to the $\lambdazero$--$\lambdak$ relation. Here, we show that residual in Fig.~\ref{fig:blove_res_mpa1}. 
In Figs.~\ref{fig:blove_res_mpa1}(a) and \ref{fig:blove_res_mpa1}(b), we compare the fitted $\lambdazero$--$\lambdak$ relation with that assuming the MPA1 EOS. Observe that the difference increases for larger $k$. 
The actual $\lambdazero$ and $\lambdak$ used in Sec.~\ref{subsec:robust_lam0k} are labeled by the orange dotted line, which corresponds to $\lambdazero=\lambdatilde_\mathrm{MPA1}(m_0)=422$, where the reference mass is $m_0=1.4\,\msun$. At this point of parameter space, the relative fitting errors of $\lambdak$ are all below $20\%$.
Using these fitted $\lambdak$ at the reference point, we reconstruct the $\lambdatilde(M)$ function through Eq.~\eqref{eq:lambda_m}. The result is compared with $\lambdatilde_\mathrm{MPA1}(M)$ in Figs.~\ref{fig:blove_res_mpa1}(c) and \ref{fig:blove_res_mpa1}(d). The relative error of this fit is up to $\sim10\%$ [$\sim20\%$] for $M\in(1.0,1.5)\,\msun$ [$M\in(0.9,1.6)\,\msun$].


\begin{thebibliography}{79}%
\makeatletter
\providecommand \@ifxundefined [1]{%
 \@ifx{#1\undefined}
}%
\providecommand \@ifnum [1]{%
 \ifnum #1\expandafter \@firstoftwo
 \else \expandafter \@secondoftwo
 \fi
}%
\providecommand \@ifx [1]{%
 \ifx #1\expandafter \@firstoftwo
 \else \expandafter \@secondoftwo
 \fi
}%
\providecommand \natexlab [1]{#1}%
\providecommand \enquote  [1]{``#1''}%
\providecommand \bibnamefont  [1]{#1}%
\providecommand \bibfnamefont [1]{#1}%
\providecommand \citenamefont [1]{#1}%
\providecommand \href@noop [0]{\@secondoftwo}%
\providecommand \href [0]{\begingroup \@sanitize@url \@href}%
\providecommand \@href[1]{\@@startlink{#1}\@@href}%
\providecommand \@@href[1]{\endgroup#1\@@endlink}%
\providecommand \@sanitize@url [0]{\catcode `\\12\catcode `\$12\catcode
  `\&12\catcode `\#12\catcode `\^12\catcode `\_12\catcode `\%12\relax}%
\providecommand \@@startlink[1]{}%
\providecommand \@@endlink[0]{}%
\providecommand \url  [0]{\begingroup\@sanitize@url \@url }%
\providecommand \@url [1]{\endgroup\@href {#1}{\urlprefix }}%
\providecommand \urlprefix  [0]{URL }%
\providecommand \Eprint [0]{\href }%
\providecommand \doibase [0]{https://doi.org/}%
\providecommand \selectlanguage [0]{\@gobble}%
\providecommand \bibinfo  [0]{\@secondoftwo}%
\providecommand \bibfield  [0]{\@secondoftwo}%
\providecommand \translation [1]{[#1]}%
\providecommand \BibitemOpen [0]{}%
\providecommand \bibitemStop [0]{}%
\providecommand \bibitemNoStop [0]{.\EOS\space}%
\providecommand \EOS [0]{\spacefactor3000\relax}%
\providecommand \BibitemShut  [1]{\csname bibitem#1\endcsname}%
\let\auto@bib@innerbib\@empty
\bibitem [{\citenamefont {Abbott}\ \emph
  {et~al.}(2016{\natexlab{a}})\citenamefont {Abbott} \emph
  {et~al.}}]{LIGOScientific:2016aoc}%
  \BibitemOpen
  \bibfield  {author} {\bibinfo {author} {\bibfnamefont {B.~P.}\ \bibnamefont
  {Abbott}} \emph {et~al.} (\bibinfo {collaboration} {LIGO Scientific,
  Virgo}),\ }\bibfield  {title} {\bibinfo {title} {{Observation of
  Gravitational Waves from a Binary Black Hole Merger}},\ }\href
  {https://doi.org/10.1103/PhysRevLett.116.061102} {\bibfield  {journal}
  {\bibinfo  {journal} {Phys. Rev. Lett.}\ }\textbf {\bibinfo {volume} {116}},\
  \bibinfo {pages} {061102} (\bibinfo {year} {2016}{\natexlab{a}})},\ \Eprint
  {https://arxiv.org/abs/1602.03837} {arXiv:1602.03837 [gr-qc]} \BibitemShut
  {NoStop}%
\bibitem [{\citenamefont {Aasi}\ \emph {et~al.}(2015)\citenamefont {Aasi},
  \citenamefont {Abbott}, \citenamefont {Abbott}, \citenamefont {Abbott},
  \citenamefont {Abernathy}, \citenamefont {Ackley}, \citenamefont {Adams},
  \citenamefont {Adams}, \citenamefont {Addesso},\ and\ \citenamefont
  {et~al.}}]{advanced_ligo}%
  \BibitemOpen
  \bibfield  {author} {\bibinfo {author} {\bibfnamefont {J.}~\bibnamefont
  {Aasi}}, \bibinfo {author} {\bibfnamefont {B.~P.}\ \bibnamefont {Abbott}},
  \bibinfo {author} {\bibfnamefont {R.}~\bibnamefont {Abbott}}, \bibinfo
  {author} {\bibfnamefont {T.}~\bibnamefont {Abbott}}, \bibinfo {author}
  {\bibfnamefont {M.~R.}\ \bibnamefont {Abernathy}}, \bibinfo {author}
  {\bibfnamefont {K.}~\bibnamefont {Ackley}}, \bibinfo {author} {\bibfnamefont
  {C.}~\bibnamefont {Adams}}, \bibinfo {author} {\bibfnamefont
  {T.}~\bibnamefont {Adams}}, \bibinfo {author} {\bibfnamefont
  {P.}~\bibnamefont {Addesso}},\ and\ \bibinfo {author} {\bibnamefont
  {et~al.}},\ }\bibfield  {title} {\bibinfo {title} {Advanced ligo},\ }\href
  {https://doi.org/10.1088/0264-9381/32/7/074001} {\bibfield  {journal}
  {\bibinfo  {journal} {Classical and Quantum Gravity}\ }\textbf {\bibinfo
  {volume} {32}},\ \bibinfo {pages} {074001} (\bibinfo {year}
  {2015})}\BibitemShut {NoStop}%
\bibitem [{\citenamefont {Acernese}\ \emph {et~al.}(2014)\citenamefont
  {Acernese}, \citenamefont {Agathos}, \citenamefont {Agatsuma}, \citenamefont
  {Aisa}, \citenamefont {Allemandou}, \citenamefont {Allocca}, \citenamefont
  {Amarni}, \citenamefont {Astone}, \citenamefont {Balestri}, \citenamefont
  {Ballardin},\ and\ \citenamefont {et~al.}}]{advanced_virgo}%
  \BibitemOpen
  \bibfield  {author} {\bibinfo {author} {\bibfnamefont {F.}~\bibnamefont
  {Acernese}}, \bibinfo {author} {\bibfnamefont {M.}~\bibnamefont {Agathos}},
  \bibinfo {author} {\bibfnamefont {K.}~\bibnamefont {Agatsuma}}, \bibinfo
  {author} {\bibfnamefont {D.}~\bibnamefont {Aisa}}, \bibinfo {author}
  {\bibfnamefont {N.}~\bibnamefont {Allemandou}}, \bibinfo {author}
  {\bibfnamefont {A.}~\bibnamefont {Allocca}}, \bibinfo {author} {\bibfnamefont
  {J.}~\bibnamefont {Amarni}}, \bibinfo {author} {\bibfnamefont
  {P.}~\bibnamefont {Astone}}, \bibinfo {author} {\bibfnamefont
  {G.}~\bibnamefont {Balestri}}, \bibinfo {author} {\bibfnamefont
  {G.}~\bibnamefont {Ballardin}},\ and\ \bibinfo {author} {\bibnamefont
  {et~al.}},\ }\bibfield  {title} {\bibinfo {title} {Advanced virgo: a
  second-generation interferometric gravitational wave detector},\ }\href
  {https://doi.org/10.1088/0264-9381/32/2/024001} {\bibfield  {journal}
  {\bibinfo  {journal} {Classical and Quantum Gravity}\ }\textbf {\bibinfo
  {volume} {32}},\ \bibinfo {pages} {024001} (\bibinfo {year}
  {2014})}\BibitemShut {NoStop}%
\bibitem [{\citenamefont {Abbott}\ \emph {et~al.}(2021)\citenamefont {Abbott}
  \emph {et~al.}}]{LIGOScientific:2021djp}%
  \BibitemOpen
  \bibfield  {author} {\bibinfo {author} {\bibfnamefont {R.}~\bibnamefont
  {Abbott}} \emph {et~al.} (\bibinfo {collaboration} {LIGO Scientific, VIRGO,
  KAGRA}),\ }\bibfield  {title} {\bibinfo {title} {{GWTC-3: Compact Binary
  Coalescences Observed by LIGO and Virgo During the Second Part of the Third
  Observing Run}},\ }\href@noop {} {\  (\bibinfo {year} {2021})},\ \Eprint
  {https://arxiv.org/abs/2111.03606} {arXiv:2111.03606 [gr-qc]} \BibitemShut
  {NoStop}%
\bibitem [{\citenamefont {Nitz}\ \emph {et~al.}(2019)\citenamefont {Nitz},
  \citenamefont {Capano}, \citenamefont {Nielsen}, \citenamefont {Reyes},
  \citenamefont {White}, \citenamefont {Brown},\ and\ \citenamefont
  {Krishnan}}]{Nitz:2018imz}%
  \BibitemOpen
  \bibfield  {author} {\bibinfo {author} {\bibfnamefont {A.~H.}\ \bibnamefont
  {Nitz}}, \bibinfo {author} {\bibfnamefont {C.}~\bibnamefont {Capano}},
  \bibinfo {author} {\bibfnamefont {A.~B.}\ \bibnamefont {Nielsen}}, \bibinfo
  {author} {\bibfnamefont {S.}~\bibnamefont {Reyes}}, \bibinfo {author}
  {\bibfnamefont {R.}~\bibnamefont {White}}, \bibinfo {author} {\bibfnamefont
  {D.~A.}\ \bibnamefont {Brown}},\ and\ \bibinfo {author} {\bibfnamefont
  {B.}~\bibnamefont {Krishnan}},\ }\bibfield  {title} {\bibinfo {title}
  {{1-OGC: The first open gravitational-wave catalog of binary mergers from
  analysis of public Advanced LIGO data}},\ }\href
  {https://doi.org/10.3847/1538-4357/ab0108} {\bibfield  {journal} {\bibinfo
  {journal} {Astrophys. J.}\ }\textbf {\bibinfo {volume} {872}},\ \bibinfo
  {pages} {195} (\bibinfo {year} {2019})},\ \Eprint
  {https://arxiv.org/abs/1811.01921} {arXiv:1811.01921 [gr-qc]} \BibitemShut
  {NoStop}%
\bibitem [{\citenamefont {Magee}\ \emph {et~al.}(2019)\citenamefont {Magee}
  \emph {et~al.}}]{Magee:2019vmb}%
  \BibitemOpen
  \bibfield  {author} {\bibinfo {author} {\bibfnamefont {R.}~\bibnamefont
  {Magee}} \emph {et~al.},\ }\bibfield  {title} {\bibinfo {title}
  {{Sub-threshold Binary Neutron Star Search in Advanced LIGO\textquoteright{}s
  First Observing Run}},\ }\href {https://doi.org/10.3847/2041-8213/ab20cf}
  {\bibfield  {journal} {\bibinfo  {journal} {Astrophys. J. Lett.}\ }\textbf
  {\bibinfo {volume} {878}},\ \bibinfo {pages} {L17} (\bibinfo {year}
  {2019})},\ \Eprint {https://arxiv.org/abs/1901.09884} {arXiv:1901.09884
  [gr-qc]} \BibitemShut {NoStop}%
\bibitem [{\citenamefont {Zackay}\ \emph {et~al.}(2019)\citenamefont {Zackay},
  \citenamefont {Venumadhav}, \citenamefont {Dai}, \citenamefont {Roulet},\
  and\ \citenamefont {Zaldarriaga}}]{Zackay:2019tzo}%
  \BibitemOpen
  \bibfield  {author} {\bibinfo {author} {\bibfnamefont {B.}~\bibnamefont
  {Zackay}}, \bibinfo {author} {\bibfnamefont {T.}~\bibnamefont {Venumadhav}},
  \bibinfo {author} {\bibfnamefont {L.}~\bibnamefont {Dai}}, \bibinfo {author}
  {\bibfnamefont {J.}~\bibnamefont {Roulet}},\ and\ \bibinfo {author}
  {\bibfnamefont {M.}~\bibnamefont {Zaldarriaga}},\ }\bibfield  {title}
  {\bibinfo {title} {{Highly spinning and aligned binary black hole merger in
  the Advanced LIGO first observing run}},\ }\href
  {https://doi.org/10.1103/PhysRevD.100.023007} {\bibfield  {journal} {\bibinfo
   {journal} {Phys. Rev. D}\ }\textbf {\bibinfo {volume} {100}},\ \bibinfo
  {pages} {023007} (\bibinfo {year} {2019})},\ \Eprint
  {https://arxiv.org/abs/1902.10331} {arXiv:1902.10331 [astro-ph.HE]}
  \BibitemShut {NoStop}%
\bibitem [{\citenamefont {Nitz}\ \emph {et~al.}(2020)\citenamefont {Nitz},
  \citenamefont {Dent}, \citenamefont {Davies}, \citenamefont {Kumar},
  \citenamefont {Capano}, \citenamefont {Harry}, \citenamefont {Mozzon},
  \citenamefont {Nuttall}, \citenamefont {Lundgren},\ and\ \citenamefont
  {T\'apai}}]{Nitz:2020oeq}%
  \BibitemOpen
  \bibfield  {author} {\bibinfo {author} {\bibfnamefont {A.~H.}\ \bibnamefont
  {Nitz}}, \bibinfo {author} {\bibfnamefont {T.}~\bibnamefont {Dent}}, \bibinfo
  {author} {\bibfnamefont {G.~S.}\ \bibnamefont {Davies}}, \bibinfo {author}
  {\bibfnamefont {S.}~\bibnamefont {Kumar}}, \bibinfo {author} {\bibfnamefont
  {C.~D.}\ \bibnamefont {Capano}}, \bibinfo {author} {\bibfnamefont
  {I.}~\bibnamefont {Harry}}, \bibinfo {author} {\bibfnamefont
  {S.}~\bibnamefont {Mozzon}}, \bibinfo {author} {\bibfnamefont
  {L.}~\bibnamefont {Nuttall}}, \bibinfo {author} {\bibfnamefont
  {A.}~\bibnamefont {Lundgren}},\ and\ \bibinfo {author} {\bibfnamefont
  {M.}~\bibnamefont {T\'apai}},\ }\bibfield  {title} {\bibinfo {title} {{2-OGC:
  Open Gravitational-wave Catalog of binary mergers from analysis of public
  Advanced LIGO and Virgo data}},\ }\href
  {https://doi.org/10.3847/1538-4357/ab733f} {\bibfield  {journal} {\bibinfo
  {journal} {Astrophys. J.}\ }\textbf {\bibinfo {volume} {891}},\ \bibinfo
  {pages} {123} (\bibinfo {year} {2020})},\ \Eprint
  {https://arxiv.org/abs/1910.05331} {arXiv:1910.05331 [astro-ph.HE]}
  \BibitemShut {NoStop}%
\bibitem [{\citenamefont {Venumadhav}\ \emph {et~al.}(2020)\citenamefont
  {Venumadhav}, \citenamefont {Zackay}, \citenamefont {Roulet}, \citenamefont
  {Dai},\ and\ \citenamefont {Zaldarriaga}}]{Venumadhav:2019lyq}%
  \BibitemOpen
  \bibfield  {author} {\bibinfo {author} {\bibfnamefont {T.}~\bibnamefont
  {Venumadhav}}, \bibinfo {author} {\bibfnamefont {B.}~\bibnamefont {Zackay}},
  \bibinfo {author} {\bibfnamefont {J.}~\bibnamefont {Roulet}}, \bibinfo
  {author} {\bibfnamefont {L.}~\bibnamefont {Dai}},\ and\ \bibinfo {author}
  {\bibfnamefont {M.}~\bibnamefont {Zaldarriaga}},\ }\bibfield  {title}
  {\bibinfo {title} {{New binary black hole mergers in the second observing run
  of Advanced LIGO and Advanced Virgo}},\ }\href
  {https://doi.org/10.1103/PhysRevD.101.083030} {\bibfield  {journal} {\bibinfo
   {journal} {Phys. Rev. D}\ }\textbf {\bibinfo {volume} {101}},\ \bibinfo
  {pages} {083030} (\bibinfo {year} {2020})},\ \Eprint
  {https://arxiv.org/abs/1904.07214} {arXiv:1904.07214 [astro-ph.HE]}
  \BibitemShut {NoStop}%
\bibitem [{\citenamefont {Zackay}\ \emph {et~al.}(2021)\citenamefont {Zackay},
  \citenamefont {Dai}, \citenamefont {Venumadhav}, \citenamefont {Roulet},\
  and\ \citenamefont {Zaldarriaga}}]{Zackay:2019btq}%
  \BibitemOpen
  \bibfield  {author} {\bibinfo {author} {\bibfnamefont {B.}~\bibnamefont
  {Zackay}}, \bibinfo {author} {\bibfnamefont {L.}~\bibnamefont {Dai}},
  \bibinfo {author} {\bibfnamefont {T.}~\bibnamefont {Venumadhav}}, \bibinfo
  {author} {\bibfnamefont {J.}~\bibnamefont {Roulet}},\ and\ \bibinfo {author}
  {\bibfnamefont {M.}~\bibnamefont {Zaldarriaga}},\ }\bibfield  {title}
  {\bibinfo {title} {{Detecting gravitational waves with disparate detector
  responses: Two new binary black hole mergers}},\ }\href
  {https://doi.org/10.1103/PhysRevD.104.063030} {\bibfield  {journal} {\bibinfo
   {journal} {Phys. Rev. D}\ }\textbf {\bibinfo {volume} {104}},\ \bibinfo
  {pages} {063030} (\bibinfo {year} {2021})},\ \Eprint
  {https://arxiv.org/abs/1910.09528} {arXiv:1910.09528 [astro-ph.HE]}
  \BibitemShut {NoStop}%
\bibitem [{\citenamefont {Nitz}\ \emph
  {et~al.}(2021{\natexlab{a}})\citenamefont {Nitz}, \citenamefont {Capano},
  \citenamefont {Kumar}, \citenamefont {Wang}, \citenamefont {Kastha},
  \citenamefont {Sch\"afer}, \citenamefont {Dhurkunde},\ and\ \citenamefont
  {Cabero}}]{Nitz:2021uxj}%
  \BibitemOpen
  \bibfield  {author} {\bibinfo {author} {\bibfnamefont {A.~H.}\ \bibnamefont
  {Nitz}}, \bibinfo {author} {\bibfnamefont {C.~D.}\ \bibnamefont {Capano}},
  \bibinfo {author} {\bibfnamefont {S.}~\bibnamefont {Kumar}}, \bibinfo
  {author} {\bibfnamefont {Y.-F.}\ \bibnamefont {Wang}}, \bibinfo {author}
  {\bibfnamefont {S.}~\bibnamefont {Kastha}}, \bibinfo {author} {\bibfnamefont
  {M.}~\bibnamefont {Sch\"afer}}, \bibinfo {author} {\bibfnamefont
  {R.}~\bibnamefont {Dhurkunde}},\ and\ \bibinfo {author} {\bibfnamefont
  {M.}~\bibnamefont {Cabero}},\ }\bibfield  {title} {\bibinfo {title} {{3-OGC:
  Catalog of Gravitational Waves from Compact-binary Mergers}},\ }\href
  {https://doi.org/10.3847/1538-4357/ac1c03} {\bibfield  {journal} {\bibinfo
  {journal} {Astrophys. J.}\ }\textbf {\bibinfo {volume} {922}},\ \bibinfo
  {pages} {76} (\bibinfo {year} {2021}{\natexlab{a}})},\ \Eprint
  {https://arxiv.org/abs/2105.09151} {arXiv:2105.09151 [astro-ph.HE]}
  \BibitemShut {NoStop}%
\bibitem [{\citenamefont {Nitz}\ \emph
  {et~al.}(2021{\natexlab{b}})\citenamefont {Nitz}, \citenamefont {Kumar},
  \citenamefont {Wang}, \citenamefont {Kastha}, \citenamefont {Wu},
  \citenamefont {Sch\"afer}, \citenamefont {Dhurkunde},\ and\ \citenamefont
  {Capano}}]{Nitz:2021zwj}%
  \BibitemOpen
  \bibfield  {author} {\bibinfo {author} {\bibfnamefont {A.~H.}\ \bibnamefont
  {Nitz}}, \bibinfo {author} {\bibfnamefont {S.}~\bibnamefont {Kumar}},
  \bibinfo {author} {\bibfnamefont {Y.-F.}\ \bibnamefont {Wang}}, \bibinfo
  {author} {\bibfnamefont {S.}~\bibnamefont {Kastha}}, \bibinfo {author}
  {\bibfnamefont {S.}~\bibnamefont {Wu}}, \bibinfo {author} {\bibfnamefont
  {M.}~\bibnamefont {Sch\"afer}}, \bibinfo {author} {\bibfnamefont
  {R.}~\bibnamefont {Dhurkunde}},\ and\ \bibinfo {author} {\bibfnamefont
  {C.~D.}\ \bibnamefont {Capano}},\ }\bibfield  {title} {\bibinfo {title}
  {{4-OGC: Catalog of gravitational waves from compact-binary mergers}},\
  }\href@noop {} {\  (\bibinfo {year} {2021}{\natexlab{b}})},\ \Eprint
  {https://arxiv.org/abs/2112.06878} {arXiv:2112.06878 [astro-ph.HE]}
  \BibitemShut {NoStop}%
\bibitem [{\citenamefont {Akutsu}\ \emph {et~al.}(2019)\citenamefont {Akutsu}
  \emph {et~al.}}]{KAGRA:2018plz}%
  \BibitemOpen
  \bibfield  {author} {\bibinfo {author} {\bibfnamefont {T.}~\bibnamefont
  {Akutsu}} \emph {et~al.} (\bibinfo {collaboration} {KAGRA}),\ }\bibfield
  {title} {\bibinfo {title} {{KAGRA: 2.5 Generation Interferometric
  Gravitational Wave Detector}},\ }\href
  {https://doi.org/10.1038/s41550-018-0658-y} {\bibfield  {journal} {\bibinfo
  {journal} {Nature Astron.}\ }\textbf {\bibinfo {volume} {3}},\ \bibinfo
  {pages} {35} (\bibinfo {year} {2019})},\ \Eprint
  {https://arxiv.org/abs/1811.08079} {arXiv:1811.08079 [gr-qc]} \BibitemShut
  {NoStop}%
\bibitem [{\citenamefont {Unnikrishnan}(2013)}]{Unnikrishnan:2013qwa}%
  \BibitemOpen
  \bibfield  {author} {\bibinfo {author} {\bibfnamefont {C.~S.}\ \bibnamefont
  {Unnikrishnan}},\ }\bibfield  {title} {\bibinfo {title} {{IndIGO and
  LIGO-India: Scope and plans for gravitational wave research and precision
  metrology in India}},\ }\href {https://doi.org/10.1142/S0218271813410101}
  {\bibfield  {journal} {\bibinfo  {journal} {Int. J. Mod. Phys. D}\ }\textbf
  {\bibinfo {volume} {22}},\ \bibinfo {pages} {1341010} (\bibinfo {year}
  {2013})},\ \Eprint {https://arxiv.org/abs/1510.06059} {arXiv:1510.06059
  [physics.ins-det]} \BibitemShut {NoStop}%
\bibitem [{\citenamefont {Abbott}\ \emph
  {et~al.}(2017{\natexlab{a}})\citenamefont {Abbott} \emph
  {et~al.}}]{LIGOScientific:2017vwq}%
  \BibitemOpen
  \bibfield  {author} {\bibinfo {author} {\bibfnamefont {B.~P.}\ \bibnamefont
  {Abbott}} \emph {et~al.} (\bibinfo {collaboration} {LIGO Scientific,
  Virgo}),\ }\bibfield  {title} {\bibinfo {title} {{GW170817: Observation of
  Gravitational Waves from a Binary Neutron Star Inspiral}},\ }\href
  {https://doi.org/10.1103/PhysRevLett.119.161101} {\bibfield  {journal}
  {\bibinfo  {journal} {Phys. Rev. Lett.}\ }\textbf {\bibinfo {volume} {119}},\
  \bibinfo {pages} {161101} (\bibinfo {year} {2017}{\natexlab{a}})},\ \Eprint
  {https://arxiv.org/abs/1710.05832} {arXiv:1710.05832 [gr-qc]} \BibitemShut
  {NoStop}%
\bibitem [{\citenamefont {Goldstein}\ \emph {et~al.}(2017)\citenamefont
  {Goldstein} \emph {et~al.}}]{Goldstein:2017mmi}%
  \BibitemOpen
  \bibfield  {author} {\bibinfo {author} {\bibfnamefont {A.}~\bibnamefont
  {Goldstein}} \emph {et~al.},\ }\bibfield  {title} {\bibinfo {title} {{An
  Ordinary Short Gamma-Ray Burst with Extraordinary Implications: Fermi-GBM
  Detection of GRB 170817A}},\ }\href
  {https://doi.org/10.3847/2041-8213/aa8f41} {\bibfield  {journal} {\bibinfo
  {journal} {Astrophys. J. Lett.}\ }\textbf {\bibinfo {volume} {848}},\
  \bibinfo {pages} {L14} (\bibinfo {year} {2017})},\ \Eprint
  {https://arxiv.org/abs/1710.05446} {arXiv:1710.05446 [astro-ph.HE]}
  \BibitemShut {NoStop}%
\bibitem [{\citenamefont {Abbott}\ \emph
  {et~al.}(2017{\natexlab{b}})\citenamefont {Abbott} \emph
  {et~al.}}]{LIGOScientific:2017zic}%
  \BibitemOpen
  \bibfield  {author} {\bibinfo {author} {\bibfnamefont {B.~P.}\ \bibnamefont
  {Abbott}} \emph {et~al.} (\bibinfo {collaboration} {LIGO Scientific, Virgo,
  Fermi-GBM, INTEGRAL}),\ }\bibfield  {title} {\bibinfo {title} {{Gravitational
  Waves and Gamma-rays from a Binary Neutron Star Merger: GW170817 and GRB
  170817A}},\ }\href {https://doi.org/10.3847/2041-8213/aa920c} {\bibfield
  {journal} {\bibinfo  {journal} {Astrophys. J. Lett.}\ }\textbf {\bibinfo
  {volume} {848}},\ \bibinfo {pages} {L13} (\bibinfo {year}
  {2017}{\natexlab{b}})},\ \Eprint {https://arxiv.org/abs/1710.05834}
  {arXiv:1710.05834 [astro-ph.HE]} \BibitemShut {NoStop}%
\bibitem [{\citenamefont {Smartt}\ \emph {et~al.}(2017)\citenamefont {Smartt}
  \emph {et~al.}}]{Smartt:2017fuw}%
  \BibitemOpen
  \bibfield  {author} {\bibinfo {author} {\bibfnamefont {S.~J.}\ \bibnamefont
  {Smartt}} \emph {et~al.},\ }\bibfield  {title} {\bibinfo {title} {{A kilonova
  as the electromagnetic counterpart to a gravitational-wave source}},\ }\href
  {https://doi.org/10.1038/nature24303} {\bibfield  {journal} {\bibinfo
  {journal} {Nature}\ }\textbf {\bibinfo {volume} {551}},\ \bibinfo {pages}
  {75} (\bibinfo {year} {2017})},\ \Eprint {https://arxiv.org/abs/1710.05841}
  {arXiv:1710.05841 [astro-ph.HE]} \BibitemShut {NoStop}%
\bibitem [{\citenamefont {Riess}\ \emph {et~al.}(2021)\citenamefont {Riess}
  \emph {et~al.}}]{Riess:2021jrx}%
  \BibitemOpen
  \bibfield  {author} {\bibinfo {author} {\bibfnamefont {A.~G.}\ \bibnamefont
  {Riess}} \emph {et~al.},\ }\bibfield  {title} {\bibinfo {title} {{A
  Comprehensive Measurement of the Local Value of the Hubble Constant with 1
  km/s/Mpc Uncertainty from the Hubble Space Telescope and the SH0ES Team}},\
  }\href@noop {} {\  (\bibinfo {year} {2021})},\ \Eprint
  {https://arxiv.org/abs/2112.04510} {arXiv:2112.04510 [astro-ph.CO]}
  \BibitemShut {NoStop}%
\bibitem [{\citenamefont {Aghanim}\ \emph {et~al.}(2020)\citenamefont {Aghanim}
  \emph {et~al.}}]{Planck:2018vyg}%
  \BibitemOpen
  \bibfield  {author} {\bibinfo {author} {\bibfnamefont {N.}~\bibnamefont
  {Aghanim}} \emph {et~al.} (\bibinfo {collaboration} {Planck}),\ }\bibfield
  {title} {\bibinfo {title} {{Planck 2018 results. VI. Cosmological
  parameters}},\ }\href {https://doi.org/10.1051/0004-6361/201833910}
  {\bibfield  {journal} {\bibinfo  {journal} {Astron. Astrophys.}\ }\textbf
  {\bibinfo {volume} {641}},\ \bibinfo {pages} {A6} (\bibinfo {year} {2020})},\
  \bibinfo {note} {[Erratum: Astron.Astrophys. 652, C4 (2021)]},\ \Eprint
  {https://arxiv.org/abs/1807.06209} {arXiv:1807.06209 [astro-ph.CO]}
  \BibitemShut {NoStop}%
\bibitem [{\citenamefont {{Schutz}}(1986)}]{schutz_1986}%
  \BibitemOpen
  \bibfield  {author} {\bibinfo {author} {\bibfnamefont {B.~F.}\ \bibnamefont
  {{Schutz}}},\ }\bibfield  {title} {\bibinfo {title} {{Determining the Hubble
  constant from gravitational wave observations}},\ }\href
  {https://doi.org/10.1038/323310a0} {\bibfield  {journal} {\bibinfo  {journal}
  {\nat}\ }\textbf {\bibinfo {volume} {323}},\ \bibinfo {pages} {310} (\bibinfo
  {year} {1986})}\BibitemShut {NoStop}%
\bibitem [{\citenamefont {{Holz}}\ and\ \citenamefont
  {{Hughes}}(2005)}]{holz_hughes_2005}%
  \BibitemOpen
  \bibfield  {author} {\bibinfo {author} {\bibfnamefont {D.~E.}\ \bibnamefont
  {{Holz}}}\ and\ \bibinfo {author} {\bibfnamefont {S.~A.}\ \bibnamefont
  {{Hughes}}},\ }\bibfield  {title} {\bibinfo {title} {{Using
  Gravitational-Wave Standard Sirens}},\ }\href
  {https://doi.org/10.1086/431341} {\bibfield  {journal} {\bibinfo  {journal}
  {\apj}\ }\textbf {\bibinfo {volume} {629}},\ \bibinfo {pages} {15} (\bibinfo
  {year} {2005})},\ \Eprint {https://arxiv.org/abs/astro-ph/0504616}
  {arXiv:astro-ph/0504616 [astro-ph]} \BibitemShut {NoStop}%
\bibitem [{\citenamefont {{Dalal}}\ \emph {et~al.}(2006)\citenamefont
  {{Dalal}}, \citenamefont {{Holz}}, \citenamefont {{Hughes}},\ and\
  \citenamefont {{Jain}}}]{2006PhRvD..74f3006D}%
  \BibitemOpen
  \bibfield  {author} {\bibinfo {author} {\bibfnamefont {N.}~\bibnamefont
  {{Dalal}}}, \bibinfo {author} {\bibfnamefont {D.~E.}\ \bibnamefont {{Holz}}},
  \bibinfo {author} {\bibfnamefont {S.~A.}\ \bibnamefont {{Hughes}}},\ and\
  \bibinfo {author} {\bibfnamefont {B.}~\bibnamefont {{Jain}}},\ }\bibfield
  {title} {\bibinfo {title} {{Short GRB and binary black hole standard sirens
  as a probe of dark energy}},\ }\href
  {https://doi.org/10.1103/PhysRevD.74.063006} {\bibfield  {journal} {\bibinfo
  {journal} {\prd}\ }\textbf {\bibinfo {volume} {74}},\ \bibinfo {eid} {063006}
  (\bibinfo {year} {2006})},\ \Eprint {https://arxiv.org/abs/astro-ph/0601275}
  {arXiv:astro-ph/0601275 [astro-ph]} \BibitemShut {NoStop}%
\bibitem [{\citenamefont {Abbott}\ \emph
  {et~al.}(2017{\natexlab{c}})\citenamefont {Abbott} \emph
  {et~al.}}]{2017Natur.551...85A}%
  \BibitemOpen
  \bibfield  {author} {\bibinfo {author} {\bibfnamefont {B.~P.}\ \bibnamefont
  {Abbott}} \emph {et~al.},\ }\bibfield  {title} {\bibinfo {title} {{A
  gravitational-wave standard siren measurement of the Hubble constant}},\
  }\href {https://doi.org/10.1038/nature24471} {\bibfield  {journal} {\bibinfo
  {journal} {\nat}\ }\textbf {\bibinfo {volume} {551}},\ \bibinfo {pages} {85}
  (\bibinfo {year} {2017}{\natexlab{c}})},\ \Eprint
  {https://arxiv.org/abs/1710.05835} {arXiv:1710.05835 [astro-ph.CO]}
  \BibitemShut {NoStop}%
\bibitem [{\citenamefont {{Taylor}}\ \emph {et~al.}(2012)\citenamefont
  {{Taylor}}, \citenamefont {{Gair}},\ and\ \citenamefont
  {{Mandel}}}]{2012PhRvD..85b3535T}%
  \BibitemOpen
  \bibfield  {author} {\bibinfo {author} {\bibfnamefont {S.~R.}\ \bibnamefont
  {{Taylor}}}, \bibinfo {author} {\bibfnamefont {J.~R.}\ \bibnamefont
  {{Gair}}},\ and\ \bibinfo {author} {\bibfnamefont {I.}~\bibnamefont
  {{Mandel}}},\ }\bibfield  {title} {\bibinfo {title} {{Cosmology using
  advanced gravitational-wave detectors alone}},\ }\href
  {https://doi.org/10.1103/PhysRevD.85.023535} {\bibfield  {journal} {\bibinfo
  {journal} {\prd}\ }\textbf {\bibinfo {volume} {85}},\ \bibinfo {eid} {023535}
  (\bibinfo {year} {2012})},\ \Eprint {https://arxiv.org/abs/1108.5161}
  {arXiv:1108.5161 [gr-qc]} \BibitemShut {NoStop}%
\bibitem [{\citenamefont {Fishbach}\ \emph {et~al.}(2019)\citenamefont
  {Fishbach}, \citenamefont {Gray}, \citenamefont {Hernandez}, \citenamefont
  {Qi}, \citenamefont {Sur}, \citenamefont {Acernese}, \citenamefont {Aiello},
  \citenamefont {Allocca}, \citenamefont {Aloy}, \citenamefont {Amato},\ and\
  \citenamefont {et~al.}}]{Fishbach_2019}%
  \BibitemOpen
  \bibfield  {author} {\bibinfo {author} {\bibfnamefont {M.}~\bibnamefont
  {Fishbach}}, \bibinfo {author} {\bibfnamefont {R.}~\bibnamefont {Gray}},
  \bibinfo {author} {\bibfnamefont {I.~M.}\ \bibnamefont {Hernandez}}, \bibinfo
  {author} {\bibfnamefont {H.}~\bibnamefont {Qi}}, \bibinfo {author}
  {\bibfnamefont {A.}~\bibnamefont {Sur}}, \bibinfo {author} {\bibfnamefont
  {F.}~\bibnamefont {Acernese}}, \bibinfo {author} {\bibfnamefont
  {L.}~\bibnamefont {Aiello}}, \bibinfo {author} {\bibfnamefont
  {A.}~\bibnamefont {Allocca}}, \bibinfo {author} {\bibfnamefont {M.~A.}\
  \bibnamefont {Aloy}}, \bibinfo {author} {\bibfnamefont {A.}~\bibnamefont
  {Amato}},\ and\ \bibinfo {author} {\bibnamefont {et~al.}},\ }\bibfield
  {title} {\bibinfo {title} {A standard siren measurement of the hubble
  constant from gw170817 without the electromagnetic counterpart},\ }\href
  {https://doi.org/10.3847/2041-8213/aaf96e} {\bibfield  {journal} {\bibinfo
  {journal} {The Astrophysical Journal}\ }\textbf {\bibinfo {volume} {871}},\
  \bibinfo {pages} {L13} (\bibinfo {year} {2019})}\BibitemShut {NoStop}%
\bibitem [{\citenamefont {Soares-Santos}\ \emph {et~al.}(2019)\citenamefont
  {Soares-Santos}, \citenamefont {Palmese}, \citenamefont {Hartley},
  \citenamefont {Annis}, \citenamefont {Garcia-Bellido}, \citenamefont {Lahav},
  \citenamefont {Doctor}, \citenamefont {Fishbach}, \citenamefont {Holz},
  \citenamefont {Lin},\ and\ \citenamefont {et~al.}}]{Soares_Santos_2019}%
  \BibitemOpen
  \bibfield  {author} {\bibinfo {author} {\bibfnamefont {M.}~\bibnamefont
  {Soares-Santos}}, \bibinfo {author} {\bibfnamefont {A.}~\bibnamefont
  {Palmese}}, \bibinfo {author} {\bibfnamefont {W.}~\bibnamefont {Hartley}},
  \bibinfo {author} {\bibfnamefont {J.}~\bibnamefont {Annis}}, \bibinfo
  {author} {\bibfnamefont {J.}~\bibnamefont {Garcia-Bellido}}, \bibinfo
  {author} {\bibfnamefont {O.}~\bibnamefont {Lahav}}, \bibinfo {author}
  {\bibfnamefont {Z.}~\bibnamefont {Doctor}}, \bibinfo {author} {\bibfnamefont
  {M.}~\bibnamefont {Fishbach}}, \bibinfo {author} {\bibfnamefont {D.~E.}\
  \bibnamefont {Holz}}, \bibinfo {author} {\bibfnamefont {H.}~\bibnamefont
  {Lin}},\ and\ \bibinfo {author} {\bibnamefont {et~al.}},\ }\bibfield  {title}
  {\bibinfo {title} {First measurement of the hubble constant from a dark
  standard siren using the dark energy survey galaxies and the ligo/virgo
  binary–black-hole merger gw170814},\ }\href
  {https://doi.org/10.3847/2041-8213/ab14f1} {\bibfield  {journal} {\bibinfo
  {journal} {The Astrophysical Journal}\ }\textbf {\bibinfo {volume} {876}},\
  \bibinfo {pages} {L7} (\bibinfo {year} {2019})}\BibitemShut {NoStop}%
\bibitem [{\citenamefont {Mukherjee}\ \emph {et~al.}(2021)\citenamefont
  {Mukherjee}, \citenamefont {Wandelt}, \citenamefont {Nissanke},\ and\
  \citenamefont {Silvestri}}]{Mukherjee:2020hyn}%
  \BibitemOpen
  \bibfield  {author} {\bibinfo {author} {\bibfnamefont {S.}~\bibnamefont
  {Mukherjee}}, \bibinfo {author} {\bibfnamefont {B.~D.}\ \bibnamefont
  {Wandelt}}, \bibinfo {author} {\bibfnamefont {S.~M.}\ \bibnamefont
  {Nissanke}},\ and\ \bibinfo {author} {\bibfnamefont {A.}~\bibnamefont
  {Silvestri}},\ }\bibfield  {title} {\bibinfo {title} {{Accurate precision
  Cosmology with redshift unknown gravitational wave sources}},\ }\href
  {https://doi.org/10.1103/PhysRevD.103.043520} {\bibfield  {journal} {\bibinfo
   {journal} {Phys. Rev. D}\ }\textbf {\bibinfo {volume} {103}},\ \bibinfo
  {pages} {043520} (\bibinfo {year} {2021})},\ \Eprint
  {https://arxiv.org/abs/2007.02943} {arXiv:2007.02943 [astro-ph.CO]}
  \BibitemShut {NoStop}%
\bibitem [{\citenamefont {{Gray}}\ \emph {et~al.}(2020)\citenamefont {{Gray}},
  \citenamefont {{Hernandez}}, \citenamefont {{Qi}}, \citenamefont {{Sur}},
  \citenamefont {{Brady}}, \citenamefont {{Chen}}, \citenamefont {{Farr}},
  \citenamefont {{Fishbach}}, \citenamefont {{Gair}}, \citenamefont {{Ghosh}},
  \citenamefont {{Holz}}, \citenamefont {{Mastrogiovanni}}, \citenamefont
  {{Messenger}}, \citenamefont {{Steer}},\ and\ \citenamefont
  {{Veitch}}}]{2020PhRvD.101l2001G}%
  \BibitemOpen
  \bibfield  {author} {\bibinfo {author} {\bibfnamefont {R.}~\bibnamefont
  {{Gray}}}, \bibinfo {author} {\bibfnamefont {I.~M.}\ \bibnamefont
  {{Hernandez}}}, \bibinfo {author} {\bibfnamefont {H.}~\bibnamefont {{Qi}}},
  \bibinfo {author} {\bibfnamefont {A.}~\bibnamefont {{Sur}}}, \bibinfo
  {author} {\bibfnamefont {P.~R.}\ \bibnamefont {{Brady}}}, \bibinfo {author}
  {\bibfnamefont {H.-Y.}\ \bibnamefont {{Chen}}}, \bibinfo {author}
  {\bibfnamefont {W.~M.}\ \bibnamefont {{Farr}}}, \bibinfo {author}
  {\bibfnamefont {M.}~\bibnamefont {{Fishbach}}}, \bibinfo {author}
  {\bibfnamefont {J.~R.}\ \bibnamefont {{Gair}}}, \bibinfo {author}
  {\bibfnamefont {A.}~\bibnamefont {{Ghosh}}}, \bibinfo {author} {\bibfnamefont
  {D.~E.}\ \bibnamefont {{Holz}}}, \bibinfo {author} {\bibfnamefont
  {S.}~\bibnamefont {{Mastrogiovanni}}}, \bibinfo {author} {\bibfnamefont
  {C.}~\bibnamefont {{Messenger}}}, \bibinfo {author} {\bibfnamefont {D.~A.}\
  \bibnamefont {{Steer}}},\ and\ \bibinfo {author} {\bibfnamefont
  {J.}~\bibnamefont {{Veitch}}},\ }\bibfield  {title} {\bibinfo {title}
  {{Cosmological inference using gravitational wave standard sirens: A mock
  data analysis}},\ }\href {https://doi.org/10.1103/PhysRevD.101.122001}
  {\bibfield  {journal} {\bibinfo  {journal} {\prd}\ }\textbf {\bibinfo
  {volume} {101}},\ \bibinfo {eid} {122001} (\bibinfo {year} {2020})},\ \Eprint
  {https://arxiv.org/abs/1908.06050} {arXiv:1908.06050 [gr-qc]} \BibitemShut
  {NoStop}%
\bibitem [{\citenamefont {{Chernoff}}\ and\ \citenamefont
  {{Finn}}(1993)}]{1993ApJ...411L...5C}%
  \BibitemOpen
  \bibfield  {author} {\bibinfo {author} {\bibfnamefont {D.~F.}\ \bibnamefont
  {{Chernoff}}}\ and\ \bibinfo {author} {\bibfnamefont {L.~S.}\ \bibnamefont
  {{Finn}}},\ }\bibfield  {title} {\bibinfo {title} {{Gravitational Radiation,
  Inspiraling Binaries, and Cosmology}},\ }\href
  {https://doi.org/10.1086/186898} {\bibfield  {journal} {\bibinfo  {journal}
  {\apjl}\ }\textbf {\bibinfo {volume} {411}},\ \bibinfo {pages} {L5} (\bibinfo
  {year} {1993})},\ \Eprint {https://arxiv.org/abs/gr-qc/9304020}
  {arXiv:gr-qc/9304020 [gr-qc]} \BibitemShut {NoStop}%
\bibitem [{\citenamefont {{Ezquiaga}}\ and\ \citenamefont
  {{Holz}}(2022)}]{2022PhRvL.129f1102E}%
  \BibitemOpen
  \bibfield  {author} {\bibinfo {author} {\bibfnamefont {J.~M.}\ \bibnamefont
  {{Ezquiaga}}}\ and\ \bibinfo {author} {\bibfnamefont {D.~E.}\ \bibnamefont
  {{Holz}}},\ }\bibfield  {title} {\bibinfo {title} {{Spectral Sirens:
  Cosmology from the Full Mass Distribution of Compact Binaries}},\ }\href
  {https://doi.org/10.1103/PhysRevLett.129.061102} {\bibfield  {journal}
  {\bibinfo  {journal} {\prl}\ }\textbf {\bibinfo {volume} {129}},\ \bibinfo
  {eid} {061102} (\bibinfo {year} {2022})},\ \Eprint
  {https://arxiv.org/abs/2202.08240} {arXiv:2202.08240 [astro-ph.CO]}
  \BibitemShut {NoStop}%
\bibitem [{\citenamefont {Freedman}(2021)}]{Freedman:2021ahq}%
  \BibitemOpen
  \bibfield  {author} {\bibinfo {author} {\bibfnamefont {W.~L.}\ \bibnamefont
  {Freedman}},\ }\bibfield  {title} {\bibinfo {title} {{Measurements of the
  Hubble Constant: Tensions in Perspective}},\ }\href
  {https://doi.org/10.3847/1538-4357/ac0e95} {\bibfield  {journal} {\bibinfo
  {journal} {Astrophys. J.}\ }\textbf {\bibinfo {volume} {919}},\ \bibinfo
  {pages} {16} (\bibinfo {year} {2021})},\ \Eprint
  {https://arxiv.org/abs/2106.15656} {arXiv:2106.15656 [astro-ph.CO]}
  \BibitemShut {NoStop}%
\bibitem [{\citenamefont {Anand}\ \emph {et~al.}(2021)\citenamefont {Anand},
  \citenamefont {Tully}, \citenamefont {Rizzi}, \citenamefont {Riess},\ and\
  \citenamefont {Yuan}}]{Anand:2021sum}%
  \BibitemOpen
  \bibfield  {author} {\bibinfo {author} {\bibfnamefont {G.~S.}\ \bibnamefont
  {Anand}}, \bibinfo {author} {\bibfnamefont {R.~B.}\ \bibnamefont {Tully}},
  \bibinfo {author} {\bibfnamefont {L.}~\bibnamefont {Rizzi}}, \bibinfo
  {author} {\bibfnamefont {A.~G.}\ \bibnamefont {Riess}},\ and\ \bibinfo
  {author} {\bibfnamefont {W.}~\bibnamefont {Yuan}},\ }\bibfield  {title}
  {\bibinfo {title} {{Comparing Tip of the Red Giant Branch Distance Scales: An
  Independent Reduction of the Carnegie-Chicago Hubble Program and the Value of
  the Hubble Constant}},\ }\href@noop {} {\  (\bibinfo {year} {2021})},\
  \Eprint {https://arxiv.org/abs/2108.00007} {arXiv:2108.00007 [astro-ph.CO]}
  \BibitemShut {NoStop}%
\bibitem [{\citenamefont {Markovic}(1993)}]{Markovic:1993cr}%
  \BibitemOpen
  \bibfield  {author} {\bibinfo {author} {\bibfnamefont {D.}~\bibnamefont
  {Markovic}},\ }\bibfield  {title} {\bibinfo {title} {{On the possibility of
  determining cosmological parameters from measurements of gravitational waves
  emitted by coalescing, compact binaries}},\ }\href
  {https://doi.org/10.1103/PhysRevD.48.4738} {\bibfield  {journal} {\bibinfo
  {journal} {Phys. Rev. D}\ }\textbf {\bibinfo {volume} {48}},\ \bibinfo
  {pages} {4738} (\bibinfo {year} {1993})}\BibitemShut {NoStop}%
\bibitem [{\citenamefont {Cutler}\ and\ \citenamefont
  {Flanagan}(1994)}]{Cutler_1994}%
  \BibitemOpen
  \bibfield  {author} {\bibinfo {author} {\bibfnamefont {C.}~\bibnamefont
  {Cutler}}\ and\ \bibinfo {author} {\bibfnamefont {E.~E.}\ \bibnamefont
  {Flanagan}},\ }\bibfield  {title} {\bibinfo {title} {{Gravitational waves
  from merging compact binaries: How accurately can one extract the binary's
  parameters from the inspiral wave form?}},\ }\href
  {https://doi.org/10.1103/PhysRevD.49.2658} {\bibfield  {journal} {\bibinfo
  {journal} {Phys. Rev. D}\ }\textbf {\bibinfo {volume} {49}},\ \bibinfo
  {pages} {2658} (\bibinfo {year} {1994})},\ \Eprint
  {https://arxiv.org/abs/gr-qc/9402014} {arXiv:gr-qc/9402014} \BibitemShut
  {NoStop}%
\bibitem [{\citenamefont {Usman}\ \emph {et~al.}(2019)\citenamefont {Usman},
  \citenamefont {Mills},\ and\ \citenamefont {Fairhurst}}]{Usman:2018imj}%
  \BibitemOpen
  \bibfield  {author} {\bibinfo {author} {\bibfnamefont {S.~A.}\ \bibnamefont
  {Usman}}, \bibinfo {author} {\bibfnamefont {J.~C.}\ \bibnamefont {Mills}},\
  and\ \bibinfo {author} {\bibfnamefont {S.}~\bibnamefont {Fairhurst}},\
  }\bibfield  {title} {\bibinfo {title} {{Constraining the Inclinations of
  Binary Mergers from Gravitational-wave Observations}},\ }\href
  {https://doi.org/10.3847/1538-4357/ab0b3e} {\bibfield  {journal} {\bibinfo
  {journal} {Astrophys. J.}\ }\textbf {\bibinfo {volume} {877}},\ \bibinfo
  {pages} {82} (\bibinfo {year} {2019})},\ \Eprint
  {https://arxiv.org/abs/1809.10727} {arXiv:1809.10727 [gr-qc]} \BibitemShut
  {NoStop}%
\bibitem [{\citenamefont {London}\ \emph {et~al.}(2018)\citenamefont {London},
  \citenamefont {Khan}, \citenamefont {Fauchon-Jones}, \citenamefont
  {Garc\'\i{}a}, \citenamefont {Hannam}, \citenamefont {Husa}, \citenamefont
  {Jim\'enez-Forteza}, \citenamefont {Kalaghatgi}, \citenamefont {Ohme},\ and\
  \citenamefont {Pannarale}}]{London:2017bcn}%
  \BibitemOpen
  \bibfield  {author} {\bibinfo {author} {\bibfnamefont {L.}~\bibnamefont
  {London}}, \bibinfo {author} {\bibfnamefont {S.}~\bibnamefont {Khan}},
  \bibinfo {author} {\bibfnamefont {E.}~\bibnamefont {Fauchon-Jones}}, \bibinfo
  {author} {\bibfnamefont {C.}~\bibnamefont {Garc\'\i{}a}}, \bibinfo {author}
  {\bibfnamefont {M.}~\bibnamefont {Hannam}}, \bibinfo {author} {\bibfnamefont
  {S.}~\bibnamefont {Husa}}, \bibinfo {author} {\bibfnamefont {X.}~\bibnamefont
  {Jim\'enez-Forteza}}, \bibinfo {author} {\bibfnamefont {C.}~\bibnamefont
  {Kalaghatgi}}, \bibinfo {author} {\bibfnamefont {F.}~\bibnamefont {Ohme}},\
  and\ \bibinfo {author} {\bibfnamefont {F.}~\bibnamefont {Pannarale}},\
  }\bibfield  {title} {\bibinfo {title} {{First higher-multipole model of
  gravitational waves from spinning and coalescing black-hole binaries}},\
  }\href {https://doi.org/10.1103/PhysRevLett.120.161102} {\bibfield  {journal}
  {\bibinfo  {journal} {Phys. Rev. Lett.}\ }\textbf {\bibinfo {volume} {120}},\
  \bibinfo {pages} {161102} (\bibinfo {year} {2018})},\ \Eprint
  {https://arxiv.org/abs/1708.00404} {arXiv:1708.00404 [gr-qc]} \BibitemShut
  {NoStop}%
\bibitem [{\citenamefont {Vitale}\ and\ \citenamefont
  {Chen}(2018)}]{Vitale:2018wlg}%
  \BibitemOpen
  \bibfield  {author} {\bibinfo {author} {\bibfnamefont {S.}~\bibnamefont
  {Vitale}}\ and\ \bibinfo {author} {\bibfnamefont {H.-Y.}\ \bibnamefont
  {Chen}},\ }\bibfield  {title} {\bibinfo {title} {{Measuring the Hubble
  constant with neutron star black hole mergers}},\ }\href
  {https://doi.org/10.1103/PhysRevLett.121.021303} {\bibfield  {journal}
  {\bibinfo  {journal} {Phys. Rev. Lett.}\ }\textbf {\bibinfo {volume} {121}},\
  \bibinfo {pages} {021303} (\bibinfo {year} {2018})},\ \Eprint
  {https://arxiv.org/abs/1804.07337} {arXiv:1804.07337 [astro-ph.CO]}
  \BibitemShut {NoStop}%
\bibitem [{\citenamefont {Varma}\ \emph {et~al.}(2014)\citenamefont {Varma},
  \citenamefont {Ajith}, \citenamefont {Husa}, \citenamefont {Bustillo},
  \citenamefont {Hannam},\ and\ \citenamefont {P\"urrer}}]{Varma:2014jxa}%
  \BibitemOpen
  \bibfield  {author} {\bibinfo {author} {\bibfnamefont {V.}~\bibnamefont
  {Varma}}, \bibinfo {author} {\bibfnamefont {P.}~\bibnamefont {Ajith}},
  \bibinfo {author} {\bibfnamefont {S.}~\bibnamefont {Husa}}, \bibinfo {author}
  {\bibfnamefont {J.~C.}\ \bibnamefont {Bustillo}}, \bibinfo {author}
  {\bibfnamefont {M.}~\bibnamefont {Hannam}},\ and\ \bibinfo {author}
  {\bibfnamefont {M.}~\bibnamefont {P\"urrer}},\ }\bibfield  {title} {\bibinfo
  {title} {{Gravitational-wave observations of binary black holes: Effect of
  nonquadrupole modes}},\ }\href {https://doi.org/10.1103/PhysRevD.90.124004}
  {\bibfield  {journal} {\bibinfo  {journal} {Phys. Rev. D}\ }\textbf {\bibinfo
  {volume} {90}},\ \bibinfo {pages} {124004} (\bibinfo {year} {2014})},\
  \Eprint {https://arxiv.org/abs/1409.2349} {arXiv:1409.2349 [gr-qc]}
  \BibitemShut {NoStop}%
\bibitem [{\citenamefont {Kidder}(1995)}]{Kidder:1995zr}%
  \BibitemOpen
  \bibfield  {author} {\bibinfo {author} {\bibfnamefont {L.~E.}\ \bibnamefont
  {Kidder}},\ }\bibfield  {title} {\bibinfo {title} {{Coalescing binary systems
  of compact objects to postNewtonian 5/2 order. 5. Spin effects}},\ }\href
  {https://doi.org/10.1103/PhysRevD.52.821} {\bibfield  {journal} {\bibinfo
  {journal} {Phys. Rev. D}\ }\textbf {\bibinfo {volume} {52}},\ \bibinfo
  {pages} {821} (\bibinfo {year} {1995})},\ \Eprint
  {https://arxiv.org/abs/gr-qc/9506022} {arXiv:gr-qc/9506022} \BibitemShut
  {NoStop}%
\bibitem [{\citenamefont {Chatterjee}\ \emph {et~al.}(2021)\citenamefont
  {Chatterjee}, \citenamefont {R.}, \citenamefont {Holder}, \citenamefont
  {Holz}, \citenamefont {Perkins}, \citenamefont {Yagi},\ and\ \citenamefont
  {Yunes}}]{Chatterjee:2021xrm}%
  \BibitemOpen
  \bibfield  {author} {\bibinfo {author} {\bibfnamefont {D.}~\bibnamefont
  {Chatterjee}}, \bibinfo {author} {\bibfnamefont {A.~H.~K.}\ \bibnamefont
  {R.}}, \bibinfo {author} {\bibfnamefont {G.}~\bibnamefont {Holder}}, \bibinfo
  {author} {\bibfnamefont {D.~E.}\ \bibnamefont {Holz}}, \bibinfo {author}
  {\bibfnamefont {S.}~\bibnamefont {Perkins}}, \bibinfo {author} {\bibfnamefont
  {K.}~\bibnamefont {Yagi}},\ and\ \bibinfo {author} {\bibfnamefont
  {N.}~\bibnamefont {Yunes}},\ }\bibfield  {title} {\bibinfo {title}
  {{Cosmology with Love: Measuring the Hubble constant using neutron star
  universal relations}},\ }\href {https://doi.org/10.1103/PhysRevD.104.083528}
  {\bibfield  {journal} {\bibinfo  {journal} {Phys. Rev. D}\ }\textbf {\bibinfo
  {volume} {104}},\ \bibinfo {pages} {083528} (\bibinfo {year} {2021})},\
  \Eprint {https://arxiv.org/abs/2106.06589} {arXiv:2106.06589 [gr-qc]}
  \BibitemShut {NoStop}%
\bibitem [{\citenamefont {{Messenger}}\ and\ \citenamefont
  {{Read}}(2012)}]{messenger_read_2012}%
  \BibitemOpen
  \bibfield  {author} {\bibinfo {author} {\bibfnamefont {C.}~\bibnamefont
  {{Messenger}}}\ and\ \bibinfo {author} {\bibfnamefont {J.}~\bibnamefont
  {{Read}}},\ }\bibfield  {title} {\bibinfo {title} {{Measuring a Cosmological
  Distance-Redshift Relationship Using Only Gravitational Wave Observations of
  Binary Neutron Star Coalescences}},\ }\href
  {https://doi.org/10.1103/PhysRevLett.108.091101} {\bibfield  {journal}
  {\bibinfo  {journal} {\prl}\ }\textbf {\bibinfo {volume} {108}},\ \bibinfo
  {eid} {091101} (\bibinfo {year} {2012})},\ \Eprint
  {https://arxiv.org/abs/1107.5725} {arXiv:1107.5725 [gr-qc]} \BibitemShut
  {NoStop}%
\bibitem [{\citenamefont {{Yagi}}\ and\ \citenamefont
  {{Yunes}}(2016)}]{yagi_yunes_2016}%
  \BibitemOpen
  \bibfield  {author} {\bibinfo {author} {\bibfnamefont {K.}~\bibnamefont
  {{Yagi}}}\ and\ \bibinfo {author} {\bibfnamefont {N.}~\bibnamefont
  {{Yunes}}},\ }\bibfield  {title} {\bibinfo {title} {{Binary Love
  relations}},\ }\href {https://doi.org/10.1088/0264-9381/33/13/13LT01}
  {\bibfield  {journal} {\bibinfo  {journal} {Classical and Quantum Gravity}\
  }\textbf {\bibinfo {volume} {33}},\ \bibinfo {eid} {13LT01} (\bibinfo {year}
  {2016})},\ \Eprint {https://arxiv.org/abs/1512.02639} {arXiv:1512.02639
  [gr-qc]} \BibitemShut {NoStop}%
\bibitem [{\citenamefont {{Yagi}}\ and\ \citenamefont
  {{Yunes}}(2017)}]{yagi_yunes_2017}%
  \BibitemOpen
  \bibfield  {author} {\bibinfo {author} {\bibfnamefont {K.}~\bibnamefont
  {{Yagi}}}\ and\ \bibinfo {author} {\bibfnamefont {N.}~\bibnamefont
  {{Yunes}}},\ }\bibfield  {title} {\bibinfo {title} {{Approximate universal
  relations among tidal parameters for neutron star binaries}},\ }\href
  {https://doi.org/10.1088/1361-6382/34/1/015006} {\bibfield  {journal}
  {\bibinfo  {journal} {Classical and Quantum Gravity}\ }\textbf {\bibinfo
  {volume} {34}},\ \bibinfo {eid} {015006} (\bibinfo {year} {2017})},\ \Eprint
  {https://arxiv.org/abs/1608.06187} {arXiv:1608.06187 [gr-qc]} \BibitemShut
  {NoStop}%
\bibitem [{\citenamefont {Carson}\ \emph {et~al.}(2019)\citenamefont {Carson},
  \citenamefont {Chatziioannou}, \citenamefont {Haster}, \citenamefont {Yagi},\
  and\ \citenamefont {Yunes}}]{Carson:2019rjx}%
  \BibitemOpen
  \bibfield  {author} {\bibinfo {author} {\bibfnamefont {Z.}~\bibnamefont
  {Carson}}, \bibinfo {author} {\bibfnamefont {K.}~\bibnamefont
  {Chatziioannou}}, \bibinfo {author} {\bibfnamefont {C.-J.}\ \bibnamefont
  {Haster}}, \bibinfo {author} {\bibfnamefont {K.}~\bibnamefont {Yagi}},\ and\
  \bibinfo {author} {\bibfnamefont {N.}~\bibnamefont {Yunes}},\ }\bibfield
  {title} {\bibinfo {title} {{Equation-of-state insensitive relations after
  GW170817}},\ }\href {https://doi.org/10.1103/PhysRevD.99.083016} {\bibfield
  {journal} {\bibinfo  {journal} {Phys. Rev. D}\ }\textbf {\bibinfo {volume}
  {99}},\ \bibinfo {pages} {083016} (\bibinfo {year} {2019})},\ \Eprint
  {https://arxiv.org/abs/1903.03909} {arXiv:1903.03909 [gr-qc]} \BibitemShut
  {NoStop}%
\bibitem [{\citenamefont {Jaranowski}\ and\ \citenamefont
  {Krolak}(1994)}]{PhysRevD.49.1723}%
  \BibitemOpen
  \bibfield  {author} {\bibinfo {author} {\bibfnamefont {P.}~\bibnamefont
  {Jaranowski}}\ and\ \bibinfo {author} {\bibfnamefont {A.}~\bibnamefont
  {Krolak}},\ }\bibfield  {title} {\bibinfo {title} {Optimal solution to the
  inverse problem for the gravitational wave signal of a coalescing compact
  binary},\ }\href {https://doi.org/10.1103/PhysRevD.49.1723} {\bibfield
  {journal} {\bibinfo  {journal} {Phys. Rev. D}\ }\textbf {\bibinfo {volume}
  {49}},\ \bibinfo {pages} {1723} (\bibinfo {year} {1994})}\BibitemShut
  {NoStop}%
\bibitem [{\citenamefont {van~der Sluys}\ \emph {et~al.}(2008)\citenamefont
  {van~der Sluys}, \citenamefont {Röver}, \citenamefont {Stroeer},
  \citenamefont {Raymond}, \citenamefont {Mandel}, \citenamefont {Christensen},
  \citenamefont {Kalogera}, \citenamefont {Meyer},\ and\ \citenamefont
  {Vecchio}}]{van_der_Sluys_2008}%
  \BibitemOpen
  \bibfield  {author} {\bibinfo {author} {\bibfnamefont {M.~V.}\ \bibnamefont
  {van~der Sluys}}, \bibinfo {author} {\bibfnamefont {C.}~\bibnamefont
  {Röver}}, \bibinfo {author} {\bibfnamefont {A.}~\bibnamefont {Stroeer}},
  \bibinfo {author} {\bibfnamefont {V.}~\bibnamefont {Raymond}}, \bibinfo
  {author} {\bibfnamefont {I.}~\bibnamefont {Mandel}}, \bibinfo {author}
  {\bibfnamefont {N.}~\bibnamefont {Christensen}}, \bibinfo {author}
  {\bibfnamefont {V.}~\bibnamefont {Kalogera}}, \bibinfo {author}
  {\bibfnamefont {R.}~\bibnamefont {Meyer}},\ and\ \bibinfo {author}
  {\bibfnamefont {A.}~\bibnamefont {Vecchio}},\ }\bibfield  {title} {\bibinfo
  {title} {Gravitational-wave astronomy with inspiral signals of spinning
  compact-object binaries},\ }\href {https://doi.org/10.1086/595279} {\bibfield
   {journal} {\bibinfo  {journal} {The Astrophysical Journal}\ }\textbf
  {\bibinfo {volume} {688}},\ \bibinfo {pages} {L61} (\bibinfo {year}
  {2008})}\BibitemShut {NoStop}%
\bibitem [{\citenamefont {Veitch}\ \emph {et~al.}(2012)\citenamefont {Veitch},
  \citenamefont {Mandel}, \citenamefont {Aylott}, \citenamefont {Farr},
  \citenamefont {Raymond}, \citenamefont {Rodriguez}, \citenamefont {van~der
  Sluys}, \citenamefont {Kalogera},\ and\ \citenamefont
  {Vecchio}}]{Veitch_2012}%
  \BibitemOpen
  \bibfield  {author} {\bibinfo {author} {\bibfnamefont {J.}~\bibnamefont
  {Veitch}}, \bibinfo {author} {\bibfnamefont {I.}~\bibnamefont {Mandel}},
  \bibinfo {author} {\bibfnamefont {B.}~\bibnamefont {Aylott}}, \bibinfo
  {author} {\bibfnamefont {B.}~\bibnamefont {Farr}}, \bibinfo {author}
  {\bibfnamefont {V.}~\bibnamefont {Raymond}}, \bibinfo {author} {\bibfnamefont
  {C.}~\bibnamefont {Rodriguez}}, \bibinfo {author} {\bibfnamefont
  {M.}~\bibnamefont {van~der Sluys}}, \bibinfo {author} {\bibfnamefont
  {V.}~\bibnamefont {Kalogera}},\ and\ \bibinfo {author} {\bibfnamefont
  {A.}~\bibnamefont {Vecchio}},\ }\bibfield  {title} {\bibinfo {title}
  {Estimating parameters of coalescing compact binaries with proposed advanced
  detector networks},\ }\bibfield  {journal} {\bibinfo  {journal} {Physical
  Review D}\ }\textbf {\bibinfo {volume} {85}},\ \href
  {https://doi.org/10.1103/physrevd.85.104045} {10.1103/physrevd.85.104045}
  (\bibinfo {year} {2012})\BibitemShut {NoStop}%
\bibitem [{\citenamefont {Veitch}\ \emph {et~al.}(2015)\citenamefont {Veitch},
  \citenamefont {Raymond}, \citenamefont {Farr}, \citenamefont {Farr},
  \citenamefont {Graff}, \citenamefont {Vitale}, \citenamefont {Aylott},
  \citenamefont {Blackburn}, \citenamefont {Christensen}, \citenamefont
  {Coughlin},\ and\ \citenamefont {et~al.}}]{Veitch_2015}%
  \BibitemOpen
  \bibfield  {author} {\bibinfo {author} {\bibfnamefont {J.}~\bibnamefont
  {Veitch}}, \bibinfo {author} {\bibfnamefont {V.}~\bibnamefont {Raymond}},
  \bibinfo {author} {\bibfnamefont {B.}~\bibnamefont {Farr}}, \bibinfo {author}
  {\bibfnamefont {W.}~\bibnamefont {Farr}}, \bibinfo {author} {\bibfnamefont
  {P.}~\bibnamefont {Graff}}, \bibinfo {author} {\bibfnamefont
  {S.}~\bibnamefont {Vitale}}, \bibinfo {author} {\bibfnamefont
  {B.}~\bibnamefont {Aylott}}, \bibinfo {author} {\bibfnamefont
  {K.}~\bibnamefont {Blackburn}}, \bibinfo {author} {\bibfnamefont
  {N.}~\bibnamefont {Christensen}}, \bibinfo {author} {\bibfnamefont
  {M.}~\bibnamefont {Coughlin}},\ and\ \bibinfo {author} {\bibnamefont
  {et~al.}},\ }\bibfield  {title} {\bibinfo {title} {Parameter estimation for
  compact binaries with ground-based gravitational-wave observations using the
  lalinference software library},\ }\bibfield  {journal} {\bibinfo  {journal}
  {Physical Review D}\ }\textbf {\bibinfo {volume} {91}},\ \href
  {https://doi.org/10.1103/physrevd.91.042003} {10.1103/physrevd.91.042003}
  (\bibinfo {year} {2015})\BibitemShut {NoStop}%
\bibitem [{\citenamefont {Blanchet}\ \emph {et~al.}(1995)\citenamefont
  {Blanchet}, \citenamefont {Damour}, \citenamefont {Iyer}, \citenamefont
  {Will},\ and\ \citenamefont {Wiseman}}]{Blanchet_1995}%
  \BibitemOpen
  \bibfield  {author} {\bibinfo {author} {\bibfnamefont {L.}~\bibnamefont
  {Blanchet}}, \bibinfo {author} {\bibfnamefont {T.}~\bibnamefont {Damour}},
  \bibinfo {author} {\bibfnamefont {B.~R.}\ \bibnamefont {Iyer}}, \bibinfo
  {author} {\bibfnamefont {C.~M.}\ \bibnamefont {Will}},\ and\ \bibinfo
  {author} {\bibfnamefont {A.~G.}\ \bibnamefont {Wiseman}},\ }\bibfield
  {title} {\bibinfo {title} {Gravitational-radiation damping of compact binary
  systems to second post-newtonian order},\ }\href
  {https://doi.org/10.1103/physrevlett.74.3515} {\bibfield  {journal} {\bibinfo
   {journal} {Physical Review Letters}\ }\textbf {\bibinfo {volume} {74}},\
  \bibinfo {pages} {3515–3518} (\bibinfo {year} {1995})}\BibitemShut
  {NoStop}%
\bibitem [{\citenamefont {{Flanagan}}\ and\ \citenamefont
  {{Hinderer}}(2008)}]{2008PhRvD..77b1502F}%
  \BibitemOpen
  \bibfield  {author} {\bibinfo {author} {\bibfnamefont {{\'E}.~{\'E}.}\
  \bibnamefont {{Flanagan}}}\ and\ \bibinfo {author} {\bibfnamefont
  {T.}~\bibnamefont {{Hinderer}}},\ }\bibfield  {title} {\bibinfo {title}
  {{Constraining neutron-star tidal Love numbers with gravitational-wave
  detectors}},\ }\href {https://doi.org/10.1103/PhysRevD.77.021502} {\bibfield
  {journal} {\bibinfo  {journal} {\prd}\ }\textbf {\bibinfo {volume} {77}},\
  \bibinfo {eid} {021502} (\bibinfo {year} {2008})},\ \Eprint
  {https://arxiv.org/abs/0709.1915} {arXiv:0709.1915 [astro-ph]} \BibitemShut
  {NoStop}%
\bibitem [{\citenamefont {Hinderer}(2008)}]{Hinderer_2008}%
  \BibitemOpen
  \bibfield  {author} {\bibinfo {author} {\bibfnamefont {T.}~\bibnamefont
  {Hinderer}},\ }\bibfield  {title} {\bibinfo {title} {Tidal love numbers of
  neutron stars},\ }\href {https://doi.org/10.1086/533487} {\bibfield
  {journal} {\bibinfo  {journal} {The Astrophysical Journal}\ }\textbf
  {\bibinfo {volume} {677}},\ \bibinfo {pages} {1216} (\bibinfo {year}
  {2008})}\BibitemShut {NoStop}%
\bibitem [{\citenamefont {Yagi}\ and\ \citenamefont
  {Yunes}(2013{\natexlab{a}})}]{Yagi_2013}%
  \BibitemOpen
  \bibfield  {author} {\bibinfo {author} {\bibfnamefont {K.}~\bibnamefont
  {Yagi}}\ and\ \bibinfo {author} {\bibfnamefont {N.}~\bibnamefont {Yunes}},\
  }\bibfield  {title} {\bibinfo {title} {I-love-q: Unexpected universal
  relations for neutron stars and quark stars},\ }\href
  {https://doi.org/10.1126/science.1236462} {\bibfield  {journal} {\bibinfo
  {journal} {Science}\ }\textbf {\bibinfo {volume} {341}},\ \bibinfo {pages}
  {365–368} (\bibinfo {year} {2013}{\natexlab{a}})}\BibitemShut {NoStop}%
\bibitem [{\citenamefont {Yagi}\ and\ \citenamefont
  {Yunes}(2013{\natexlab{b}})}]{Yagi_2013_PRD}%
  \BibitemOpen
  \bibfield  {author} {\bibinfo {author} {\bibfnamefont {K.}~\bibnamefont
  {Yagi}}\ and\ \bibinfo {author} {\bibfnamefont {N.}~\bibnamefont {Yunes}},\
  }\bibfield  {title} {\bibinfo {title} {I-love-q relations in neutron stars
  and their applications to astrophysics, gravitational waves, and fundamental
  physics},\ }\bibfield  {journal} {\bibinfo  {journal} {Physical Review D}\
  }\textbf {\bibinfo {volume} {88}},\ \href
  {https://doi.org/10.1103/physrevd.88.023009} {10.1103/physrevd.88.023009}
  (\bibinfo {year} {2013}{\natexlab{b}})\BibitemShut {NoStop}%
\bibitem [{\citenamefont {Yagi}\ and\ \citenamefont
  {Yunes}(2017)}]{Yagi:2016bkt}%
  \BibitemOpen
  \bibfield  {author} {\bibinfo {author} {\bibfnamefont {K.}~\bibnamefont
  {Yagi}}\ and\ \bibinfo {author} {\bibfnamefont {N.}~\bibnamefont {Yunes}},\
  }\bibfield  {title} {\bibinfo {title} {{Approximate Universal Relations for
  Neutron Stars and Quark Stars}},\ }\href
  {https://doi.org/10.1016/j.physrep.2017.03.002} {\bibfield  {journal}
  {\bibinfo  {journal} {Phys. Rept.}\ }\textbf {\bibinfo {volume} {681}},\
  \bibinfo {pages} {1} (\bibinfo {year} {2017})},\ \Eprint
  {https://arxiv.org/abs/1608.02582} {arXiv:1608.02582 [gr-qc]} \BibitemShut
  {NoStop}%
\bibitem [{\citenamefont {Doneva}\ and\ \citenamefont
  {Pappas}(2018)}]{Doneva:2017jop}%
  \BibitemOpen
  \bibfield  {author} {\bibinfo {author} {\bibfnamefont {D.~D.}\ \bibnamefont
  {Doneva}}\ and\ \bibinfo {author} {\bibfnamefont {G.}~\bibnamefont
  {Pappas}},\ }\bibfield  {title} {\bibinfo {title} {{Universal Relations and
  Alternative Gravity Theories}},\ }\href
  {https://doi.org/10.1007/978-3-319-97616-7_13} {\bibfield  {journal}
  {\bibinfo  {journal} {Astrophys. Space Sci. Libr.}\ }\textbf {\bibinfo
  {volume} {457}},\ \bibinfo {pages} {737} (\bibinfo {year} {2018})},\ \Eprint
  {https://arxiv.org/abs/1709.08046} {arXiv:1709.08046 [gr-qc]} \BibitemShut
  {NoStop}%
\bibitem [{\citenamefont {Yunes}\ \emph {et~al.}(2022)\citenamefont {Yunes},
  \citenamefont {Miller},\ and\ \citenamefont {Yagi}}]{Yunes:2022ldq}%
  \BibitemOpen
  \bibfield  {author} {\bibinfo {author} {\bibfnamefont {N.}~\bibnamefont
  {Yunes}}, \bibinfo {author} {\bibfnamefont {M.~C.}\ \bibnamefont {Miller}},\
  and\ \bibinfo {author} {\bibfnamefont {K.}~\bibnamefont {Yagi}},\ }\bibfield
  {title} {\bibinfo {title} {{Gravitational-wave and X-ray probes of the
  neutron star equation of state}},\ }\href
  {https://doi.org/10.1038/s42254-022-00420-y} {\bibfield  {journal} {\bibinfo
  {journal} {Nature Rev. Phys.}\ }\textbf {\bibinfo {volume} {4}},\ \bibinfo
  {pages} {237} (\bibinfo {year} {2022})},\ \Eprint
  {https://arxiv.org/abs/2202.04117} {arXiv:2202.04117 [gr-qc]} \BibitemShut
  {NoStop}%
\bibitem [{\citenamefont {Abbott}\ \emph {et~al.}(2018)\citenamefont {Abbott}
  \emph {et~al.}}]{LIGOScientific:2018cki}%
  \BibitemOpen
  \bibfield  {author} {\bibinfo {author} {\bibfnamefont {B.~P.}\ \bibnamefont
  {Abbott}} \emph {et~al.} (\bibinfo {collaboration} {LIGO Scientific,
  Virgo}),\ }\bibfield  {title} {\bibinfo {title} {{GW170817: Measurements of
  neutron star radii and equation of state}},\ }\href
  {https://doi.org/10.1103/PhysRevLett.121.161101} {\bibfield  {journal}
  {\bibinfo  {journal} {Phys. Rev. Lett.}\ }\textbf {\bibinfo {volume} {121}},\
  \bibinfo {pages} {161101} (\bibinfo {year} {2018})},\ \Eprint
  {https://arxiv.org/abs/1805.11581} {arXiv:1805.11581 [gr-qc]} \BibitemShut
  {NoStop}%
\bibitem [{\citenamefont {Ivezi{\'c}}\ \emph {et~al.}(2019)\citenamefont
  {Ivezi{\'c}}, \citenamefont {Kahn}, \citenamefont {Tyson}, \citenamefont
  {Abel}, \citenamefont {Acosta}, \citenamefont {Allsman}, \citenamefont
  {Alonso}, \citenamefont {AlSayyad}, \citenamefont {Anderson}, \citenamefont
  {Andrew},\ and\ \citenamefont {et~al.}}]{Ivezic_2019}%
  \BibitemOpen
  \bibfield  {author} {\bibinfo {author} {\bibfnamefont {v.}~\bibnamefont
  {Ivezi{\'c}}}, \bibinfo {author} {\bibfnamefont {S.~M.}\ \bibnamefont
  {Kahn}}, \bibinfo {author} {\bibfnamefont {J.~A.}\ \bibnamefont {Tyson}},
  \bibinfo {author} {\bibfnamefont {B.}~\bibnamefont {Abel}}, \bibinfo {author}
  {\bibfnamefont {E.}~\bibnamefont {Acosta}}, \bibinfo {author} {\bibfnamefont
  {R.}~\bibnamefont {Allsman}}, \bibinfo {author} {\bibfnamefont
  {D.}~\bibnamefont {Alonso}}, \bibinfo {author} {\bibfnamefont
  {Y.}~\bibnamefont {AlSayyad}}, \bibinfo {author} {\bibfnamefont {S.~F.}\
  \bibnamefont {Anderson}}, \bibinfo {author} {\bibfnamefont {J.}~\bibnamefont
  {Andrew}},\ and\ \bibinfo {author} {\bibnamefont {et~al.}},\ }\bibfield
  {title} {\bibinfo {title} {Lsst: From science drivers to reference design and
  anticipated data products},\ }\href
  {https://doi.org/10.3847/1538-4357/ab042c} {\bibfield  {journal} {\bibinfo
  {journal} {The Astrophysical Journal}\ }\textbf {\bibinfo {volume} {873}},\
  \bibinfo {pages} {111} (\bibinfo {year} {2019})}\BibitemShut {NoStop}%
\bibitem [{cre(2011)}]{creighton2011gravitational.ch7}%
  \BibitemOpen
  \bibinfo {title} {Gravitational-wave data analysis},\ in\ \href
  {https://doi.org/https://doi.org/10.1002/9783527636037.ch7} {\emph {\bibinfo
  {booktitle} {Gravitational‐Wave Physics and Astronomy}}}\ (\bibinfo
  {publisher} {John Wiley \& Sons, Ltd},\ \bibinfo {year} {2011})\
  Chap.~\bibinfo {chapter} {7}, pp.\ \bibinfo {pages} {269--347}\BibitemShut
  {NoStop}%
\bibitem [{\citenamefont {{Astropy Collaboration}}\ \emph
  {et~al.}(2013)\citenamefont {{Astropy Collaboration}}, \citenamefont
  {{Robitaille}}, \citenamefont {{Tollerud}}, \citenamefont {{Greenfield}},
  \citenamefont {{Droettboom}}, \citenamefont {{Bray}}, \citenamefont
  {{Aldcroft}}, \citenamefont {{Davis}}, \citenamefont {{Ginsburg}},
  \citenamefont {{Price-Whelan}}, \citenamefont {{Kerzendorf}}, \citenamefont
  {{Conley}}, \citenamefont {{Crighton}}, \citenamefont {{Barbary}},
  \citenamefont {{Muna}}, \citenamefont {{Ferguson}}, \citenamefont
  {{Grollier}}, \citenamefont {{Parikh}}, \citenamefont {{Nair}}, \citenamefont
  {{Unther}}, \citenamefont {{Deil}}, \citenamefont {{Woillez}}, \citenamefont
  {{Conseil}}, \citenamefont {{Kramer}}, \citenamefont {{Turner}},
  \citenamefont {{Singer}}, \citenamefont {{Fox}}, \citenamefont {{Weaver}},
  \citenamefont {{Zabalza}}, \citenamefont {{Edwards}}, \citenamefont {{Azalee
  Bostroem}}, \citenamefont {{Burke}}, \citenamefont {{Casey}}, \citenamefont
  {{Crawford}}, \citenamefont {{Dencheva}}, \citenamefont {{Ely}},
  \citenamefont {{Jenness}}, \citenamefont {{Labrie}}, \citenamefont {{Lim}},
  \citenamefont {{Pierfederici}}, \citenamefont {{Pontzen}}, \citenamefont
  {{Ptak}}, \citenamefont {{Refsdal}}, \citenamefont {{Servillat}},\ and\
  \citenamefont {{Streicher}}}]{astropy:2013}%
  \BibitemOpen
  \bibfield  {author} {\bibinfo {author} {\bibnamefont {{Astropy
  Collaboration}}}, \bibinfo {author} {\bibfnamefont {T.~P.}\ \bibnamefont
  {{Robitaille}}}, \bibinfo {author} {\bibfnamefont {E.~J.}\ \bibnamefont
  {{Tollerud}}}, \bibinfo {author} {\bibfnamefont {P.}~\bibnamefont
  {{Greenfield}}}, \bibinfo {author} {\bibfnamefont {M.}~\bibnamefont
  {{Droettboom}}}, \bibinfo {author} {\bibfnamefont {E.}~\bibnamefont
  {{Bray}}}, \bibinfo {author} {\bibfnamefont {T.}~\bibnamefont {{Aldcroft}}},
  \bibinfo {author} {\bibfnamefont {M.}~\bibnamefont {{Davis}}}, \bibinfo
  {author} {\bibfnamefont {A.}~\bibnamefont {{Ginsburg}}}, \bibinfo {author}
  {\bibfnamefont {A.~M.}\ \bibnamefont {{Price-Whelan}}}, \bibinfo {author}
  {\bibfnamefont {W.~E.}\ \bibnamefont {{Kerzendorf}}}, \bibinfo {author}
  {\bibfnamefont {A.}~\bibnamefont {{Conley}}}, \bibinfo {author}
  {\bibfnamefont {N.}~\bibnamefont {{Crighton}}}, \bibinfo {author}
  {\bibfnamefont {K.}~\bibnamefont {{Barbary}}}, \bibinfo {author}
  {\bibfnamefont {D.}~\bibnamefont {{Muna}}}, \bibinfo {author} {\bibfnamefont
  {H.}~\bibnamefont {{Ferguson}}}, \bibinfo {author} {\bibfnamefont
  {F.}~\bibnamefont {{Grollier}}}, \bibinfo {author} {\bibfnamefont {M.~M.}\
  \bibnamefont {{Parikh}}}, \bibinfo {author} {\bibfnamefont {P.~H.}\
  \bibnamefont {{Nair}}}, \bibinfo {author} {\bibfnamefont {H.~M.}\
  \bibnamefont {{Unther}}}, \bibinfo {author} {\bibfnamefont {C.}~\bibnamefont
  {{Deil}}}, \bibinfo {author} {\bibfnamefont {J.}~\bibnamefont {{Woillez}}},
  \bibinfo {author} {\bibfnamefont {S.}~\bibnamefont {{Conseil}}}, \bibinfo
  {author} {\bibfnamefont {R.}~\bibnamefont {{Kramer}}}, \bibinfo {author}
  {\bibfnamefont {J.~E.~H.}\ \bibnamefont {{Turner}}}, \bibinfo {author}
  {\bibfnamefont {L.}~\bibnamefont {{Singer}}}, \bibinfo {author}
  {\bibfnamefont {R.}~\bibnamefont {{Fox}}}, \bibinfo {author} {\bibfnamefont
  {B.~A.}\ \bibnamefont {{Weaver}}}, \bibinfo {author} {\bibfnamefont
  {V.}~\bibnamefont {{Zabalza}}}, \bibinfo {author} {\bibfnamefont {Z.~I.}\
  \bibnamefont {{Edwards}}}, \bibinfo {author} {\bibfnamefont {K.}~\bibnamefont
  {{Azalee Bostroem}}}, \bibinfo {author} {\bibfnamefont {D.~J.}\ \bibnamefont
  {{Burke}}}, \bibinfo {author} {\bibfnamefont {A.~R.}\ \bibnamefont
  {{Casey}}}, \bibinfo {author} {\bibfnamefont {S.~M.}\ \bibnamefont
  {{Crawford}}}, \bibinfo {author} {\bibfnamefont {N.}~\bibnamefont
  {{Dencheva}}}, \bibinfo {author} {\bibfnamefont {J.}~\bibnamefont {{Ely}}},
  \bibinfo {author} {\bibfnamefont {T.}~\bibnamefont {{Jenness}}}, \bibinfo
  {author} {\bibfnamefont {K.}~\bibnamefont {{Labrie}}}, \bibinfo {author}
  {\bibfnamefont {P.~L.}\ \bibnamefont {{Lim}}}, \bibinfo {author}
  {\bibfnamefont {F.}~\bibnamefont {{Pierfederici}}}, \bibinfo {author}
  {\bibfnamefont {A.}~\bibnamefont {{Pontzen}}}, \bibinfo {author}
  {\bibfnamefont {A.}~\bibnamefont {{Ptak}}}, \bibinfo {author} {\bibfnamefont
  {B.}~\bibnamefont {{Refsdal}}}, \bibinfo {author} {\bibfnamefont
  {M.}~\bibnamefont {{Servillat}}},\ and\ \bibinfo {author} {\bibfnamefont
  {O.}~\bibnamefont {{Streicher}}},\ }\bibfield  {title} {\bibinfo {title}
  {{Astropy: A community Python package for astronomy}},\ }\href
  {https://doi.org/10.1051/0004-6361/201322068} {\bibfield  {journal} {\bibinfo
   {journal} {\aap}\ }\textbf {\bibinfo {volume} {558}},\ \bibinfo {eid} {A33}
  (\bibinfo {year} {2013})},\ \Eprint {https://arxiv.org/abs/1307.6212}
  {arXiv:1307.6212 [astro-ph.IM]} \BibitemShut {NoStop}%
\bibitem [{\citenamefont {{Astropy Collaboration}}\ \emph
  {et~al.}(2018)\citenamefont {{Astropy Collaboration}}, \citenamefont
  {{Price-Whelan}}, \citenamefont {{Sip{\H{o}}cz}}, \citenamefont
  {{G{\"u}nther}}, \citenamefont {{Lim}}, \citenamefont {{Crawford}},
  \citenamefont {{Conseil}}, \citenamefont {{Shupe}}, \citenamefont {{Craig}},
  \citenamefont {{Dencheva}}, \citenamefont {{Ginsburg}}, \citenamefont {{Vand
  erPlas}}, \citenamefont {{Bradley}}, \citenamefont {{P{\'e}rez-Su{\'a}rez}},
  \citenamefont {{de Val-Borro}}, \citenamefont {{Aldcroft}}, \citenamefont
  {{Cruz}}, \citenamefont {{Robitaille}}, \citenamefont {{Tollerud}},
  \citenamefont {{Ardelean}}, \citenamefont {{Babej}}, \citenamefont {{Bach}},
  \citenamefont {{Bachetti}}, \citenamefont {{Bakanov}}, \citenamefont
  {{Bamford}}, \citenamefont {{Barentsen}}, \citenamefont {{Barmby}},
  \citenamefont {{Baumbach}}, \citenamefont {{Berry}}, \citenamefont
  {{Biscani}}, \citenamefont {{Boquien}}, \citenamefont {{Bostroem}},
  \citenamefont {{Bouma}}, \citenamefont {{Brammer}}, \citenamefont {{Bray}},
  \citenamefont {{Breytenbach}}, \citenamefont {{Buddelmeijer}}, \citenamefont
  {{Burke}}, \citenamefont {{Calderone}}, \citenamefont {{Cano
  Rodr{\'\i}guez}}, \citenamefont {{Cara}}, \citenamefont {{Cardoso}},
  \citenamefont {{Cheedella}}, \citenamefont {{Copin}}, \citenamefont
  {{Corrales}}, \citenamefont {{Crichton}}, \citenamefont {{D'Avella}},
  \citenamefont {{Deil}}, \citenamefont {{Depagne}}, \citenamefont
  {{Dietrich}}, \citenamefont {{Donath}}, \citenamefont {{Droettboom}},
  \citenamefont {{Earl}}, \citenamefont {{Erben}}, \citenamefont {{Fabbro}},
  \citenamefont {{Ferreira}}, \citenamefont {{Finethy}}, \citenamefont {{Fox}},
  \citenamefont {{Garrison}}, \citenamefont {{Gibbons}}, \citenamefont
  {{Goldstein}}, \citenamefont {{Gommers}}, \citenamefont {{Greco}},
  \citenamefont {{Greenfield}}, \citenamefont {{Groener}}, \citenamefont
  {{Grollier}}, \citenamefont {{Hagen}}, \citenamefont {{Hirst}}, \citenamefont
  {{Homeier}}, \citenamefont {{Horton}}, \citenamefont {{Hosseinzadeh}},
  \citenamefont {{Hu}}, \citenamefont {{Hunkeler}}, \citenamefont
  {{Ivezi{\'c}}}, \citenamefont {{Jain}}, \citenamefont {{Jenness}},
  \citenamefont {{Kanarek}}, \citenamefont {{Kendrew}}, \citenamefont {{Kern}},
  \citenamefont {{Kerzendorf}}, \citenamefont {{Khvalko}}, \citenamefont
  {{King}}, \citenamefont {{Kirkby}}, \citenamefont {{Kulkarni}}, \citenamefont
  {{Kumar}}, \citenamefont {{Lee}}, \citenamefont {{Lenz}}, \citenamefont
  {{Littlefair}}, \citenamefont {{Ma}}, \citenamefont {{Macleod}},
  \citenamefont {{Mastropietro}}, \citenamefont {{McCully}}, \citenamefont
  {{Montagnac}}, \citenamefont {{Morris}}, \citenamefont {{Mueller}},
  \citenamefont {{Mumford}}, \citenamefont {{Muna}}, \citenamefont {{Murphy}},
  \citenamefont {{Nelson}}, \citenamefont {{Nguyen}}, \citenamefont {{Ninan}},
  \citenamefont {{N{\"o}the}}, \citenamefont {{Ogaz}}, \citenamefont {{Oh}},
  \citenamefont {{Parejko}}, \citenamefont {{Parley}}, \citenamefont
  {{Pascual}}, \citenamefont {{Patil}}, \citenamefont {{Patil}}, \citenamefont
  {{Plunkett}}, \citenamefont {{Prochaska}}, \citenamefont {{Rastogi}},
  \citenamefont {{Reddy Janga}}, \citenamefont {{Sabater}}, \citenamefont
  {{Sakurikar}}, \citenamefont {{Seifert}}, \citenamefont {{Sherbert}},
  \citenamefont {{Sherwood-Taylor}}, \citenamefont {{Shih}}, \citenamefont
  {{Sick}}, \citenamefont {{Silbiger}}, \citenamefont {{Singanamalla}},
  \citenamefont {{Singer}}, \citenamefont {{Sladen}}, \citenamefont {{Sooley}},
  \citenamefont {{Sornarajah}}, \citenamefont {{Streicher}}, \citenamefont
  {{Teuben}}, \citenamefont {{Thomas}}, \citenamefont {{Tremblay}},
  \citenamefont {{Turner}}, \citenamefont {{Terr{\'o}n}}, \citenamefont {{van
  Kerkwijk}}, \citenamefont {{de la Vega}}, \citenamefont {{Watkins}},
  \citenamefont {{Weaver}}, \citenamefont {{Whitmore}}, \citenamefont
  {{Woillez}}, \citenamefont {{Zabalza}},\ and\ \citenamefont {{Astropy
  Contributors}}}]{astropy:2018}%
  \BibitemOpen
  \bibfield  {author} {\bibinfo {author} {\bibnamefont {{Astropy
  Collaboration}}}, \bibinfo {author} {\bibfnamefont {A.~M.}\ \bibnamefont
  {{Price-Whelan}}}, \bibinfo {author} {\bibfnamefont {B.~M.}\ \bibnamefont
  {{Sip{\H{o}}cz}}}, \bibinfo {author} {\bibfnamefont {H.~M.}\ \bibnamefont
  {{G{\"u}nther}}}, \bibinfo {author} {\bibfnamefont {P.~L.}\ \bibnamefont
  {{Lim}}}, \bibinfo {author} {\bibfnamefont {S.~M.}\ \bibnamefont
  {{Crawford}}}, \bibinfo {author} {\bibfnamefont {S.}~\bibnamefont
  {{Conseil}}}, \bibinfo {author} {\bibfnamefont {D.~L.}\ \bibnamefont
  {{Shupe}}}, \bibinfo {author} {\bibfnamefont {M.~W.}\ \bibnamefont
  {{Craig}}}, \bibinfo {author} {\bibfnamefont {N.}~\bibnamefont {{Dencheva}}},
  \bibinfo {author} {\bibfnamefont {A.}~\bibnamefont {{Ginsburg}}}, \bibinfo
  {author} {\bibfnamefont {J.~T.}\ \bibnamefont {{Vand erPlas}}}, \bibinfo
  {author} {\bibfnamefont {L.~D.}\ \bibnamefont {{Bradley}}}, \bibinfo {author}
  {\bibfnamefont {D.}~\bibnamefont {{P{\'e}rez-Su{\'a}rez}}}, \bibinfo {author}
  {\bibfnamefont {M.}~\bibnamefont {{de Val-Borro}}}, \bibinfo {author}
  {\bibfnamefont {T.~L.}\ \bibnamefont {{Aldcroft}}}, \bibinfo {author}
  {\bibfnamefont {K.~L.}\ \bibnamefont {{Cruz}}}, \bibinfo {author}
  {\bibfnamefont {T.~P.}\ \bibnamefont {{Robitaille}}}, \bibinfo {author}
  {\bibfnamefont {E.~J.}\ \bibnamefont {{Tollerud}}}, \bibinfo {author}
  {\bibfnamefont {C.}~\bibnamefont {{Ardelean}}}, \bibinfo {author}
  {\bibfnamefont {T.}~\bibnamefont {{Babej}}}, \bibinfo {author} {\bibfnamefont
  {Y.~P.}\ \bibnamefont {{Bach}}}, \bibinfo {author} {\bibfnamefont
  {M.}~\bibnamefont {{Bachetti}}}, \bibinfo {author} {\bibfnamefont {A.~V.}\
  \bibnamefont {{Bakanov}}}, \bibinfo {author} {\bibfnamefont {S.~P.}\
  \bibnamefont {{Bamford}}}, \bibinfo {author} {\bibfnamefont {G.}~\bibnamefont
  {{Barentsen}}}, \bibinfo {author} {\bibfnamefont {P.}~\bibnamefont
  {{Barmby}}}, \bibinfo {author} {\bibfnamefont {A.}~\bibnamefont
  {{Baumbach}}}, \bibinfo {author} {\bibfnamefont {K.~L.}\ \bibnamefont
  {{Berry}}}, \bibinfo {author} {\bibfnamefont {F.}~\bibnamefont {{Biscani}}},
  \bibinfo {author} {\bibfnamefont {M.}~\bibnamefont {{Boquien}}}, \bibinfo
  {author} {\bibfnamefont {K.~A.}\ \bibnamefont {{Bostroem}}}, \bibinfo
  {author} {\bibfnamefont {L.~G.}\ \bibnamefont {{Bouma}}}, \bibinfo {author}
  {\bibfnamefont {G.~B.}\ \bibnamefont {{Brammer}}}, \bibinfo {author}
  {\bibfnamefont {E.~M.}\ \bibnamefont {{Bray}}}, \bibinfo {author}
  {\bibfnamefont {H.}~\bibnamefont {{Breytenbach}}}, \bibinfo {author}
  {\bibfnamefont {H.}~\bibnamefont {{Buddelmeijer}}}, \bibinfo {author}
  {\bibfnamefont {D.~J.}\ \bibnamefont {{Burke}}}, \bibinfo {author}
  {\bibfnamefont {G.}~\bibnamefont {{Calderone}}}, \bibinfo {author}
  {\bibfnamefont {J.~L.}\ \bibnamefont {{Cano Rodr{\'\i}guez}}}, \bibinfo
  {author} {\bibfnamefont {M.}~\bibnamefont {{Cara}}}, \bibinfo {author}
  {\bibfnamefont {J.~V.~M.}\ \bibnamefont {{Cardoso}}}, \bibinfo {author}
  {\bibfnamefont {S.}~\bibnamefont {{Cheedella}}}, \bibinfo {author}
  {\bibfnamefont {Y.}~\bibnamefont {{Copin}}}, \bibinfo {author} {\bibfnamefont
  {L.}~\bibnamefont {{Corrales}}}, \bibinfo {author} {\bibfnamefont
  {D.}~\bibnamefont {{Crichton}}}, \bibinfo {author} {\bibfnamefont
  {D.}~\bibnamefont {{D'Avella}}}, \bibinfo {author} {\bibfnamefont
  {C.}~\bibnamefont {{Deil}}}, \bibinfo {author} {\bibfnamefont
  {{\'E}.}~\bibnamefont {{Depagne}}}, \bibinfo {author} {\bibfnamefont {J.~P.}\
  \bibnamefont {{Dietrich}}}, \bibinfo {author} {\bibfnamefont
  {A.}~\bibnamefont {{Donath}}}, \bibinfo {author} {\bibfnamefont
  {M.}~\bibnamefont {{Droettboom}}}, \bibinfo {author} {\bibfnamefont
  {N.}~\bibnamefont {{Earl}}}, \bibinfo {author} {\bibfnamefont
  {T.}~\bibnamefont {{Erben}}}, \bibinfo {author} {\bibfnamefont
  {S.}~\bibnamefont {{Fabbro}}}, \bibinfo {author} {\bibfnamefont {L.~A.}\
  \bibnamefont {{Ferreira}}}, \bibinfo {author} {\bibfnamefont
  {T.}~\bibnamefont {{Finethy}}}, \bibinfo {author} {\bibfnamefont {R.~T.}\
  \bibnamefont {{Fox}}}, \bibinfo {author} {\bibfnamefont {L.~H.}\ \bibnamefont
  {{Garrison}}}, \bibinfo {author} {\bibfnamefont {S.~L.~J.}\ \bibnamefont
  {{Gibbons}}}, \bibinfo {author} {\bibfnamefont {D.~A.}\ \bibnamefont
  {{Goldstein}}}, \bibinfo {author} {\bibfnamefont {R.}~\bibnamefont
  {{Gommers}}}, \bibinfo {author} {\bibfnamefont {J.~P.}\ \bibnamefont
  {{Greco}}}, \bibinfo {author} {\bibfnamefont {P.}~\bibnamefont
  {{Greenfield}}}, \bibinfo {author} {\bibfnamefont {A.~M.}\ \bibnamefont
  {{Groener}}}, \bibinfo {author} {\bibfnamefont {F.}~\bibnamefont
  {{Grollier}}}, \bibinfo {author} {\bibfnamefont {A.}~\bibnamefont {{Hagen}}},
  \bibinfo {author} {\bibfnamefont {P.}~\bibnamefont {{Hirst}}}, \bibinfo
  {author} {\bibfnamefont {D.}~\bibnamefont {{Homeier}}}, \bibinfo {author}
  {\bibfnamefont {A.~J.}\ \bibnamefont {{Horton}}}, \bibinfo {author}
  {\bibfnamefont {G.}~\bibnamefont {{Hosseinzadeh}}}, \bibinfo {author}
  {\bibfnamefont {L.}~\bibnamefont {{Hu}}}, \bibinfo {author} {\bibfnamefont
  {J.~S.}\ \bibnamefont {{Hunkeler}}}, \bibinfo {author} {\bibfnamefont
  {{\v{Z}}.}~\bibnamefont {{Ivezi{\'c}}}}, \bibinfo {author} {\bibfnamefont
  {A.}~\bibnamefont {{Jain}}}, \bibinfo {author} {\bibfnamefont
  {T.}~\bibnamefont {{Jenness}}}, \bibinfo {author} {\bibfnamefont
  {G.}~\bibnamefont {{Kanarek}}}, \bibinfo {author} {\bibfnamefont
  {S.}~\bibnamefont {{Kendrew}}}, \bibinfo {author} {\bibfnamefont {N.~S.}\
  \bibnamefont {{Kern}}}, \bibinfo {author} {\bibfnamefont {W.~E.}\
  \bibnamefont {{Kerzendorf}}}, \bibinfo {author} {\bibfnamefont
  {A.}~\bibnamefont {{Khvalko}}}, \bibinfo {author} {\bibfnamefont
  {J.}~\bibnamefont {{King}}}, \bibinfo {author} {\bibfnamefont
  {D.}~\bibnamefont {{Kirkby}}}, \bibinfo {author} {\bibfnamefont {A.~M.}\
  \bibnamefont {{Kulkarni}}}, \bibinfo {author} {\bibfnamefont
  {A.}~\bibnamefont {{Kumar}}}, \bibinfo {author} {\bibfnamefont
  {A.}~\bibnamefont {{Lee}}}, \bibinfo {author} {\bibfnamefont
  {D.}~\bibnamefont {{Lenz}}}, \bibinfo {author} {\bibfnamefont {S.~P.}\
  \bibnamefont {{Littlefair}}}, \bibinfo {author} {\bibfnamefont
  {Z.}~\bibnamefont {{Ma}}}, \bibinfo {author} {\bibfnamefont {D.~M.}\
  \bibnamefont {{Macleod}}}, \bibinfo {author} {\bibfnamefont {M.}~\bibnamefont
  {{Mastropietro}}}, \bibinfo {author} {\bibfnamefont {C.}~\bibnamefont
  {{McCully}}}, \bibinfo {author} {\bibfnamefont {S.}~\bibnamefont
  {{Montagnac}}}, \bibinfo {author} {\bibfnamefont {B.~M.}\ \bibnamefont
  {{Morris}}}, \bibinfo {author} {\bibfnamefont {M.}~\bibnamefont {{Mueller}}},
  \bibinfo {author} {\bibfnamefont {S.~J.}\ \bibnamefont {{Mumford}}}, \bibinfo
  {author} {\bibfnamefont {D.}~\bibnamefont {{Muna}}}, \bibinfo {author}
  {\bibfnamefont {N.~A.}\ \bibnamefont {{Murphy}}}, \bibinfo {author}
  {\bibfnamefont {S.}~\bibnamefont {{Nelson}}}, \bibinfo {author}
  {\bibfnamefont {G.~H.}\ \bibnamefont {{Nguyen}}}, \bibinfo {author}
  {\bibfnamefont {J.~P.}\ \bibnamefont {{Ninan}}}, \bibinfo {author}
  {\bibfnamefont {M.}~\bibnamefont {{N{\"o}the}}}, \bibinfo {author}
  {\bibfnamefont {S.}~\bibnamefont {{Ogaz}}}, \bibinfo {author} {\bibfnamefont
  {S.}~\bibnamefont {{Oh}}}, \bibinfo {author} {\bibfnamefont {J.~K.}\
  \bibnamefont {{Parejko}}}, \bibinfo {author} {\bibfnamefont {N.}~\bibnamefont
  {{Parley}}}, \bibinfo {author} {\bibfnamefont {S.}~\bibnamefont {{Pascual}}},
  \bibinfo {author} {\bibfnamefont {R.}~\bibnamefont {{Patil}}}, \bibinfo
  {author} {\bibfnamefont {A.~A.}\ \bibnamefont {{Patil}}}, \bibinfo {author}
  {\bibfnamefont {A.~L.}\ \bibnamefont {{Plunkett}}}, \bibinfo {author}
  {\bibfnamefont {J.~X.}\ \bibnamefont {{Prochaska}}}, \bibinfo {author}
  {\bibfnamefont {T.}~\bibnamefont {{Rastogi}}}, \bibinfo {author}
  {\bibfnamefont {V.}~\bibnamefont {{Reddy Janga}}}, \bibinfo {author}
  {\bibfnamefont {J.}~\bibnamefont {{Sabater}}}, \bibinfo {author}
  {\bibfnamefont {P.}~\bibnamefont {{Sakurikar}}}, \bibinfo {author}
  {\bibfnamefont {M.}~\bibnamefont {{Seifert}}}, \bibinfo {author}
  {\bibfnamefont {L.~E.}\ \bibnamefont {{Sherbert}}}, \bibinfo {author}
  {\bibfnamefont {H.}~\bibnamefont {{Sherwood-Taylor}}}, \bibinfo {author}
  {\bibfnamefont {A.~Y.}\ \bibnamefont {{Shih}}}, \bibinfo {author}
  {\bibfnamefont {J.}~\bibnamefont {{Sick}}}, \bibinfo {author} {\bibfnamefont
  {M.~T.}\ \bibnamefont {{Silbiger}}}, \bibinfo {author} {\bibfnamefont
  {S.}~\bibnamefont {{Singanamalla}}}, \bibinfo {author} {\bibfnamefont
  {L.~P.}\ \bibnamefont {{Singer}}}, \bibinfo {author} {\bibfnamefont {P.~H.}\
  \bibnamefont {{Sladen}}}, \bibinfo {author} {\bibfnamefont {K.~A.}\
  \bibnamefont {{Sooley}}}, \bibinfo {author} {\bibfnamefont {S.}~\bibnamefont
  {{Sornarajah}}}, \bibinfo {author} {\bibfnamefont {O.}~\bibnamefont
  {{Streicher}}}, \bibinfo {author} {\bibfnamefont {P.}~\bibnamefont
  {{Teuben}}}, \bibinfo {author} {\bibfnamefont {S.~W.}\ \bibnamefont
  {{Thomas}}}, \bibinfo {author} {\bibfnamefont {G.~R.}\ \bibnamefont
  {{Tremblay}}}, \bibinfo {author} {\bibfnamefont {J.~E.~H.}\ \bibnamefont
  {{Turner}}}, \bibinfo {author} {\bibfnamefont {V.}~\bibnamefont
  {{Terr{\'o}n}}}, \bibinfo {author} {\bibfnamefont {M.~H.}\ \bibnamefont {{van
  Kerkwijk}}}, \bibinfo {author} {\bibfnamefont {A.}~\bibnamefont {{de la
  Vega}}}, \bibinfo {author} {\bibfnamefont {L.~L.}\ \bibnamefont {{Watkins}}},
  \bibinfo {author} {\bibfnamefont {B.~A.}\ \bibnamefont {{Weaver}}}, \bibinfo
  {author} {\bibfnamefont {J.~B.}\ \bibnamefont {{Whitmore}}}, \bibinfo
  {author} {\bibfnamefont {J.}~\bibnamefont {{Woillez}}}, \bibinfo {author}
  {\bibfnamefont {V.}~\bibnamefont {{Zabalza}}},\ and\ \bibinfo {author}
  {\bibnamefont {{Astropy Contributors}}},\ }\bibfield  {title} {\bibinfo
  {title} {{The Astropy Project: Building an Open-science Project and Status of
  the v2.0 Core Package}},\ }\href {https://doi.org/10.3847/1538-3881/aabc4f}
  {\bibfield  {journal} {\bibinfo  {journal} {\aj}\ }\textbf {\bibinfo {volume}
  {156}},\ \bibinfo {eid} {123} (\bibinfo {year} {2018})},\ \Eprint
  {https://arxiv.org/abs/1801.02634} {arXiv:1801.02634 [astro-ph.IM]}
  \BibitemShut {NoStop}%
\bibitem [{\citenamefont {Abbott}\ \emph {et~al.}(2019)\citenamefont {Abbott}
  \emph {et~al.}}]{LIGOScientific:2018hze}%
  \BibitemOpen
  \bibfield  {author} {\bibinfo {author} {\bibfnamefont {B.~P.}\ \bibnamefont
  {Abbott}} \emph {et~al.} (\bibinfo {collaboration} {LIGO Scientific,
  Virgo}),\ }\bibfield  {title} {\bibinfo {title} {{Properties of the binary
  neutron star merger GW170817}},\ }\href
  {https://doi.org/10.1103/PhysRevX.9.011001} {\bibfield  {journal} {\bibinfo
  {journal} {Phys. Rev. X}\ }\textbf {\bibinfo {volume} {9}},\ \bibinfo {pages}
  {011001} (\bibinfo {year} {2019})},\ \Eprint
  {https://arxiv.org/abs/1805.11579} {arXiv:1805.11579 [gr-qc]} \BibitemShut
  {NoStop}%
\bibitem [{\citenamefont {Smith}\ \emph {et~al.}(2020)\citenamefont {Smith},
  \citenamefont {Ashton}, \citenamefont {Vajpeyi},\ and\ \citenamefont
  {Talbot}}]{Smith_2020}%
  \BibitemOpen
  \bibfield  {author} {\bibinfo {author} {\bibfnamefont {R.~J.~E.}\
  \bibnamefont {Smith}}, \bibinfo {author} {\bibfnamefont {G.}~\bibnamefont
  {Ashton}}, \bibinfo {author} {\bibfnamefont {A.}~\bibnamefont {Vajpeyi}},\
  and\ \bibinfo {author} {\bibfnamefont {C.}~\bibnamefont {Talbot}},\
  }\bibfield  {title} {\bibinfo {title} {Massively parallel bayesian inference
  for transient gravitational-wave astronomy},\ }\href
  {https://doi.org/10.1093/mnras/staa2483} {\bibfield  {journal} {\bibinfo
  {journal} {Monthly Notices of the Royal Astronomical Society}\ }\textbf
  {\bibinfo {volume} {498}},\ \bibinfo {pages} {4492–4502} (\bibinfo {year}
  {2020})}\BibitemShut {NoStop}%
\bibitem [{\citenamefont {Dietrich}\ \emph {et~al.}(2019)\citenamefont
  {Dietrich}, \citenamefont {Khan}, \citenamefont {Dudi}, \citenamefont
  {Kapadia}, \citenamefont {Kumar}, \citenamefont {Nagar}, \citenamefont
  {Ohme}, \citenamefont {Pannarale}, \citenamefont {Samajdar}, \citenamefont
  {Bernuzzi}, \citenamefont {Carullo}, \citenamefont {Del~Pozzo}, \citenamefont
  {Haney}, \citenamefont {Markakis}, \citenamefont {P\"urrer}, \citenamefont
  {Riemenschneider}, \citenamefont {Setyawati}, \citenamefont {Tsang},\ and\
  \citenamefont {Van Den~Broeck}}]{PhysRevD.99.024029}%
  \BibitemOpen
  \bibfield  {author} {\bibinfo {author} {\bibfnamefont {T.}~\bibnamefont
  {Dietrich}}, \bibinfo {author} {\bibfnamefont {S.}~\bibnamefont {Khan}},
  \bibinfo {author} {\bibfnamefont {R.}~\bibnamefont {Dudi}}, \bibinfo {author}
  {\bibfnamefont {S.~J.}\ \bibnamefont {Kapadia}}, \bibinfo {author}
  {\bibfnamefont {P.}~\bibnamefont {Kumar}}, \bibinfo {author} {\bibfnamefont
  {A.}~\bibnamefont {Nagar}}, \bibinfo {author} {\bibfnamefont
  {F.}~\bibnamefont {Ohme}}, \bibinfo {author} {\bibfnamefont {F.}~\bibnamefont
  {Pannarale}}, \bibinfo {author} {\bibfnamefont {A.}~\bibnamefont {Samajdar}},
  \bibinfo {author} {\bibfnamefont {S.}~\bibnamefont {Bernuzzi}}, \bibinfo
  {author} {\bibfnamefont {G.}~\bibnamefont {Carullo}}, \bibinfo {author}
  {\bibfnamefont {W.}~\bibnamefont {Del~Pozzo}}, \bibinfo {author}
  {\bibfnamefont {M.}~\bibnamefont {Haney}}, \bibinfo {author} {\bibfnamefont
  {C.}~\bibnamefont {Markakis}}, \bibinfo {author} {\bibfnamefont
  {M.}~\bibnamefont {P\"urrer}}, \bibinfo {author} {\bibfnamefont
  {G.}~\bibnamefont {Riemenschneider}}, \bibinfo {author} {\bibfnamefont
  {Y.~E.}\ \bibnamefont {Setyawati}}, \bibinfo {author} {\bibfnamefont {K.~W.}\
  \bibnamefont {Tsang}},\ and\ \bibinfo {author} {\bibfnamefont
  {C.}~\bibnamefont {Van Den~Broeck}},\ }\bibfield  {title} {\bibinfo {title}
  {Matter imprints in waveform models for neutron star binaries: Tidal and
  self-spin effects},\ }\href {https://doi.org/10.1103/PhysRevD.99.024029}
  {\bibfield  {journal} {\bibinfo  {journal} {Phys. Rev. D}\ }\textbf {\bibinfo
  {volume} {99}},\ \bibinfo {pages} {024029} (\bibinfo {year}
  {2019})}\BibitemShut {NoStop}%
\bibitem [{\citenamefont {{Speagle}}(2020)}]{dynesty}%
  \BibitemOpen
  \bibfield  {author} {\bibinfo {author} {\bibfnamefont {J.~S.}\ \bibnamefont
  {{Speagle}}},\ }\bibfield  {title} {\bibinfo {title} {{DYNESTY: a dynamic
  nested sampling package for estimating Bayesian posteriors and evidences}},\
  }\href {https://doi.org/10.1093/mnras/staa278} {\bibfield  {journal}
  {\bibinfo  {journal} {\mnras}\ }\textbf {\bibinfo {volume} {493}},\ \bibinfo
  {pages} {3132} (\bibinfo {year} {2020})},\ \Eprint
  {https://arxiv.org/abs/1904.02180} {arXiv:1904.02180 [astro-ph.IM]}
  \BibitemShut {NoStop}%
\bibitem [{\citenamefont {Abbott}\ \emph {et~al.}(2020)\citenamefont {Abbott},
  \citenamefont {Abbott}, \citenamefont {Abbott}, \citenamefont {Abraham},
  \citenamefont {Acernese}, \citenamefont {Ackley}, \citenamefont {Adams},
  \citenamefont {Adya}, \citenamefont {Affeldt},\ and\ \citenamefont
  {et~al.}}]{observing_scenarios}%
  \BibitemOpen
  \bibfield  {author} {\bibinfo {author} {\bibfnamefont {B.~P.}\ \bibnamefont
  {Abbott}}, \bibinfo {author} {\bibfnamefont {R.}~\bibnamefont {Abbott}},
  \bibinfo {author} {\bibfnamefont {T.~D.}\ \bibnamefont {Abbott}}, \bibinfo
  {author} {\bibfnamefont {S.}~\bibnamefont {Abraham}}, \bibinfo {author}
  {\bibfnamefont {F.}~\bibnamefont {Acernese}}, \bibinfo {author}
  {\bibfnamefont {K.}~\bibnamefont {Ackley}}, \bibinfo {author} {\bibfnamefont
  {C.}~\bibnamefont {Adams}}, \bibinfo {author} {\bibfnamefont {V.~B.}\
  \bibnamefont {Adya}}, \bibinfo {author} {\bibfnamefont {C.}~\bibnamefont
  {Affeldt}},\ and\ \bibinfo {author} {\bibnamefont {et~al.}},\ }\bibfield
  {title} {\bibinfo {title} {Prospects for observing and localizing
  gravitational-wave transients with advanced ligo, advanced virgo and kagra},\
  }\bibfield  {journal} {\bibinfo  {journal} {Living Reviews in Relativity}\
  }\textbf {\bibinfo {volume} {23}},\ \href
  {https://doi.org/10.1007/s41114-020-00026-9} {10.1007/s41114-020-00026-9}
  (\bibinfo {year} {2020})\BibitemShut {NoStop}%
\bibitem [{\citenamefont {Abbott}\ \emph
  {et~al.}(2017{\natexlab{d}})\citenamefont {Abbott}, \citenamefont {Abbott},
  \citenamefont {Abbott}, \citenamefont {Abernathy}, \citenamefont {Ackley},
  \citenamefont {Adams}, \citenamefont {Addesso}, \citenamefont {Adhikari},
  \citenamefont {Adya}, \citenamefont {Affeldt},\ and\ \citenamefont
  {et~al.}}]{Abbott_2017}%
  \BibitemOpen
  \bibfield  {author} {\bibinfo {author} {\bibfnamefont {B.~P.}\ \bibnamefont
  {Abbott}}, \bibinfo {author} {\bibfnamefont {R.}~\bibnamefont {Abbott}},
  \bibinfo {author} {\bibfnamefont {T.~D.}\ \bibnamefont {Abbott}}, \bibinfo
  {author} {\bibfnamefont {M.~R.}\ \bibnamefont {Abernathy}}, \bibinfo {author}
  {\bibfnamefont {K.}~\bibnamefont {Ackley}}, \bibinfo {author} {\bibfnamefont
  {C.}~\bibnamefont {Adams}}, \bibinfo {author} {\bibfnamefont
  {P.}~\bibnamefont {Addesso}}, \bibinfo {author} {\bibfnamefont {R.~X.}\
  \bibnamefont {Adhikari}}, \bibinfo {author} {\bibfnamefont {V.~B.}\
  \bibnamefont {Adya}}, \bibinfo {author} {\bibfnamefont {C.}~\bibnamefont
  {Affeldt}},\ and\ \bibinfo {author} {\bibnamefont {et~al.}},\ }\bibfield
  {title} {\bibinfo {title} {Exploring the sensitivity of next generation
  gravitational wave detectors},\ }\href
  {https://doi.org/10.1088/1361-6382/aa51f4} {\bibfield  {journal} {\bibinfo
  {journal} {Classical and Quantum Gravity}\ }\textbf {\bibinfo {volume}
  {34}},\ \bibinfo {pages} {044001} (\bibinfo {year}
  {2017}{\natexlab{d}})}\BibitemShut {NoStop}%
\bibitem [{\citenamefont {Abbott}\ \emph
  {et~al.}(2016{\natexlab{b}})\citenamefont {Abbott} \emph
  {et~al.}}]{LIGOScientific:2016vlm}%
  \BibitemOpen
  \bibfield  {author} {\bibinfo {author} {\bibfnamefont {B.~P.}\ \bibnamefont
  {Abbott}} \emph {et~al.} (\bibinfo {collaboration} {LIGO Scientific,
  Virgo}),\ }\bibfield  {title} {\bibinfo {title} {{Properties of the Binary
  Black Hole Merger GW150914}},\ }\href
  {https://doi.org/10.1103/PhysRevLett.116.241102} {\bibfield  {journal}
  {\bibinfo  {journal} {Phys. Rev. Lett.}\ }\textbf {\bibinfo {volume} {116}},\
  \bibinfo {pages} {241102} (\bibinfo {year} {2016}{\natexlab{b}})},\ \Eprint
  {https://arxiv.org/abs/1602.03840} {arXiv:1602.03840 [gr-qc]} \BibitemShut
  {NoStop}%
\bibitem [{\citenamefont {Punturo}\ \emph {et~al.}(2010)\citenamefont {Punturo}
  \emph {et~al.}}]{Punturo:2010zza}%
  \BibitemOpen
  \bibfield  {author} {\bibinfo {author} {\bibfnamefont {M.}~\bibnamefont
  {Punturo}} \emph {et~al.},\ }\bibfield  {title} {\bibinfo {title} {{The third
  generation of gravitational wave observatories and their science reach}},\
  }\href {https://doi.org/10.1088/0264-9381/27/8/084007} {\bibfield  {journal}
  {\bibinfo  {journal} {Class. Quant. Grav.}\ }\textbf {\bibinfo {volume}
  {27}},\ \bibinfo {pages} {084007} (\bibinfo {year} {2010})}\BibitemShut
  {NoStop}%
\bibitem [{\citenamefont {Schutz}(2011)}]{schutz_2011}%
  \BibitemOpen
  \bibfield  {author} {\bibinfo {author} {\bibfnamefont {B.~F.}\ \bibnamefont
  {Schutz}},\ }\bibfield  {title} {\bibinfo {title} {Networks of gravitational
  wave detectors and three figures of merit},\ }\href
  {https://doi.org/10.1088/0264-9381/28/12/125023} {\bibfield  {journal}
  {\bibinfo  {journal} {Classical and Quantum Gravity}\ }\textbf {\bibinfo
  {volume} {28}},\ \bibinfo {pages} {125023} (\bibinfo {year}
  {2011})}\BibitemShut {NoStop}%
\bibitem [{\citenamefont {Chen}\ \emph {et~al.}(2019)\citenamefont {Chen},
  \citenamefont {Vitale},\ and\ \citenamefont {Narayan}}]{Chen:2018omi}%
  \BibitemOpen
  \bibfield  {author} {\bibinfo {author} {\bibfnamefont {H.-Y.}\ \bibnamefont
  {Chen}}, \bibinfo {author} {\bibfnamefont {S.}~\bibnamefont {Vitale}},\ and\
  \bibinfo {author} {\bibfnamefont {R.}~\bibnamefont {Narayan}},\ }\bibfield
  {title} {\bibinfo {title} {{Viewing angle of binary neutron star mergers}},\
  }\href {https://doi.org/10.1103/PhysRevX.9.031028} {\bibfield  {journal}
  {\bibinfo  {journal} {Phys. Rev. X}\ }\textbf {\bibinfo {volume} {9}},\
  \bibinfo {pages} {031028} (\bibinfo {year} {2019})},\ \Eprint
  {https://arxiv.org/abs/1807.05226} {arXiv:1807.05226 [astro-ph.HE]}
  \BibitemShut {NoStop}%
\bibitem [{\citenamefont {Chatziioannou}\ \emph {et~al.}(2018)\citenamefont
  {Chatziioannou}, \citenamefont {Haster},\ and\ \citenamefont
  {Zimmerman}}]{Chatziioannou:2018vzf}%
  \BibitemOpen
  \bibfield  {author} {\bibinfo {author} {\bibfnamefont {K.}~\bibnamefont
  {Chatziioannou}}, \bibinfo {author} {\bibfnamefont {C.-J.}\ \bibnamefont
  {Haster}},\ and\ \bibinfo {author} {\bibfnamefont {A.}~\bibnamefont
  {Zimmerman}},\ }\bibfield  {title} {\bibinfo {title} {{Measuring the neutron
  star tidal deformability with equation-of-state-independent relations and
  gravitational waves}},\ }\href {https://doi.org/10.1103/PhysRevD.97.104036}
  {\bibfield  {journal} {\bibinfo  {journal} {Phys. Rev. D}\ }\textbf {\bibinfo
  {volume} {97}},\ \bibinfo {pages} {104036} (\bibinfo {year} {2018})},\
  \Eprint {https://arxiv.org/abs/1804.03221} {arXiv:1804.03221 [gr-qc]}
  \BibitemShut {NoStop}%
\bibitem [{\citenamefont {Andreoni}\ \emph {et~al.}(2022)\citenamefont
  {Andreoni} \emph {et~al.}}]{Andreoni:2021epw}%
  \BibitemOpen
  \bibfield  {author} {\bibinfo {author} {\bibfnamefont {I.}~\bibnamefont
  {Andreoni}} \emph {et~al.},\ }\bibfield  {title} {\bibinfo {title}
  {{Target-of-opportunity Observations of Gravitational-wave Events with Vera
  C. Rubin Observatory}},\ }\href {https://doi.org/10.3847/1538-4365/ac617c}
  {\bibfield  {journal} {\bibinfo  {journal} {Astrophys. J. Supp.}\ }\textbf
  {\bibinfo {volume} {260}},\ \bibinfo {pages} {18} (\bibinfo {year} {2022})},\
  \Eprint {https://arxiv.org/abs/2111.01945} {arXiv:2111.01945 [astro-ph.HE]}
  \BibitemShut {NoStop}%
\bibitem [{\citenamefont {Borhanian}(2020)}]{Borhanian:2020ypi}%
  \BibitemOpen
  \bibfield  {author} {\bibinfo {author} {\bibfnamefont {S.}~\bibnamefont
  {Borhanian}},\ }\bibfield  {title} {\bibinfo {title} {{gwbench: a novel
  Fisher information package for gravitational-wave benchmarking}},\
  }\href@noop {} {\  (\bibinfo {year} {2020})},\ \Eprint
  {https://arxiv.org/abs/2010.15202} {arXiv:2010.15202 [gr-qc]} \BibitemShut
  {NoStop}%
\bibitem [{\citenamefont {Vallisneri}(2008)}]{Vallisneri:2007ev}%
  \BibitemOpen
  \bibfield  {author} {\bibinfo {author} {\bibfnamefont {M.}~\bibnamefont
  {Vallisneri}},\ }\bibfield  {title} {\bibinfo {title} {{Use and abuse of the
  Fisher information matrix in the assessment of gravitational-wave
  parameter-estimation prospects}},\ }\href
  {https://doi.org/10.1103/PhysRevD.77.042001} {\bibfield  {journal} {\bibinfo
  {journal} {Phys. Rev. D}\ }\textbf {\bibinfo {volume} {77}},\ \bibinfo
  {pages} {042001} (\bibinfo {year} {2008})},\ \Eprint
  {https://arxiv.org/abs/gr-qc/0703086} {arXiv:gr-qc/0703086} \BibitemShut
  {NoStop}%
\bibitem [{\citenamefont {{Farah}}\ \emph {et~al.}(2020)\citenamefont
  {{Farah}}, \citenamefont {{Essick}}, \citenamefont {{Doctor}}, \citenamefont
  {{Fishbach}},\ and\ \citenamefont {{Holz}}}]{2020ApJ...895..108F}%
  \BibitemOpen
  \bibfield  {author} {\bibinfo {author} {\bibfnamefont {A.}~\bibnamefont
  {{Farah}}}, \bibinfo {author} {\bibfnamefont {R.}~\bibnamefont {{Essick}}},
  \bibinfo {author} {\bibfnamefont {Z.}~\bibnamefont {{Doctor}}}, \bibinfo
  {author} {\bibfnamefont {M.}~\bibnamefont {{Fishbach}}},\ and\ \bibinfo
  {author} {\bibfnamefont {D.~E.}\ \bibnamefont {{Holz}}},\ }\bibfield  {title}
  {\bibinfo {title} {{Counting on Short Gamma-Ray Bursts: Gravitational-wave
  Constraints of Jet Geometry}},\ }\href
  {https://doi.org/10.3847/1538-4357/ab8d26} {\bibfield  {journal} {\bibinfo
  {journal} {\apj}\ }\textbf {\bibinfo {volume} {895}},\ \bibinfo {eid} {108}
  (\bibinfo {year} {2020})},\ \Eprint {https://arxiv.org/abs/1912.04906}
  {arXiv:1912.04906 [astro-ph.HE]} \BibitemShut {NoStop}%
\bibitem [{\citenamefont {Calder\'on~Bustillo}\ \emph
  {et~al.}(2021)\citenamefont {Calder\'on~Bustillo}, \citenamefont {Leong},
  \citenamefont {Dietrich},\ and\ \citenamefont
  {Lasky}}]{CalderonBustillo:2020kcg}%
  \BibitemOpen
  \bibfield  {author} {\bibinfo {author} {\bibfnamefont {J.}~\bibnamefont
  {Calder\'on~Bustillo}}, \bibinfo {author} {\bibfnamefont {S.~H.~W.}\
  \bibnamefont {Leong}}, \bibinfo {author} {\bibfnamefont {T.}~\bibnamefont
  {Dietrich}},\ and\ \bibinfo {author} {\bibfnamefont {P.~D.}\ \bibnamefont
  {Lasky}},\ }\bibfield  {title} {\bibinfo {title} {{Mapping the Universe
  Expansion: Enabling Percent-level Measurements of the Hubble Constant with a
  Single Binary Neutron-star Merger Detection}},\ }\href
  {https://doi.org/10.3847/2041-8213/abf502} {\bibfield  {journal} {\bibinfo
  {journal} {Astrophys. J. Lett.}\ }\textbf {\bibinfo {volume} {912}},\
  \bibinfo {pages} {L10} (\bibinfo {year} {2021})},\ \Eprint
  {https://arxiv.org/abs/2006.11525} {arXiv:2006.11525 [gr-qc]} \BibitemShut
  {NoStop}%
\bibitem [{\citenamefont {Tan}\ \emph {et~al.}(2022)\citenamefont {Tan},
  \citenamefont {Dexheimer}, \citenamefont {Noronha-Hostler},\ and\
  \citenamefont {Yunes}}]{Tan:2021nat}%
  \BibitemOpen
  \bibfield  {author} {\bibinfo {author} {\bibfnamefont {H.}~\bibnamefont
  {Tan}}, \bibinfo {author} {\bibfnamefont {V.}~\bibnamefont {Dexheimer}},
  \bibinfo {author} {\bibfnamefont {J.}~\bibnamefont {Noronha-Hostler}},\ and\
  \bibinfo {author} {\bibfnamefont {N.}~\bibnamefont {Yunes}},\ }\bibfield
  {title} {\bibinfo {title} {{Finding Structure in the Speed of Sound of
  Supranuclear Matter from Binary Love Relations}},\ }\href
  {https://doi.org/10.1103/PhysRevLett.128.161101} {\bibfield  {journal}
  {\bibinfo  {journal} {Phys. Rev. Lett.}\ }\textbf {\bibinfo {volume} {128}},\
  \bibinfo {pages} {161101} (\bibinfo {year} {2022})},\ \Eprint
  {https://arxiv.org/abs/2111.10260} {arXiv:2111.10260 [astro-ph.HE]}
  \BibitemShut {NoStop}%
\end{thebibliography}
%

\end{document}